%% file: preprint_main.tex
\def\artifactpreprintlayout{1}
\documentclass{artifactpreprint}

\usepackage{tikz}
\usetikzlibrary{trees,positioning,arrows.meta,calc,shapes.geometric,decorations.pathreplacing,fit,backgrounds,shadows}
\usepackage{multicol}
\usepackage{pifont}
\usepackage{fontawesome5}
\usepackage{colortbl}
\usepackage{float}
\usepackage{placeins}

\definecolor{cPlan}{HTML}{FFF3E0}
\definecolor{cExec}{HTML}{E3F2FD}
\definecolor{cObs}{HTML}{F3E5F5}
\definecolor{cGoal}{HTML}{FFF8E1}
\definecolor{cArtifact}{HTML}{E8F5E9}
\definecolor{cMem}{HTML}{FFFDE7}
\definecolor{cHard}{HTML}{C8E6C9}
\definecolor{cSoft}{HTML}{FFCDD2}
\definecolor{cAccent}{HTML}{1565C0}
\definecolor{cAccent2}{HTML}{00897B}
\definecolor{cDark}{HTML}{263238}
\definecolor{cLight}{HTML}{F5F5F5}
\definecolor{cWarn}{HTML}{FF8F00}
\definecolor{cRowAlt}{HTML}{EFF5FB}

\tikzset{
  survey-box/.style={draw=gray!50, rounded corners=5pt, minimum height=12mm, minimum width=26mm,
    align=center, font=\small, line width=0.6pt,
    drop shadow={shadow xshift=0.4mm, shadow yshift=-0.4mm, fill=gray!15, opacity=0.4}},
  survey-step/.style={draw=gray!45, rounded corners=4pt, minimum height=10mm, minimum width=20mm,
    align=center, font=\scriptsize, line width=0.5pt,
    drop shadow={shadow xshift=0.3mm, shadow yshift=-0.3mm, fill=gray!12, opacity=0.35}},
  survey-tool/.style={draw=cAccent!35, rounded corners=4pt, fill=cExec, minimum height=10mm,
    minimum width=20mm, align=center, font=\scriptsize, line width=0.5pt,
    drop shadow={shadow xshift=0.3mm, shadow yshift=-0.3mm, fill=gray!12, opacity=0.35}},
  survey-arr/.style={-{Stealth[length=2.5mm, width=2mm]}, line width=0.7pt, gray!70},
  survey-bigarr/.style={-{Stealth[length=3mm, width=2.5mm]}, line width=1pt, cAccent!65},
  survey-feedback/.style={-{Stealth[length=2mm, width=1.5mm]}, line width=0.5pt, dashed, cAccent2!55},
  survey-seclbl/.style={font=\footnotesize\bfseries\sffamily, text=cDark},
  survey-prop/.style={font=\scriptsize, rounded corners=3pt, inner sep=3pt, line width=0.4pt},
}

\newcommand{\cmark}{\textcolor{cAccent2}{\ding{51}}}

\newcommand{\affiliationlogo}[1]{%
  \raisebox{-0.18ex}{\includegraphics[height=0.33cm]{#1}}%
  \nobreak\hspace{0.27em}%
}

\title{Agentic Artifact Creation: Systems, Evaluation, Principles, and Opportunities}

\ifdefined\preprintanonymous
\author{Anonymous Authors}
\preprintmeta{Anonymous Manuscript}
\else
\author{%
  Tianfu Wang\textsuperscript{1,\textdagger} \and Zhezheng Hao\textsuperscript{2,\textdagger} \and
  Xilin Xia\textsuperscript{3,\textdagger} \and Lixin Liu\textsuperscript{4} \and
  Mengkang Hu\textsuperscript{5} \and Hongzhang Liu\textsuperscript{6}\\[0.2em]
  Xi Chen\textsuperscript{3} \and Ziyan Liu\textsuperscript{3} \and
  Xiankun Lin\textsuperscript{7} \and Weijia Zhang\textsuperscript{1} \and
  Nicholas Jing Yuan\textsuperscript{1,*} \and
  Hui Xiong\textsuperscript{1,*}
}

\preprintaffiliations{%
  {\fontsize{10}{12}\selectfont
    \textsuperscript{1}\affiliationlogo{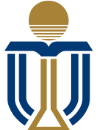}The Hong Kong University of Science and Technology (Guangzhou)%
    \hspace{0.75em}\textsuperscript{2}\affiliationlogo{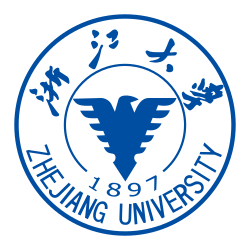}Zhejiang University\\[0.15em]
    \textsuperscript{3}\affiliationlogo{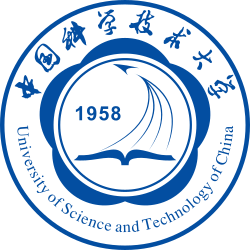}University of Science and Technology of China%
    \hspace{0.75em}\textsuperscript{4}\affiliationlogo{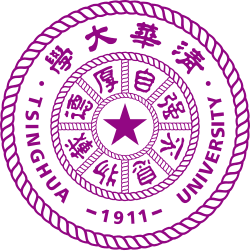}Tsinghua University\\[0.15em]
    \textsuperscript{5}\affiliationlogo{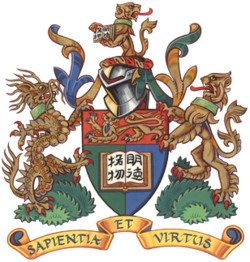}The University of Hong Kong%
    \hspace{0.75em}\textsuperscript{6}\affiliationlogo{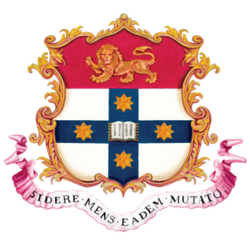}The University of Sydney%
    \hspace{0.75em}\textsuperscript{7}\affiliationlogo{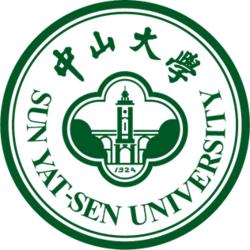}Sun Yat-sen University
  }\\[0.5em]
  {\fontsize{10}{12}\selectfont
    \textsuperscript{\textdagger}Equal contribution
    \quad \textsuperscript{*}Corresponding authors
    \quad \textcolor{preprintAccentTwo}{\faEnvelope}\hspace{0.35em}%
    \href{mailto:tianfuwang.cs@gmail.com}{tianfuwang.cs@gmail.com},
    \href{mailto:xionghui@ust.hk}{xionghui@ust.hk}%
  }
}
\fi

\preprintlinks{%
  {\fontsize{10}{12}\selectfont
    \textcolor{preprintAccentTwo}{\faHome}\enspace
    \textbf{Project Website:}\enspace
    \href{https://agentic-creation.github.io}{agentic-creation.github.io}
    \qquad
    \textcolor{preprintAccentTwo}{\faGithub}\enspace
    \textbf{GitHub Repo:}\enspace
    \href{https://github.com/GeminiLight/awesome-agentic-artifact-creation}{awesome-agentic-artifact-creation}%
  }
}

\preprintbrand{\includegraphics[height=1.30cm,trim=100 100 100 100,clip]{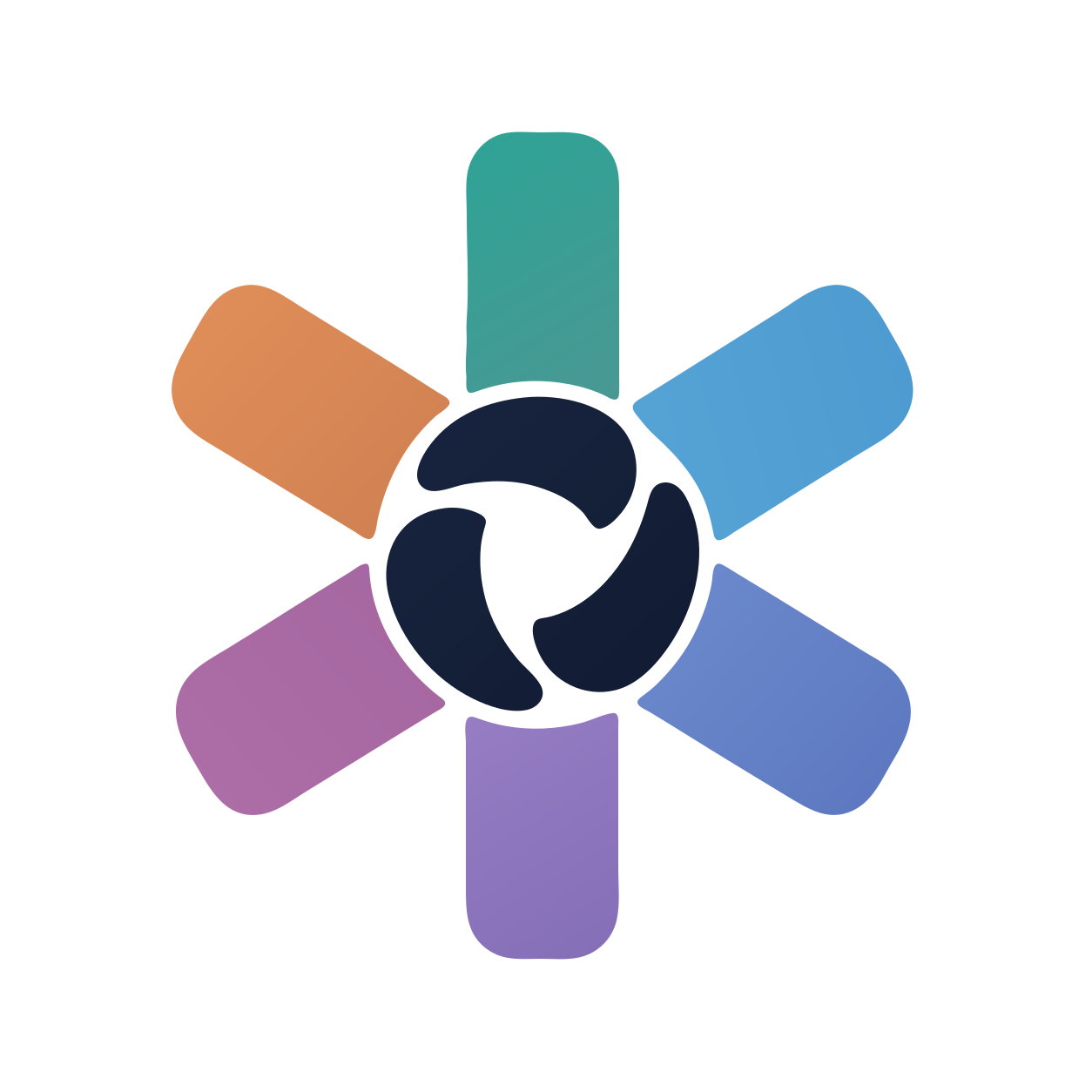}}

\preprintabstract{\input{sections/abstract}}

\ifdefined\preprintanonymous
\hypersetup{
  pdftitle={Agentic Artifact Creation: Systems, Evaluation, Principles, and Opportunities},
  pdfauthor={Anonymous Authors},
  pdfsubject={Anonymous manuscript on agentic artifact creation},
}
\else
\hypersetup{
  pdftitle={Agentic Artifact Creation: Systems, Evaluation, Principles, and Opportunities},
  pdfauthor={Tianfu Wang et al.},
  pdfsubject={Survey of agentic artifact creation},
}
\fi

\begin{document}

\maketitle

\input{sections/01_introduction}
\input{sections/02_background}
\input{sections/02_formalizing}
\input{sections/03_design_space}
\input{sections/04_landscape}
\input{sections/04b_applications}
\input{sections/04c_verification}
\input{sections/05_methodology}
\input{sections/06_challenges}
\input{sections/08_conclusion}

\bibliographystyle{unsrtnat-no-url}
\bibliography{references}

\end{document}

%% file: sections/abstract.tex
Generative models can turn natural-language prompts into images, text, code, and other content, lowering the cost of producing drafts and components.
Their practical impact increasingly depends on whether those pieces can become complete, dependable deliverables.
Such deliverables pose a different challenge because their requirements interact, and failures visible in the final output may be difficult to trace or repair.
This survey examines \emph{agentic artifact creation}, which we define as stateful construction in which an AI system materially constructs or revises a deliverable and intermediate observations redirect later work.
Functionally, the process links an operational representation of the artifact, a construction policy, and runtime verification whose feedback can redirect later actions.
This control structure can expose dependencies and support targeted revision, but only when observations identify failures at a scope that the available actions can repair.
We reviewed 259 works available through August 20, 2026: 230 systems meeting this definition and 29 benchmarks of agentic artifact construction.
We compare six artifact families, then analyze application settings and evaluation practice as separate dimensions.
Across families, construction challenges reflect not only modality but also how tightly decisions are coupled and whether failures become visible while they remain repairable.
Decomposition can reduce local complexity while increasing coordination and reassembly costs.
Learned judges may add little independent evidence when they share the generator's preferences or blind spots.
We formulate principles for keeping commitments and responsibility explicit, turning feedback into targeted repair, and revalidating affected state after change.
We also identify challenges and opportunities in sustaining coherent, accountable control as artifacts, creator intent, and construction systems evolve, particularly when failures are difficult to diagnose or several outcomes may be valid.
We maintain a curated list of papers on agentic artifact creation \anon[as anonymized supplementary material]{at \url{https://github.com/GeminiLight/awesome-agentic-artifact-creation}}.

%% file: sections/01_introduction.tex
\section{Introduction}
\label{sec:intro}

\subsection{From Direct Generation to Agentic Creation}

Generative models can now produce a wide range of content from natural-language prompts~\citep{Rombach2022_StableDiffusion,OpenAI2023_GPT4}.
By lowering the cost of producing drafts and components, this capability is entering a wider range of creative and professional workflows; controlled and field studies already report productivity gains in professional writing and customer support~\citep{Noy2023_GenAIProductivity,Brynjolfsson2025_GenerativeAIAtWork}.
In these settings, success increasingly depends on complete artifacts whose parts must work together rather than isolated outputs.
Here, \emph{artifact} denotes an intentionally produced, purpose-directed deliverable~\citep{Preston2018Artifact}.
Direct generation works well for bounded tasks whose outputs are easy to inspect and inexpensive to regenerate.
It is less reliable when the deliverable is governed by several acceptance criteria.
Paper2Poster~\citep{Poster_Paper2Poster2025}, for example, illustrates how a scientific poster must preserve source content, fit a constrained page, and communicate the paper clearly.
These requirements call for different kinds of evidence; a plausible image or scalar quality score cannot establish that the poster is ready to deliver.

Such deliverables are difficult to construct because their decisions are interdependent, so changing content can alter layout, behavior, or downstream constraints.
Their validity is also only partially observable because checks cover different criteria, may arrive late, and can become stale after revision.
SWE-bench~\citep{Code_SWEBench2024} and SWE-agent~\citep{Code_SWEAgent2024} make this concrete for software: agents must edit repository state and interpret test and execution feedback rather than emit isolated code.
Even when a check detects a failure, broad edit operations may make local repair impossible.
The next useful action therefore depends on the current state, available evidence, and feasible repair scope.
A construction process must expose the state and evidence needed to localize failures and direct revision before delivery.

Recent language-model agents~\citep{Gao2026_SelfEvolvingAgents} combine tool-mediated reasoning in ReAct~\citep{Yao2023_ReAct}, iterative feedback in Self-Refine~\citep{Madaan2023_SelfRefine} and Reflexion~\citep{Shinn2024_Reflexion}, and specialized construction roles in MetaGPT~\citep{Hong2024_MetaGPT} and ChatDev~\citep{Qian2024_ChatDev}.
These mechanisms support stateful construction, but we reserve \emph{Agentic Artifact Creation} for episodes in which an AI system materially constructs or revises a deliverable and intermediate observations redirect later work.
Figure~\ref{fig:paradigm} visualizes the resulting shift from a one-way generation pipeline to a broader construction process in which generation becomes one possible operation.

\input{figures/generation_vs_creation_image}

\ifdefined\artifactpreprintlayout\else
\input{figures/corpus_overview}
\fi

\subsection{Functional Overview of Agentic Creation}

Figure~\ref{fig:paradigm} presents \emph{Agentic Artifact Creation} as a recurrent process in which observations of intermediate results are used to redirect later construction decisions.
Three functional roles make this redirection operational.
The \emph{Operational Representation} carries the current artifact state across steps and exposes the edits that can be applied to it.
The \emph{Construction Policy} interprets task requirements in light of that state and accumulated feedback, then selects what to change, where to act, or whether to continue.
\emph{Runtime Verification} examines the consequences of each change against relevant criteria and turns observations into feedback for the policy.
Later decisions can therefore draw on both the artifact's current state and what earlier actions revealed.

These functions become most useful when they describe, edit, and assess the same construction units.
A representation that separates meaningful parts allows them to be produced and coordinated independently, supporting \emph{Composability}.
Records that connect requirements, actions, artifact states, and observations provide \emph{Traceability} across construction.
When verification associates a failure with the affected parts, the policy can revise those parts and recheck their dependencies without discarding accepted work, supporting \emph{Revisability}.
Together, these affordances support controlled construction over persistent, inspectable state.

\subsection{Review Scope and Method}

The operational boundary also defines the survey scope.
We include systems meeting the definition, including mixed-initiative systems in which human input changes later system-mediated artifact work~\citep{Yannakakis2014_MixedInitiativeDesign,Liapis2016_MixedInitiative}, and exclude generic problem-solving agents and assistants that only consume or evaluate artifacts.
Prior surveys typically organize models and media, agent capabilities, domains, or processes.
As Table~\ref{tab:prior-surveys} shows, the Delivered Artifact is not their primary unit of analysis; Section~\ref{sec:background} develops this positioning.

\begin{samepage}
This gap motivates four review questions.

\begin{enumerate}
  \item[\textbf{RQ1}] What distinguishes agentic artifact creation, and how is its construction organized? (\S\ref{sec:formalizing})
  \item[\textbf{RQ2}] How do deliverable requirements and application contexts shape construction? (\S\S\ref{sec:landscape}--\ref{sec:applications})
  \item[\textbf{RQ3}] How is agentic artifact creation evaluated through evidence and protocols? (\S\ref{sec:verification})
  \item[\textbf{RQ4}] What principles guide agentic creation across artifacts and applications? (\S\ref{sec:patterns})
\end{enumerate}
\end{samepage}

We searched arXiv, Google Scholar, Semantic Scholar, ACM DL, and IEEE Xplore from January 2023 through August 20, 2026 using agentic-process terms, including \emph{agentic creation}, \emph{iterative artifact generation}, and \emph{tool-augmented generation}, crossed with terms for the six artifact families; citation tracing and targeted venue audits supplemented retrieval.
After title- and identifier-based deduplication, candidates were screened against this stateful-construction definition.
We retained 230 systems and 29 construction benchmarks; direct generators, observation-independent workflows, and artifact-consuming or evaluation-only systems were excluded.
Records were coded by corpus role, primary family, subtype, cross-cutting facets, and provenance, with ambiguous cases reconciled iteratively.
Database-specific query exports, per-source yields, and pre-reconciliation labels were not retained; released files document the resulting coding and limitations.

\ifdefined\artifactpreprintlayout
\input{figures/corpus_overview}
\fi

\subsection{Our Contributions and Organization}

This survey makes four contributions.
\begin{itemize}
  \item We characterize \emph{Agentic Artifact Creation} as stateful construction and provide a functional model for analyzing how intermediate evidence guides revision.
  \item We assemble and release a coded corpus of more than 200 works and compare six artifact families, highlighting recurring differences in decision interdependence, failure observability, and repairability.
  \item We analyze application context and evaluation independently of artifact family, separating evidence about the delivered artifact from evidence about the trajectory and system.
  \item We synthesize four principles for inspectable construction control and use the problems they expose to structure a research agenda.
\end{itemize}

The remainder of the survey proceeds from foundations and the agentic creation paradigm (Sections~\ref{sec:background}--\ref{sec:formalizing}) to artifact families, applications, and evaluation (Sections~\ref{sec:landscape}--\ref{sec:verification}), followed by cross-cutting principles, challenges and opportunities, and conclusions (Sections~\ref{sec:patterns}--\ref{sec:conclusion}).

%% file: figures/generation_vs_creation_image.tex
\begin{figure}[!t]
  \centering
  \includegraphics[width=\textwidth]{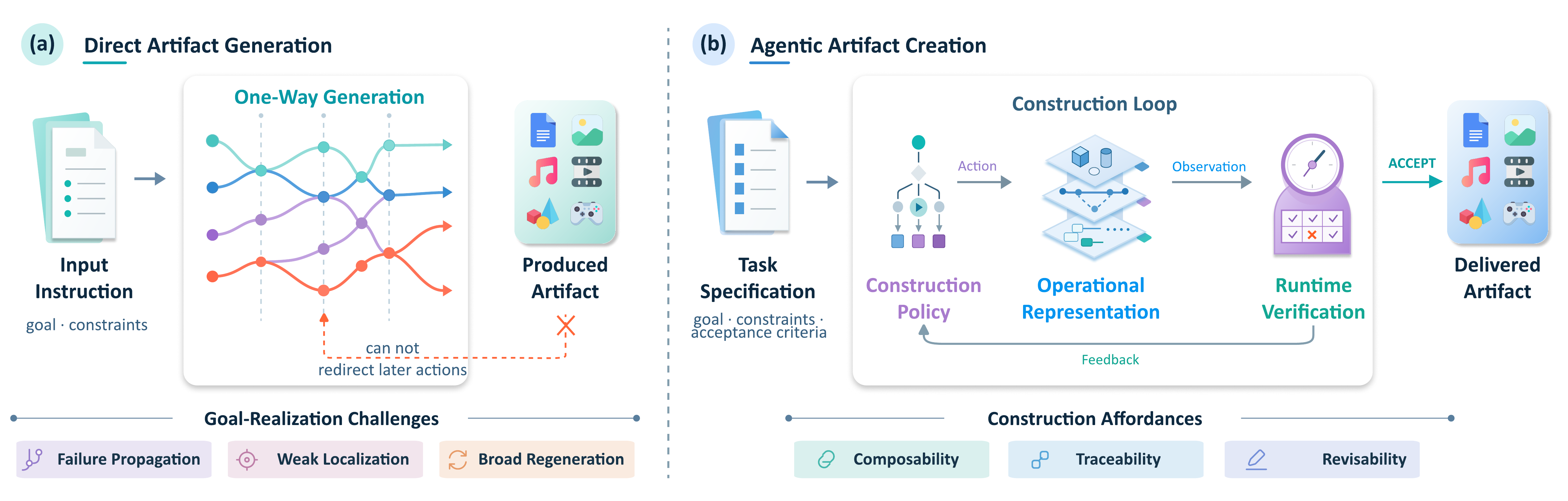}
  \caption{From Direct Artifact Generation to Agentic Artifact Creation.
  \textbf{(a)}~Direct generation follows a fixed control schedule in which intermediate observations do not redirect later actions, leaving final inspection to expose failure propagation, weak localization, and broad regeneration.
  \textbf{(b)}~Agentic creation couples a Construction Policy, an Operational Representation, and Runtime Verification so feedback can redirect later decisions before acceptance. Suitable construction units, interfaces, and records may support composability, traceability, and revisability.}
  \Description{A two-panel comparison. The left panel shows an Input Instruction entering a fixed one-way generation schedule represented by intertwined dependency paths and producing a Produced Artifact. Intermediate observations cannot redirect later actions; a final inspection can trigger regeneration. A lower annotation row identifies failure propagation, weak localization, and broad regeneration as goal-realization challenges. The right panel shows a Task Specification entering a construction loop composed of a Construction Policy, an Operational Representation, and Runtime Verification. The policy issues an Action to the representation, an Observation enters Runtime Verification, and a lower Feedback path returns to the policy. An Accept branch releases the Delivered Artifact. A lower annotation row presents composability, traceability, and revisability as potential construction affordances.}
  \label{fig:paradigm}
\end{figure}

%% file: figures/corpus_overview.tex
\begin{figure}[!t]
  \centering
  \includegraphics[width=\textwidth]{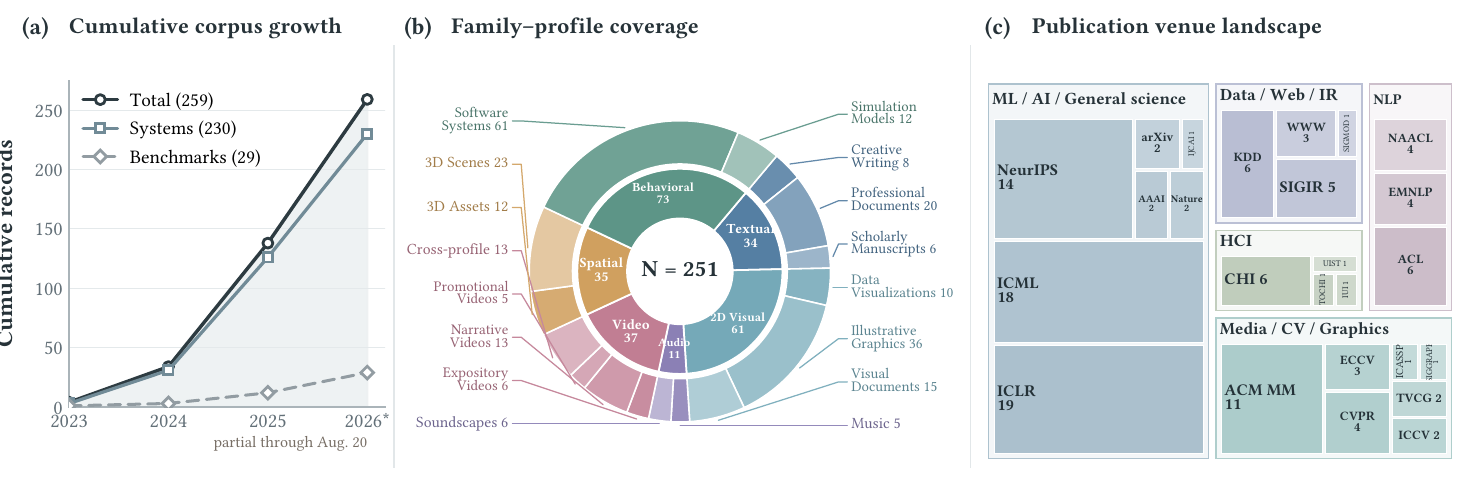}
  \caption{Corpus overview after the August 28, 2026 accounting update.
  \textbf{(a)}~Cumulative strict construction records by year, with 2026 partial through August 20.
  \textbf{(b)}~Family and Chapter~4 profile coverage for 251 family-assigned records. Thirteen video records remain Cross-profile, Spoken Audio has no strict record, and eight application-workflow systems are omitted from this view.
  \textbf{(c)}~Community and normalized-venue composition of the 120 published or accepted strict records; rectangle area encodes record count.}
  \Description{A three-panel corpus overview. The left panel shows cumulative growth from 2023 through a partial 2026, ending at 259 strict records: 230 systems and 29 benchmarks. The center panel is a nested radial view of 251 family-assigned records across textual, 2D visual, audio, video, spatial, and behavioral artifact families and the current Chapter 4 analytical profiles, with a separate cross-profile residual for video records that are not assigned a primary communicative-purpose profile. The right panel is a treemap of 120 published or accepted strict records grouped into ML and AI or general science, multimedia and graphics, data and web, natural language processing, and human-computer interaction communities, with every venue tile labeled by name and count.}
  \label{fig:corpus-overview}
\end{figure}

%% file: sections/02_background.tex
\section{Background}
\label{sec:background}

Agentic artifact creation builds on generative foundation models for content operations, agentic AI architectures for decision-making and tool use, and prior research on artifact production for explicit representations, constraints, and feedback.
We review these foundations before positioning our artifact-centered perspective against adjacent surveys.

Figure~\ref{fig:survey-organization} previews the survey's progression from these foundations and the agentic paradigm to artifact families, applications, evaluation, principles, and opportunities.
\input{figures/survey_organization}

\subsection{Generative Foundation Models}
\label{sec:rw-generation}

Generative foundation models provide reusable operations for generating, editing, and converting content across language, code, images, video, audio, and 3D~\citep{OpenAI2023_GPT4,Cao2024_AIGCSurvey,Zhu2024_VideoGenSurvey,Li2024_3DGenSurvey}.
The result of a content operation may be delivered directly or incorporated into an intermediate artifact for later inspection and revision.
Foundation models supply content operations, but they do not by themselves maintain construction state, select the next action, incorporate observations, or decide when the artifact is complete.
These responsibilities motivate the agentic control mechanisms reviewed next.

\subsection{Agentic AI Architectures}
\label{sec:rw-agents}

An LLM-based agent places control around one or more foundation models for planning, state or memory, tool use, observation, and stopping~\citep{Xi2023_RiseAndPotential,Wang2024_AutonomousAgentsSurvey,Masterman2024_AgentLandscape}.
ReAct~\citep{Yao2023_ReAct} interleaves tool-grounded action and observation, while Reflexion~\citep{Shinn2024_Reflexion} retains feedback across attempts.
Control may remain with one agent, pass through a central orchestrator, or be divided among specialized peers, with MetaGPT and ChatDev illustrating role division among agents~\citep{Hong2024_MetaGPT,Qian2024_ChatDev,Guo2024_MultiAgentSurvey}.
The consequential choices are what state is shared, who selects later actions, how observations shape control, and which decisions remain under human authority.
General agents may answer questions, use tools, or coordinate roles without materially constructing a delivered artifact.
For agentic artifact creation, these mechanisms must preserve artifact state and carry observations into later artifact-level actions or stopping; the unit of analysis is the delivered artifact and its acceptance criteria, not the number of agents.

\subsection{Artifact Production Processes}
\label{sec:rw-processes}

Earlier research on artifact production already made representation, action, feedback, and human authority explicit through domain-specific mechanisms.
Procedural content generation searches candidate game content under fitness or feasibility criteria~\citep{Togelius2011_PCG,Shaker2016_PCGBook,Game_ProceduralContent2024}.
High Dimensional Procedural Content Generation~\citep{HDPCG2025} expands the search space with gameplay-relevant dimensions, while LogicEnvGen~\citep{LogicEnvGen2025} checks the physical plausibility of generated environments.
Program synthesis and sketch-based synthesis connect specifications and partial artifacts to completion constraints~\citep{SolarLezama2008_Sketch,Gulwani2017_ProgramSynthesis}, while shape grammars express visual design through compositional rules~\citep{Stiny1980_ShapeGrammars}.
Across these approaches, specifications, structured action spaces, and partial artifacts connect global goals to local edits.
Mixed-initiative co-creation places human judgment inside iterative processes for game and visual design~\citep{Yannakakis2014_MixedInitiativeDesign,Liapis2016_MixedInitiative}.
The creator can redirect intermediate choices, preserve commitments, and decide which requirements need more work, instead of supplying only an initial prompt or final score.
Across these fields, represented artifacts and feedback influence later production decisions, providing domain-specific precedents while leaving open which relations transfer across artifact families.

\input{sections/related_surveys_table}

\subsection{Positioning Against Adjacent Surveys}
\label{sec:rw-positioning}

Adjacent surveys organize the field around different primary objects.
General agent surveys emphasize architectures, planning, memory, and collaboration~\citep{Xi2023_RiseAndPotential,Wang2024_AutonomousAgentsSurvey,Guo2024_MultiAgentSurvey,Masterman2024_AgentLandscape,Tran2025_MultiAgentCollabSurvey}, while broad generative-AI surveys organize model families and media~\citep{Cao2024_AIGCSurvey}.
Closer neighbors center creative refinement~\citep{Lin2025_CreativeMASSurvey}, software construction and repair~\citep{Liu2026_AgentsSESurvey}, multi-tool trajectories~\citep{Xu2026_MultiToolOrchestrationSurvey}, human feedback and control~\citep{Zou2026_HumanAgentSurvey}, or agent evaluation~\citep{Yehudai2026_AgentEvaluationSurvey}.
Modality-specific generation surveys organize model families and generation tasks~\citep{Zhu2024_VideoGenSurvey,Li2024_3DGenSurvey}, while agentic reinforcement learning and agentic world modeling center learned behavior and predictive environment models~\citep{Zhang2026_AgenticRLSurvey,Chu2026_AgenticWorldModeling}.
Our organizing unit is the Delivered Artifact.
This unit makes acceptance criteria, editable state, observations, and repair scope comparable across artifact families.
Table~\ref{tab:prior-surveys} positions our artifact-centered lens against these emphases.

%% file: figures/survey_organization.tex
\begin{figure}[!t]
  \centering
  \includegraphics[width=\textwidth]{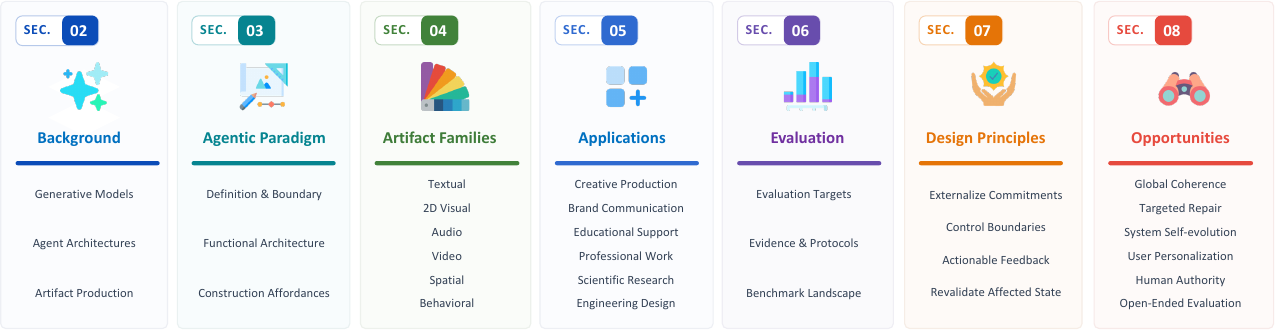}
  \caption{Survey organization. The survey moves from background and the agentic paradigm (\S\S~2--3) through artifact families, applications, and evaluation (\S\S~4--6), then distills design principles (\S~7) and six challenge--opportunity directions (\S~8).}
  \Description{A single-row organization map with seven rounded panels labeled Sections 2 through 8: Background, Agentic Paradigm, Artifact Families, Applications, Evaluation, Design Principles, and Opportunities. Each panel lists its major subsections, and the final panel lists six challenge--opportunity directions.}
  \label{fig:survey-organization}
\end{figure}

%% file: sections/related_surveys_table.tex
\newcommand{\partialcoverage}{%
  \tikz[baseline=-0.25ex]{%
    \fill (0,0.5ex) arc[start angle=90,end angle=270,radius=0.5ex] -- cycle;
    \draw[line width=0.3pt] (0,0) circle[radius=0.5ex];
  }%
}
\begin{table}[!t]
\caption{Comparison of adjacent surveys by organizing emphasis and scope.}
\label{tab:prior-surveys}
\centering
\small
\setlength{\tabcolsep}{3.0pt}
\renewcommand{\arraystretch}{1.14}
\resizebox{\textwidth}{!}{%
\begin{tabular}{
  >{\raggedright\arraybackslash}p{3.35cm}
  >{\raggedright\arraybackslash}p{3.45cm}
  >{\raggedright\arraybackslash}p{3.05cm}
  >{\centering\arraybackslash}p{2.45cm}
  >{\centering\arraybackslash}p{2.15cm}
  >{\raggedright\arraybackslash}p{3.20cm}
}
\toprule
\multicolumn{1}{c}{\shortstack[c]{\textbf{Survey}\\\textbf{Work}}} &
\multicolumn{1}{c}{\shortstack[c]{\textbf{Primary}\\\textbf{Lens}}} &
\multicolumn{1}{c}{\shortstack[c]{\textbf{Artifact}\\\textbf{Scope}}} &
\multicolumn{1}{c}{\shortstack[c]{\textbf{Construction}\\\textbf{State}}} &
\multicolumn{1}{c}{\shortstack[c]{\textbf{Feedback}\\\textbf{Control}}} &
\multicolumn{1}{c}{\shortstack[c]{\textbf{Evaluation}\\\textbf{Level}}} \\
\midrule
\citet{Cao2024_AIGCSurvey} & Model families & Multimodal content & -- & -- & Artifact \\
\rowcolor{cRowAlt}
\citet{Wang2024_AutonomousAgentsSurvey} & Agent architectures & General tasks & \partialcoverage & \partialcoverage & Task \\
\addlinespace[2pt]
\citet{Lin2025_CreativeMASSurvey} & Creative workflows & Text and image & \partialcoverage & \partialcoverage & Artifact \\
\rowcolor{cRowAlt}
\citet{Liu2026_AgentsSESurvey} & Software agents & Software & \partialcoverage & \partialcoverage & Artifact \\
\addlinespace[2pt]
\citet{Xu2026_MultiToolOrchestrationSurvey} & Tool trajectories & Digital tasks & \(\bullet\) & \(\bullet\) & Trajectory \\
\rowcolor{cRowAlt}
\citet{Zou2026_HumanAgentSurvey} & Human collaboration & General tasks & -- & \(\bullet\) & Interaction \\
\citet{Yehudai2026_AgentEvaluationSurvey} & Evaluation methods & Agent tasks & \partialcoverage & -- & Multiple levels \\
\midrule
\rowcolor{cAccent!10}
\textbf{Ours} & \textbf{Delivered Artifact} & \textbf{Six families} & \(\boldsymbol{\bullet}\) & \(\boldsymbol{\bullet}\) & \textbf{Multiple levels} \\
\bottomrule
\end{tabular}
}
\vspace{2pt}

\parbox{0.98\linewidth}{\centering\footnotesize \(\bullet\) central organizing axis; \partialcoverage\ partial coverage; -- not an organizing axis.\par}
\end{table}

%% file: sections/02_formalizing.tex
\section{The Agentic Creation Paradigm}
\label{sec:formalizing}
\label{sec:paradigm}

Agentic Artifact Creation treats the delivered artifact as state that can be inspected and changed during production.
Its defining property is that represented artifact or process state guides later decisions, irrespective of the model family or number of model calls.

\subsection{Conceptual Definition}

To apply this distinction consistently across artifact families, we define the paradigm at the level of a single artifact-creation episode.
This boundary specifies the minimum control behavior without prescribing a particular system architecture.
A \emph{creative artifact} is an artifact whose content, form, behavior, or arrangement permits multiple goal-satisfying realizations.
Agentic creation therefore characterizes the construction episode, not a separate artifact type.

\begin{quote}
\textbf{Definition} (\emph{Agentic Artifact Creation}).
An artifact-creation episode counts as Agentic Artifact Creation when an AI system materially constructs or revises the artifact to be delivered, carries artifact or process state across construction decisions, and uses at least one intermediate observation to redirect subsequent artifact-related work.
This redirection may change the next action, revision target, active branch, or stopping decision.
\end{quote}

Figure~\ref{fig:paradigm} contrasts a direct pipeline, whose observations do not redirect later actions, with Agentic Artifact Creation, which uses represented state and observations to redirect work before or after failure.
This stateful-construction boundary forms the screening rule used in our review.
Iterative and mixed-initiative agents fall within this definition when these conditions hold; the next subsection decomposes the required behavior into three functional roles.

\subsection{Functional Architecture}
\label{sec:hard-soft}
\label{sec:construction-loop}

At the functional level, Agentic Artifact Creation is organized by three roles that form a recurrent construction process.
The relations below anchor each role to a specific transition without prescribing its implementation.

Figure~\ref{fig:loop} instantiates these roles and the objects exchanged among them.
\input{figures/construction_loop_imagegen}

\emph{Operational Representation.}
At step $t$, the Operational Representation exposes the current artifact-side state $R_t$ through an Intermediate Form and provides the Edit Interface by which that state can be changed.
Applying the selected Action $a_t$ through this interface yields $R_{t+1} = U(R_t,a_t)$, where $U$ denotes the resulting state update.
The representation therefore determines what persists across steps and what changes can be expressed.

\emph{Construction Policy.}
Given the Task Specification $T$, current representation $R_t$, and available Feedback $f_t$, the policy $\pi$ selects the next Action as $a_t = \pi(T,R_t,f_t)$.
This decision may specify a revision target, operation, work allocation, branch, or stopping choice.

\emph{Runtime Verification.}
Its role is summarized by $f_{t+1} = V(T,R_{t+1},o_{t+1})$.
Here, $o_{t+1}$ is an Observation of the updated state, and $V$ evaluates the available evidence against the task's acceptance criteria.
If it accepts $R_\tau$ at step $\tau$, the delivery operation $D$ releases the Delivered Artifact as $A = D(R_\tau)$; otherwise, Feedback informs revision, escalation, or termination through the Construction Policy.

Policy selection, interface-mediated update, and verification repeat until acceptance or termination.
External evaluation protocols are examined separately in Section~\ref{sec:verification}.

\label{sec:design-space}
\label{sec:analytical-views}
\label{sec:taxonomy}
\label{sec:design-dims}
Concrete realizations vary in their intermediate forms and edit interfaces, decision control and agent topology, and observation sources and feedback functions.
Artifact family, external evaluation protocol, and application context remain separate axes of analysis.

%% file: figures/construction_loop_imagegen.tex
\begin{figure}[!t]
  \centering
  \includegraphics[width=\textwidth]{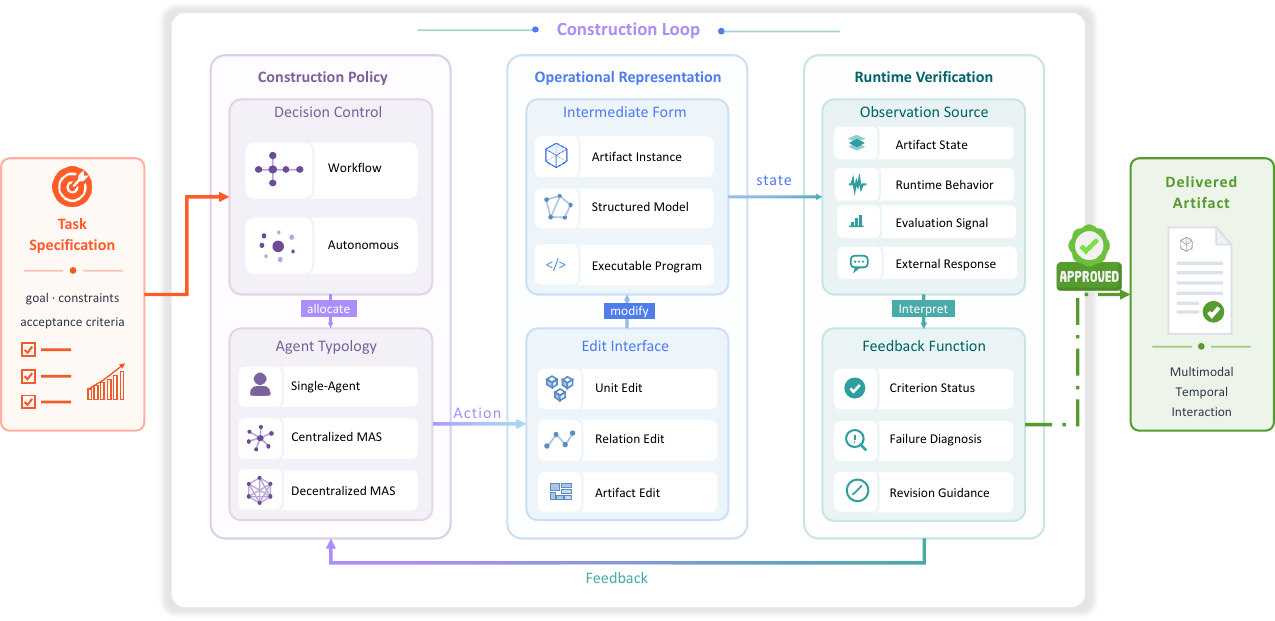}
  \caption{Functional architecture of Agentic Artifact Creation. A Task Specification informs Decision Control, while Agent Topology allocates selected work. Actions modify the Intermediate Form through the Operational Representation's Edit Interface. Runtime Verification turns observations into feedback, which returns to the Construction Policy unless Acceptance releases the Delivered Artifact. Component interiors show recurring, nonexclusive realizations.}
  \Description{A wide component-anatomy diagram. A Task Specification enters Decision Control in the Construction Policy. Decision Control contains Workflow and Autonomous modes, while Agent Topology contains Single-Agent, Centralized Multi-Agent, and Decentralized Multi-Agent realizations. An Action enters the Edit Interface of the Operational Representation, whose edit scopes are Unit Edit, Relation Edit, and Whole-Artifact Edit; the interface modifies an Intermediate Form realized as an Artifact Instance, Structured Model, or Executable Program. An Observation sampled from the Intermediate Form enters Runtime Verification. Its Observation Source lists Artifact State, Runtime Behavior, Evaluation Signal, and External Response, which an interpret arrow passes to a Feedback Function containing Criterion Status, Failure Diagnosis, and Revision Guidance. Criterion Status leads through Accept to the Delivered Artifact. A lower Feedback path returns to the bottom of the Construction Policy. Pale alignment cues connect diagnosis and guidance to addressable and editable artifact scope.}
  \label{fig:loop}
\end{figure}

%% file: sections/03_design_space.tex

\subsubsection{Operational Representation}

Operational Representation exposes the evolving artifact through an Intermediate Form and an Edit Interface.

\textbf{Intermediate Form.} An Intermediate Form is the artifact-side representation through which an evolving artifact is constructed or revised before acceptance.
We classify it by the authoritative substrate to which the Construction Policy commits revisions.
\begin{itemize}
  \item \emph{Artifact Instance} stores the current realization directly, without an intervening model or program, as in an evolving draft, image, audio track, video clip, or mesh.
  \item \emph{Structured Model} uses a declarative organization of units, properties, relations, or constraints, as in an outline, SVG object tree, symbolic score, timeline, scene graph, CAD model, or DOM tree.
  \item \emph{Executable Program} uses runnable procedures or rules, as in visualization code, animation scripts, or a source repository.
\end{itemize}
The classification concerns representational role rather than syntax or delivery status: an SVG or CAD model may be retained in the deliverable, and a software repository may itself be delivered.
When several forms coexist, the primary class follows the authoritative revision substrate, with additional synchronized forms reported explicitly.
What a form exposes determines which parts can be targeted and which dependencies can remain intact; for example, a bitmap supports perceptual review but offers less support for object-level revision than an SVG or scene graph.

\textbf{Edit Interface.} An Edit Interface specifies the operations through which Actions can address and modify the Intermediate Form.
We classify these operations by their primary target.
\begin{itemize}
  \item \emph{Unit Edit} changes an addressable unit or bounded set of units, such as a passage, vector element, timeline segment, scene object, or source file.
  \item \emph{Relation Edit} changes the organization, dependency, or constraint among units, such as ordering, layout, timing, object relations, or program dependencies.
  \item \emph{Whole-Artifact Edit} treats the current artifact as a single target, as in full rewriting or regeneration.
\end{itemize}
These operations may be exposed through human-facing interfaces, such as direct manipulation or natural-language revision, and machine-facing interfaces, such as APIs, code, or tool calls; both may act on the same state in mixed-initiative construction.
An interface may expose more than one class, but feasible repair depends on alignment between its granularity and the representation.
An object-structured form paired only with whole-artifact regeneration still provides little local control, while fine-grained operations help only when the form exposes the responsible unit and its dependencies.
The interface defines the available changes; the Construction Policy decides which one to invoke.

\subsubsection{Construction Policy}

Construction Policy turns the Task Specification, represented state, and feedback into construction decisions.
Its realization varies in who or what controls those decisions and how the responsible agents are organized.
It may consult policy-side state such as plans, action and feedback history, budgets, approvals, memory, or an internal world model.
This context remains distinct from the artifact-side Operational Representation unless the model itself is constructed and validated as the deliverable, in which case it belongs to Simulation Models~\citep{Chu2026_AgenticWorldModeling,WorldModel_Agent2World2025}.

\textbf{Decision Control.} Two recurring modes determine how a Construction Policy advances.
\emph{Workflow} control uses predefined stages, transitions, and checks to structure progression, whereas \emph{Autonomous} control lets an agent select revision targets, actions, tools, branches, and stopping from represented state and feedback.
The modes are composable, and a fixed outer workflow may contain autonomous spans.
Human approval can constrain either mode and therefore does not define a separate Decision Control category.
Planning, decomposition, tool use, and stopping may therefore be specified in advance or resolved during construction.

\textbf{Agent Topology.} Agent Topology describes how the agents implementing a Construction Policy are organized, independently of its Decision Control mechanisms.
\begin{itemize}
  \item In a \emph{Single-Agent} topology, one agent maintains the main decision trajectory even when it invokes tools or models.
  \item A \emph{Centralized Multi-Agent} topology delegates work through a manager, orchestrator, or shared global plan.
  \item A \emph{Decentralized Multi-Agent} topology coordinates peer agents through messages, negotiation, shared state, or environmental signals.
\end{itemize}
Decomposition can reduce local complexity and expose specialist interfaces, but it also creates coordination costs and dependency-bearing handoffs.
The important questions are who maintains global commitments, how conflicts are resolved, and what state survives each boundary, rather than simply how many model calls the system makes.

\subsubsection{Runtime Verification}

Runtime Verification assesses observations of Action consequences against the task criteria and returns feedback that may redirect construction.
Its realization varies in Observation Source and Feedback Function.

\textbf{Observation Source.} Observation Source identifies the evidence made available to Runtime Verification.
\begin{itemize}
  \item \emph{Artifact State} exposes the current content, structure, constraints, dependencies, or rendered state of the artifact.
  \item \emph{Runtime Behavior} exposes temporal or input-conditioned consequences through playback, execution traces, interaction trajectories, or simulation rollouts.
  \item \emph{Evaluation Signal} supplies results produced expressly to assess the artifact or its behavior, such as rule violations, test verdicts, metric scores, or model judgments.
  \item \emph{External Response} records consequences at the system boundary, such as user actions, approval decisions, or environment reactions.
\end{itemize}
The availability of an observation does not establish its validity; Section~\ref{sec:verification} separately examines evidence and judgment under external protocols.

\textbf{Feedback Function.} Feedback Function describes the control-relevant information that Runtime Verification derives from observations.
\begin{itemize}
  \item \emph{Criterion Status} reports whether and how strongly the current state satisfies acceptance criteria, supporting acceptance, further verification, or escalation.
  \item \emph{Failure Diagnosis} links a violation to an artifact unit, constraint, Action, or dependency that explains where or why the failure occurred.
  \item \emph{Revision Guidance} proposes a revision target or direction that may address the diagnosed failure.
\end{itemize}
These functions may co-occur and vary in granularity, but the Construction Policy decides whether and how to act on them.
Feedback must arrive while the implicated state or decision remains relevant.
An Evaluation Signal is therefore an input to Runtime Verification, whereas Criterion Status, Failure Diagnosis, and Revision Guidance describe how Runtime Verification interprets available observations for control.

These dimensions matter jointly: the Intermediate Form must expose targets that the Edit Interface can modify, observations must reveal relevant consequences, and feedback must support a feasible policy response.
Misalignment can neutralize an otherwise capable component; detailed diagnosis, for example, cannot support local repair when available actions remain broad.
Greater autonomy therefore remains limited by verifier coverage, reversibility, expertise, and decision authority.

\subsection{Construction Affordances}
\label{sec:construction-affordances}

The functional architecture can help before repair is needed: plans establish global structure, specialized tools handle bounded decisions, persistent records preserve commitments, and intermediate outputs enable inspection before composition.
Together, these mechanisms provide three related affordances.

\begin{itemize}
    \item \emph{Composability.} Explicit steps create candidate interfaces for replacement, reordering, and upgrading.
    When interfaces preserve downstream dependencies, a fact-checking step can be inserted before rendering or a compatible component replaced without rebuilding the rest of the system.

    \item \emph{Traceability.} Recorded steps provide links among requirements, sources, artifact regions, actions, observations, and decisions.
    If a poster's text overflows, the system can inspect content length, layout decisions, typography, and rendering state, narrowing the search for a repair without proving a unique cause.

    \item \emph{Revisability.} Explicit edit units and dependencies provide intervention points for both humans and automated systems.
    When the representation separates properties and records their dependencies, a user can request \emph{keep the layout, change only the color scheme}, and the system can change the targeted property before revalidating affected dependencies.
\end{itemize}

These system-level affordances emerge from alignment among represented artifact structure, construction decisions, interfaces, records, and runtime evidence; no functional component is sufficient in isolation.
Because planning, coordination, tool execution, and verification introduce overhead, direct generation can remain appropriate for simple artifacts whose failures are cheap to detect and regenerate.

%% file: sections/04_landscape.tex
\section{Landscape of Artifact Families}
\label{sec:landscape}

\begin{figure}[!t]
\centering
\includegraphics[width=\textwidth]{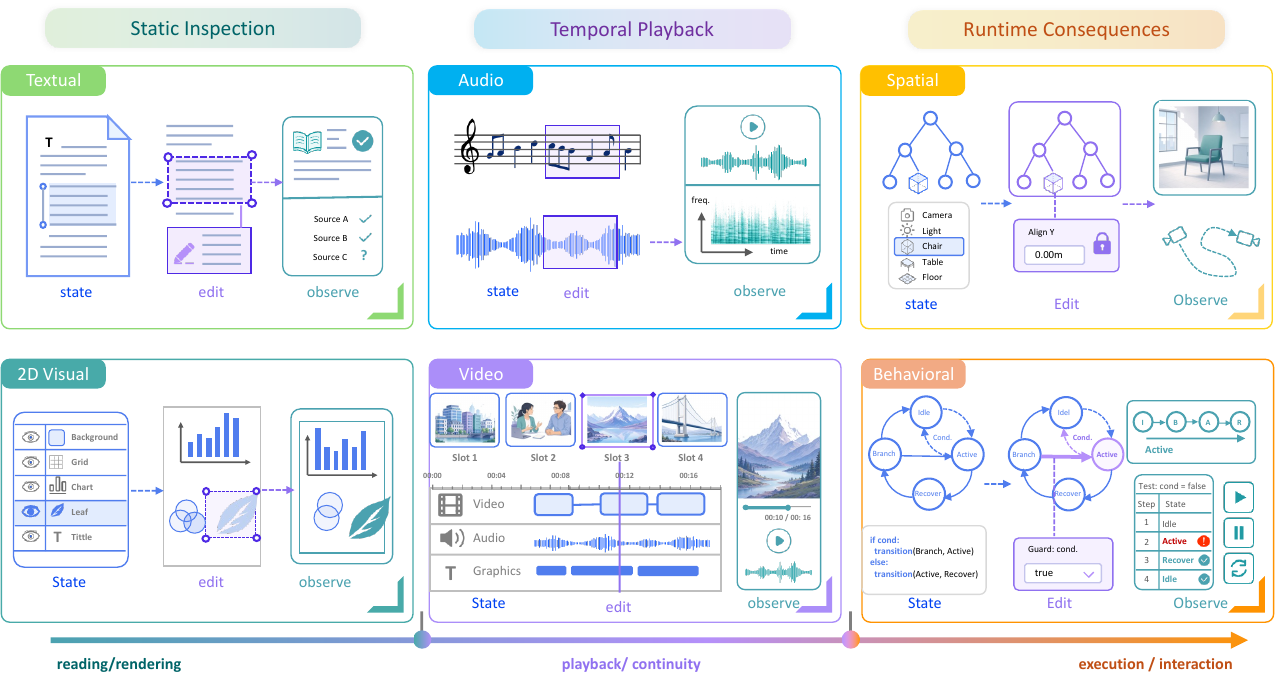}
\caption{Observability landscape of six artifact families. Each vignette follows a shared state--edit--observe grammar. The columns highlight dominant failure-observation points across static inspection for textual and 2D visual artifacts, temporal playback for audio and video, and runtime consequences for spatial and behavioral artifacts. These are tendencies rather than exclusive modes.}
\Description{A three-column by two-row comparison of six artifact families. The columns are Static Inspection, Temporal Playback, and Runtime Consequences. Textual and 2D Visual appear in the first column, Audio and Video in the second, and Spatial and Behavioral in the third. Within each vignette, blue state representations lead to violet edit targets and then teal observations. A horizontal arrow at the bottom progresses from reading and rendering, through playback and continuity, to traversal, execution, and interaction.}
\label{fig:artifact-family-landscape}
\end{figure}

We group the literature by the primary independently accepted artifact form: textual, 2D visual, audio, video, spatial, or behavioral.
Behavioral artifacts take executable, state-dependent responses as their defining delivered form.
Assignment follows the requirement whose failure would make the deliverable unacceptable and the artifact state needed to repair it; embedded components are treated separately only when delivered under distinct acceptance criteria.
The presentation moves from language- and surface-centered artifacts through temporally assembled media to persistent spatial structures and state-dependent behavioral artifacts, reflecting how state is organized and when consequential failures become observable.
For each family, we examine how its acceptance criteria shape Operational Representation, Construction Policy, and Runtime Verification, focusing on maintained state, available repairs, and observations that can redirect construction.
Table~\ref{tab:artifact-landscape} summarizes family-level dependency regimes and observation modes, while Table~\ref{tab:design-dims} gives representative component realizations.
Figures~\ref{fig:artifact-family-landscape} and~\ref{fig:artifact-taxonomy-tree} visualize the six families and their 16 profiles.

\begin{table}[t]
\caption{Compact family signatures of the dependency regimes and observation modes that shape agentic construction.}
\label{tab:artifact-landscape}
\centering
\scriptsize
\newcommand{\smark}{\textcolor{cDark!55}{\(\circ\)}}
\setlength{\tabcolsep}{3.0pt}
\renewcommand{\arraystretch}{1.18}
\begin{tabular*}{\textwidth}{@{\extracolsep{\fill}}l*{5}{c}@{\hspace{1.2em}}*{4}{c}}
\toprule
\multicolumn{1}{c}{\textbf{Artifact}} & \multicolumn{5}{c}{\textbf{Dependency Regimes}} & \multicolumn{4}{c}{\textbf{Observation Modes}} \\[-2pt]
\cmidrule(lr){2-6}\cmidrule(lr){7-10}
\multicolumn{1}{c}{\textbf{Family}} & \textbf{Semantic} & \textbf{Perceptual} & \textbf{Spatial} & \textbf{Temporal} & \textbf{Dynamic} & \textbf{Reading} & \textbf{Rendering} & \textbf{Playback} & \textbf{Interaction} \\
\midrule
Textual & \cmark & & & \smark & & \cmark & & & \\
\rowcolor{cRowAlt}
2D visual & \smark & \cmark & \cmark & & & \smark & \cmark & & \\
Audio & \smark & \cmark & & \cmark & & \smark & & \cmark & \\
\rowcolor{cRowAlt}
Video & \smark & \cmark & \cmark & \cmark & & \smark & \cmark & \cmark & \\
Spatial & \smark & \smark & \cmark & & \smark & & \cmark & & \smark \\
\rowcolor{cRowAlt}
Behavioral & \smark & \smark & \smark & \smark & \cmark & \smark & \smark & & \cmark \\
\bottomrule
\end{tabular*}
\vspace{2pt}

\raggedright\scriptsize\textit{Legend:} \cmark{} denotes a dominant dependency regime or observation mode that governs the family-level acceptance criterion; \smark{} denotes secondary involvement in substantial subtypes without determining family assignment; a blank denotes no typical family-level role.
\end{table}

\ifdefined\artifactpreprintlayout
\begin{table}[!t]
\else
\begin{table}[t]
\fi
\caption{Representative systems across artifact families. Intermediate Form uses a compact form-to-artifact path, \(X\xrightarrow{m}Y\), with the mediating renderer, executor, model, or toolchain above the arrow; the remaining columns use the functional categories in Section~\ref{sec:hard-soft}.}
\label{tab:design-dims}
\centering
\footnotesize
\setlength{\tabcolsep}{2.6pt}
\renewcommand{\arraystretch}{1.04}
\resizebox{\textwidth}{!}{%
\newcommand{\iformpath}[3]{\(\text{#1}\xrightarrow{\text{\scriptsize #2}}\text{#3}\)}
\begin{tabular}{
>{\raggedright\arraybackslash}p{2.30cm}
>{\raggedright\arraybackslash}p{1.25cm}
>{\raggedright\arraybackslash}p{3.25cm}
>{\raggedright\arraybackslash}p{2.15cm}
>{\raggedright\arraybackslash}p{1.55cm}
>{\raggedright\arraybackslash}p{2.10cm}
>{\raggedright\arraybackslash}p{3.45cm}
>{\raggedright\arraybackslash}p{2.80cm}}
\toprule
\multicolumn{1}{c}{\textbf{Agentic}} & \multicolumn{1}{c}{\textbf{Artifact}} &
\multicolumn{2}{c}{\textbf{Operational Representation}} &
\multicolumn{2}{c}{\textbf{Construction Policy}} &
\multicolumn{2}{c}{\textbf{Runtime Verification}} \\[-2pt]
\cmidrule(lr){3-4}\cmidrule(lr){5-6}\cmidrule(lr){7-8}
\multicolumn{1}{c}{\textbf{System}} & \multicolumn{1}{c}{\textbf{Family}} & \multicolumn{1}{c}{\textbf{\scriptsize Intermediate Form}} & \multicolumn{1}{c}{\textbf{\scriptsize Edit Interface}} & \multicolumn{1}{c}{\textbf{\scriptsize Decision Control}} & \multicolumn{1}{c}{\textbf{\scriptsize Agent Topology}} & \multicolumn{1}{c}{\textbf{\scriptsize Observation Source}} & \multicolumn{1}{c}{\textbf{\scriptsize Feedback Function}} \\
\midrule
STORM~\citep{Text_STORM2024} & Textual & \iformpath{Outline}{Agents}{Article} & Unit + Relation & Workflow & Centralized MAS & NR & NR \\
\rowcolor{cRowAlt}
AI Scientist~\citep{Text_AIScientist2024} & Textual & \iformpath{Manuscript}{LaTeX}{Paper} & Unit & Workflow + Autonomous & Single-Agent & Runtime + Metric & Status + Diagnosis + Guidance \\
MCQG-SRefine~\citep{Text_MCQGSRefine2025} & Textual & \iformpath{Draft}{Critique}{Questions} & Unit & Workflow & Single-Agent & State + Metric & Status + Diagnosis + Guidance \\
\rowcolor{cRowAlt}
Constella~\citep{Text_Constella2025} & Textual & \iformpath{Character Graph}{Agents}{Character Set} & Unit + Relation & NR & Decentralized MAS & State + Response & Guidance \\
PaperDebugger~\citep{Text_PaperDebugger2025} & Textual & \iformpath{LaTeX State}{Editor}{Paper} & Unit + Relation & Workflow & Centralized MAS & State + Metric + Response & Diagnosis + Guidance \\
\midrule
Data Formulator 2~\citep{DataViz_DataFormulator2_2025} & 2D visual & \iformpath{Data + Spec}{Renderer}{Chart} & Unit + Relation & Workflow & Single-Agent & State + Response & Guidance \\
\rowcolor{cRowAlt}
GenArtist~\citep{Image_GenArtist2024} & 2D visual & \iformpath{Image}{Models}{Image} & Unit + Artifact & Autonomous & Single-Agent & State + Metric & Diagnosis + Guidance \\
EvoDiagram~\citep{Diagram_EvoDiagram2026} & 2D visual & \iformpath{Canvas Schema}{Renderer}{Diagram} & Unit + Relation & Autonomous & Centralized MAS & State & Diagnosis + Guidance \\
\rowcolor{cRowAlt}
PosterAgent~\citep{Poster_Paper2Poster2025} & 2D visual & \iformpath{Layout Tree}{Painter}{Poster} & Unit + Relation & Workflow & Centralized MAS & State + Metric & Status + Diagnosis + Guidance \\
EvoPresent~\citep{Presentation_EvoPresent2026} & 2D visual & \iformpath{Deck State}{Renderer}{Slides} & Unit + Relation & Workflow & Centralized MAS & State + Metric & Status + Diagnosis + Guidance \\
\midrule
CoComposer~\citep{Music_CoComposer2025} & Audio & \iformpath{Score}{Synth}{Music} & Unit + Relation & Workflow & Centralized MAS & State + Metric & Diagnosis + Guidance \\
\rowcolor{cRowAlt}
MusicSwarm~\citep{Music_MusicSwarm2025} & Audio & \iformpath{Bar Cues}{Agents}{Score} & Unit + Relation & Autonomous & Decentralized MAS & State + Metric & Status + Guidance \\
WavCraft~\citep{Audio_WavCraft2024} & Audio & \iformpath{Edit Script}{Tools}{Audio} & Unit + Relation + Artifact & Workflow & Single-Agent & State + Response & Guidance \\
\rowcolor{cRowAlt}
LVAS-Agent~\citep{Audio_LVASAgent2025} & Audio & \iformpath{Audio Script}{Models}{Soundtrack} & Unit + Relation & Workflow & Centralized MAS & State + Metric & Diagnosis + Guidance \\
AudioGenie~\citep{Audio_AudioGenie2025} & Audio & \iformpath{Event Plan}{Models}{Audio} & Unit + Relation & Workflow + Autonomous & Centralized MAS & State + Metric & Status + Diagnosis + Guidance \\
\midrule
PaperTalker~\citep{Video_Paper2Video2025} & Video & \iformpath{Beamer + Timeline}{Tools}{Video} & Unit + Relation & Workflow & Centralized MAS & State + Runtime + Metric & Status + Diagnosis \\
\rowcolor{cRowAlt}
StoryAgent~\citep{Video_StoryAgent2024} & Video & \iformpath{Storyboard}{Models}{Video} & Unit + Relation & Workflow & Centralized MAS & State + Metric & Status + Guidance \\
GenMAC~\citep{Video_GenMAC2024} & Video & \iformpath{Controls}{Models}{Video} & Unit + Relation & Workflow + Autonomous & Centralized MAS & State + Metric & Diagnosis + Guidance \\
\rowcolor{cRowAlt}
ATVG~\citep{Video_ATVG2026} & Video & NR & NR & NR & NR & NR & NR \\
EditDuet~\citep{D3_EditDuet2025} & Video & \iformpath{Timeline}{Editor}{Video} & Unit + Relation & Autonomous & Centralized MAS & State + Metric & Status + Diagnosis + Guidance \\
\midrule
ShapeCraft~\citep{3D_ShapeCraft2025} & Spatial & \iformpath{GPS/Code}{Blender}{Asset} & Unit + Relation & Autonomous & Centralized MAS & State + Metric & Diagnosis + Guidance \\
\rowcolor{cRowAlt}
SceneCraft~\citep{D3_SceneCraft2024} & Spatial & \iformpath{Graph + Code}{Blender}{Scene} & Unit + Relation & Autonomous & Single-Agent & State + Metric & Diagnosis + Guidance \\
SceneWeaver~\citep{D3_SceneWeaver2025} & Spatial & \iformpath{Scene Graph}{Engine}{Scene} & Unit + Relation & Autonomous & Single-Agent & State + Metric & Diagnosis + Guidance \\
\rowcolor{cRowAlt}
SPADA~\citep{CAD_SPADA2026} & Spatial & \iformpath{Assembly Code}{CAD Kernel}{Assembly} & Unit + Relation & Autonomous & Single-Agent & Runtime + Metric & Status + Diagnosis + Guidance \\
Sketch2BIM~\citep{D3_Sketch2BIM2025} & Spatial & \iformpath{JSON Layout}{BIM Scripts}{BIM Model} & Unit + Relation & Workflow & Centralized MAS & State + Runtime + Metric + Response & Status + Diagnosis + Guidance \\
\midrule
SWE-agent~\citep{Code_SWEAgent2024} & Behavioral & \iformpath{Repository}{Shell}{Patch} & Unit & Autonomous & Single-Agent & State + Runtime + Metric & Status + Diagnosis \\
\rowcolor{cRowAlt}
CodeChain~\citep{Code_CodeChain2024} & Behavioral & \iformpath{Candidate Set}{LLM}{Program} & Artifact & Workflow & Single-Agent & State & Guidance \\
DuetUI~\citep{UI_DuetUI2026} & Behavioral & \iformpath{Task + Schema}{Renderer}{Interface} & Unit + Relation & Workflow + Autonomous & Centralized MAS & State + Response & Guidance \\
\rowcolor{cRowAlt}
DreamGarden~\citep{Game_DreamGarden2025} & Behavioral & \iformpath{Plan Tree + Code}{Unreal}{Game} & Unit + Relation & Autonomous & Centralized MAS & State + Runtime + Metric + Response & Status + Diagnosis + Guidance \\
Agent2World~\citep{WorldModel_Agent2World2025} & Behavioral & \iformpath{PDDL / Code}{Executor}{Simulator} & Unit + Relation & Workflow + Autonomous & Centralized MAS & Runtime + Metric & Status + Diagnosis + Guidance \\
\bottomrule
\end{tabular}}
\vspace{2pt}

\raggedright\scriptsize\textit{Note:} Cells summarize implementation details reported by the cited system papers. Each path names the maintained form, its mediating substrate, and the realized artifact. The MusicSwarm row reports its decentralized configuration. Multiple labels indicate co-occurring categories; MAS denotes multi-agent system; NR denotes details not reported in the cited source.
\end{table}

\input{figures/artifact_taxonomy_tree}

\subsection{Textual Artifacts}
\label{sec:landscape-text}

Textual artifacts are judged by their linguistic content, discourse organization, and fit to a communicative purpose.
Their represented state may include outlines, evidence links, drafts, character or discourse memory, and revision histories, supporting choices among retrieval, local editing, reorganization, and broader rewriting.
Source and schema checks provide early feedback; reader, creator, or domain-expert responses arrive later and address different criteria.

\subsubsection{Creative Writing}
Creative writing develops expressive commitments whose validity is not exhausted by source coverage or task compliance.
Narratives organize those commitments through an evolving story world, whereas performative texts depend more directly on form, voice, timing, and anticipated reception.

\paragraph{Narratives.}
Narratives depend on long-range commitments to plot, character, voice, and payoff, and violations may surface many passages after the decision that caused them.
Agents' Room~\citep{Text_AgentsRoom2025}, StoryBox~\citep{Text_StoryBox2025}, and CreAgentive~\citep{Text_CreAgentive2025} assign planning and revision to separate roles.
BookWorld~\citep{Text_BookWorld2025} and Constella~\citep{Text_Constella2025} instead carry character or social state across the writing process, a pattern also explored in multi-agent character simulation~\citep{Text_CharacterSimulation2025}.
A persona compresses constraints into a generation condition~\citep{Text_PersonasToPlot2026}; queryable narrative memory records event order, relationships, and unresolved payoffs~\citep{Text_NarrativeWorldModel2026}.
The second representation supports more specific repair when a later inconsistency can be traced to the event or relationship that caused it.
Fluency and preference feedback rarely provide that trace.

\paragraph{Performative Texts.}
Performative texts must realize an expressive effect through language intended to be voiced, staged, or received by an audience, even when they do not construct a persistent story world.
Comedy systems retain discussion, timing, setup--punchline structure, callbacks, or performance cues across revision~\citep{Text_ComedyClub2026,Text_OpenMic2026}.
Multi-agent poetry generation~\citep{Text_Poetry2024} and EchoVoices~\citep{Text_EchoVoices2025} provide adjacent evidence for stylistic diversity, persona, and voice, but do not expose the same artifact-revision loop.
The repair unit may be a line, stanza, setup, punchline, callback, or delivery cue.
Audience response supplies acceptance evidence, but a preference score rarely identifies which expressive commitment should change.

\subsubsection{Professional Documents}
Professional documents are task-facing texts accepted against source, domain, audience, or workflow obligations.
Informational reports organize external evidence into an audience-facing account, whereas functional documents bind text to the source, rule, audience, or workflow that governs its use.
Scholarly manuscripts are separated below because contribution-level argument changes the dominant acceptance criterion.

\paragraph{Informational Reports.}
Informational reports are accepted for their coverage, traceability, and organization of external evidence, with a claim--source link or evidence-backed section as the usual repair unit.
Report systems stage research, planning, drafting, or simulated feedback in different combinations~\citep{Text_STORM2024}.
Longer workflows carry intermediate evidence through several stages~\citep{Report_DeepResearchAgent2026,Code_AgenticDeepResearch2025}.
LateralReader~\citep{Report_LateralReader2026} makes evidence sufficiency a control decision before drafting, while the DRAGUN resources~\citep{Report_DRAGUNResources2026} expose attributed report construction as a benchmarkable task.
Recent benchmarks cover undermeasured revision behavior.
Mr.~Dre~\citep{Text_MrDre2026} tests repeated feedback-driven revision and identifies regressions in previously accepted content and citations.
ResearchAgent~\citep{Text_ResearchAgent2024} complements these systems at the upstream ideation stage.

\paragraph{Functional Documents.}
Functional documents are accepted for how reliably they perform an intended task under source, domain, audience, or workflow constraints.
CRMAgent~\citep{Text_CRMAgent2025}, MADS~\citep{Text_MADS2025}, and AgentCTG~\citep{Text_AgentCTG2025} condition messaging or dialogue on audience and control requirements; stylized comment generation applies a related pattern to short-video audiences~\citep{Text_LaughRelateEngage2025}.
Other systems produce patient-facing and knowledge-base content~\citep{Data_PAMEAI2025,Data_AIKnowledgeAssist2025}.
Systems such as AutoManual~\citep{Text_AutoManual2024}, FormAct~\citep{Text_FormAct2026}, ProfiliTable~\citep{Text_ProfiliTable2026}, and SheetAgent~\citep{Text_SheetAgent2025} retain more executable document state, including interaction-derived rules, rendered HTML, tabular-processing code, or workbook state that can redirect subsequent edits.
MCQG-SRefine~\citep{Text_MCQGSRefine2025} applies the same principle to assessment documents by retaining critiques, corrections, and comparison feedback across question-revision rounds.
Translation maintains paired source and target state under localization constraints~\citep{Text_TransAgents2024}; code documentation derives acceptance criteria from repository structure.
Workflow records support targeted repair when document regions remain linked to the sources, rules, audiences, or repository units that govern them.
These links help the policy revise the text, its paired source, or an upstream constraint at the appropriate scope.

\subsubsection{Scholarly Manuscripts}
Scholarly manuscripts specialize evidence-grounded professional writing through contribution-level acceptance criteria that align the document's claims with its stated methods, results, prior work, and contribution.
The artifact-local repair units are an argument or section, a claim--evidence link, or an inconsistency between a reported method and result.
PaperDebugger~\citep{Text_PaperDebugger2025} supports targeted in-editor manuscript revision.
The AI Scientist~\citep{Text_AIScientist2024} provides an end-to-end research-to-manuscript case.
MLR-Bench~\citep{Text_MLRBench2025} and SurGE~\citep{Text_SurGE2026} broaden evaluation from isolated prose quality to research trajectories and survey construction grounded in literature and experiments.
DIAGPaper~\citep{Text_DIAGPaper2026} and MARG~\citep{Text_MARG2024} localize suspect content through diagnostic and review feedback, while end-to-end systems retain broader context across the manuscript and research process.
These signals guide claim-, argument-, and section-level revision; changes to research design, code, or experiments belong to the broader scientific workflow in \S\ref{sec:applications-research}.

\subsubsection*{Summary}

Textual control becomes harder as commitments extend beyond the passage being edited.
Creative writing carries story-world or performative commitments, professional documents link text to external evidence and workflow obligations, and scholarly manuscripts align contribution-level arguments with methods, results, and prior work.
Their representations expose repair units ranging from an event or expressive unit to a claim--source link, rule-bound region, or manuscript argument.
Reader and domain judgment remain necessary because coherence, novelty, persuasion, and professional fitness are only partly observable during construction; repair remains bounded only when those judgments can be traced to the responsible event, claim, source, rule, or argument.

\subsection{2D Visual Artifacts}
\label{sec:landscape-visuals}

The 2D visual family is defined by acceptance on a rendered surface or page sequence.
Its state ranges from raster layers to SVG or TikZ, chart code, layout trees, and deck structures, with edits targeting pixels, vectors, code, spatial constraints, pages, or cross-page commitments.
Rendering supplies a common feedback channel, but each profile authorizes the surface through different evidence, whether recorded data, a perceptual or semantic specification, or source content organized as a visual document.
DrawAI~\citep{Image_DrawAI2026} spans scientific figures, presentations, posters, and diagrams and illustrates the shared challenge of recovering editable structure from flattened images.

\subsubsection{Data Visualizations}
Data visualizations have two central acceptance criteria.
The composition must communicate effectively, and every visual mark must remain faithful to source data and recorded transformations.
AMACE~\citep{DataViz_AMACE2025}, PlotGen~\citep{DataViz_PlotGen2025}, and CoDA~\citep{DataViz_CoDA2025} preserve an executable transformation path and inspect its render.
Data Formulator 2~\citep{DataViz_DataFormulator2_2025} keeps data transformation, chart specification, and conversational revision state synchronized so that user feedback can redirect both the data and its visual encoding.
SVG editing makes marks and geometry directly addressable~\citep{Data_DataWink2025}.
Other designs coordinate data reasoning, constraints, visual encoding, and presentation planning~\citep{DataVis_MultiVisAgent2026,DataViz_A2PVis2025}.
DV-World~\citep{DataViz_DVWorld2026} broadens evaluation to native spreadsheet creation and repair, adaptation of visualizations to new data, and interaction under underspecified intent.
Code retains the transformation path but can make small visual edits indirect, while SVG supports local manipulation but can sever marks from their data.
Targeted repair therefore depends on links from rendered marks to encodings, transformations, and source data, since appearance alone does not establish numerical or semantic correctness.

\subsubsection{Illustrative Graphics}
Illustrative graphics construct a bounded visual composition from a perceptual or semantic specification.
Images emphasize perceptual composition and localized appearance; diagrams depend more heavily on recoverable entities and relations.
Both forms require editable structure beneath the rendered surface if feedback is to support targeted revision.

\paragraph{Images.}
Images are judged on a rendered surface, but a visible defect may not reveal which region, layer, object, or prior operation should change.
Prompt interpretation and sequential sketches specify composition before rendering~\citep{Image_T2ICopilot2025,Image_SketchAgent2025}.
GenArtist~\citep{Image_GenArtist2024} and FaSTA*~\citep{Image_FaSTA2026} preserve tool choices or mined subroutines across image construction; earlier neural-painting and sketching agents retain strokes and regions~\citep{Image_CompositionalNeuralPainter2023,Image_DualAgentSketching2024}.
MIMO~\citep{Image_MirrorAd2025} diagnoses an initial render, while Agent Banana~\citep{Image_AgentBanana2026}, Agentic Retoucher~\citep{Image_AgenticRetoucher2026}, and CAMEO~\citep{Image_CAMEO2026} localize defects in an existing result.
Systems such as 4KAgent~\citep{Image_4KAgent2025}, CREA~\citep{Image_CREA2025}, G-Refine~\citep{Image_GRefine2024}, and CEARI~\citep{Image_CEARI2025} retain intermediate images together with quality or consistency feedback during restoration, editing, refinement, or completion.
Idea2Img~\citep{Image_Idea2Img2024} and Culture-TRIP~\citep{Image_CultureTRIP2025} instead use multimodal evaluation to revise prompts before the next generation round, including culturally grounded criteria in the latter case.
UniEdit-I~\citep{Image_UniEditI2026} couples understanding, editing, and verification within one iterative policy, whereas multi-turn consistent editing~\citep{Image_MultiTurnConsistent2025} and preference-guided prompt optimization~\citep{Image_APPO2026} preserve user instructions or pairwise preferences across successive revisions.
Application-oriented systems encode domain constraints before visual detail~\citep{Marketing_BannerAgency2025,Image_ImageGeneration2025}.
CanvasAgent~\citep{Image_CanvasAgent2026} preserves tool-level links among localization, segmentation, compositing, text recognition, and enhancement.
ReDesign~\citep{Image_ReDesign2026} recovers an editable layer hierarchy from raster designs and accepts, prunes, or retries local components to preserve successful structure.
Early planning exposes global constraints but can commit the artifact to a poor structure, whereas post-generation repair preserves flexibility at the cost of weaker edit localization.
Maintaining region, layer, or object identity across operations lets critique target the responsible edit.

\paragraph{Diagrams.}
Diagrams, including scientific figures and engineering schematics, must preserve recoverable objects and relations after rendering.
VisPainter~\citep{Image_PixelsToPaths2025}, AutoFigure~\citep{Image_AutoFigure2026}, and PaperBanana~\citep{Image_PaperBanana2026} construct vector or scientific figures from existing material.
Crafter~\citep{Diagram_Crafter2026} maintains an evolving figure specification across diverse input conditions, uses directive critique and typed revision, and pairs figure generation with raster-to-SVG conversion for downstream editing.
SketchAgent~\citep{Diagram_SketchAgent2025}, Sketch2Diagram~\citep{Diagram_Sketch2Diagram2025}, and GenAI-DrawIO-Creator~\citep{Diagram_GenAIDrawIOCreator2026} encode editable structure from sketches or specifications, while text-driven diagram construction provides a complementary route~\citep{Diagram_Text2Diagram2024}.
AutomaTikZ~\citep{Diagram_AutomaTikZ2024} and SciFig~\citep{Diagram_SciFig2026} use executable code or hierarchical layouts to preserve a replayable construction path and support structural checks.
SVG objects and graph state make direct manipulation and relation-level revision easier~\citep{Diagram_DuetSVG2025,Diagram_SAGEGraphEditing2026}.
EvoDiagram~\citep{Diagram_EvoDiagram2026} separates artifact state from process state.
Its canvas schema connects semantic topology, visual styling, and spatial layout, while coordinated agents separate semantic intent from rendering logic.
Design-knowledge evolution then distills execution traces into a hierarchical memory of domain guidelines.
The schema keeps delivered objects revisable; the memory informs later design decisions.
Table reconstruction provides adjacent structured-graphic evidence when code is an intermediate representation for a visually evaluated table~\citep{Code_Table2LaTeXRL2025}.
Engineering schematics add electrical constraints that rendering alone cannot establish~\citep{Hardware_CircuitLM2026,Hardware_PCBSchemaGen2026}.
Code can make a small visual edit indirect, while object or graph state can drift from its render.
Targeted revision needs both a revisable semantic object and a synchronized visual observation.

\subsubsection{Visual Documents}
Visual documents organize source-backed content into one or more visually composed pages.
Posters concentrate hierarchy, capacity, and legibility on a single page; presentations extend these commitments across a sequence and a discourse-level narrative.

\paragraph{Posters.}
A poster must preserve source fidelity, select the right content, establish visual hierarchy, remain legible, and fit one page.
Improving one requirement can invalidate another.
Direct paper-to-poster systems map source material to the page in one formulation~\citep{Poster_P2P2025,Poster_Paper2Poster2025}.
Staged systems separate content, layout, rendering, and review~\citep{Poster_PosterGen2025,Poster_PosterForest2025}.
PosterAgent~\citep{Poster_PosterAgent2026} instead learns stage-aware incremental actions over an editable draft.
PosterMELD~\citep{Poster_PosterMELD2026} organizes content into capacity-aware slots, routes deterministic and visual-check failures into bounded repair, and produces editable print-ready outputs.
AutoDesign~\citep{Poster_AutoDesign2026} retains an editable HTML poster through rule-based and visual-critic repair, while a separate meta-harness uses accumulated rollout evidence to propose and gate versioned changes to the construction system itself.
Unified or semantics-aware systems keep more of these decisions together~\citep{Poster_PosterCraft2025,Poster_EfficientPosterGen2026}.
Scientific posters inherit claim and source obligations.
Commercial and product posters add audience, brand, and product constraints~\citep{Poster_PosterVerse2026,Poster_AutoPP2026}.
Staging helps distinguish content failures from layout failures, but a late content edit can still invalidate page fit.
Choosing between content and layout repair requires links among source claims, content units, page regions, and rendering checks.

\paragraph{Presentations.}
Presentation construction must satisfy both page-level composition and deck-level commitments concerning narrative order, terminology, style, and cross-slide consistency.
Role-based systems coordinate specialized production steps~\citep{Slide_SlideGen2025,Slide_SlideBot2026}.
PPTAgent~\citep{Slide_PPTAgent2025} uses references and rendered feedback to revise the deck.
Interactive systems expose slide-level edits to the user~\citep{Slide_AutoSlides2025}, while EvoPresent~\citep{Presentation_EvoPresent2026} uses aesthetic diagnoses and comparisons to drive repeated deck improvement.
ArcDeck~\citep{Slide_ArcDeck2026} represents discourse and shared commitments across slides.
A local fix for overflow or imbalance can weaken narrative continuity elsewhere.
Deck-level state carries a claim, terminology, or design change beyond the slide where feedback first appeared.

\subsubsection*{Summary}

The three profiles differ in what authorizes the rendered surface.
Data visualizations bind marks to recorded transformations, illustrative graphics bind regions and relations to an intended composition, and visual documents bind source content to pages and deck-wide discourse.
Their repair units accordingly range from an encoding or transformation to a visual object, content--layout link, page area, or deck commitment.
Direct surface edits offer precise visual control but can sever data or source links, while structured code and object state preserve those links at the cost of making some visual changes indirect.
A valid local render does not establish that larger dependencies survived the edit, so bounded repair depends on preserving links from visible defects to their data, source, and cross-page commitments.

\subsection{Audio Artifacts}
\label{sec:landscape-music}

Audio artifacts are delivered as symbolic or rendered sound organized over time, with editable state in notation, lyrics, production graphs, stems, or waveforms.
Construction may revise notes, phrases, synthesis settings, effects, or a mix, while theory, signal measurements, tool responses, and listening address different criteria at different stages.
The three profiles differ in whether the audio timeline is validated through musical relations, source--script--speaker links, or scene-level event alignment.

\subsubsection{Music}
Music must keep note-, phrase-, and lyric-level decisions compatible with long-range harmony, form, and performance intent.
Role-decomposed systems assign melody, harmony, arrangement, and critique to different agents~\citep{Music_CoComposer2025,Music_HarmonyAgent2025}.
MusicSwarm~\citep{Music_MusicSwarm2025} coordinates bar-level contributions through shared signals.
Aligned lyric--melody state connects text, notation, and audio~\citep{Music_WeaveMuse2025,Music_SongComposer2025}, and an explicit grammar represents voices, bars, form, and variation~\citep{Music_Libretto2026}.
RIME~\citep{Music_RIME2026} extends the editable action space into music post-production by coordinating structured edit instructions with audio tools and iterative listening feedback.
These representations support bar-, phrase-, or production-level decisions only while dependencies across parts remain intact.
Symbolic checks can redirect structural repair early; timbre and expression still require listening after rendering.

\subsubsection{Spoken Audio}
Spoken audio must remain intelligible while keeping source content, scripts, speakers, timing, and rendered segments aligned.
No dedicated spoken-audio construction system appears in the mapped full-publication corpus; the closest evidence comes from condition-decomposed text- or video-to-audio generation~\citep{Audio_AudioAgent2024} and podcast-production pipelines with explicit discussion, voice, and synthesis stages~\citep{Audio_PodAgent2025}.
AudioToolAgent~\citep{Audio_AudioToolAgent2025} routes audio question-answering and speech-to-text tools but does not maintain an editable audio artifact.
These modular stages support targeted revision only when source units, scripts, voices, and rendered clips remain linked; otherwise, late listening feedback tends to trigger whole-segment regeneration.

\subsubsection{Soundscapes}
Soundscapes organize acoustic events, sources, temporal placement, and mixing around an intended scene.
Their state includes event plans, audio descriptions, source libraries, executable edit programs, and rendered clips.
LVAS-Agent~\citep{Audio_LVASAgent2025} coordinates scene segmentation, script generation, sound design, and synthesis, with correction and retrieval-based optimization between stages; WavCraft~\citep{Audio_WavCraft2024} retains descriptions, edit programs, and interaction history for local revision; and AudioRAG+~\citep{Audio_AudioRAGPlus2026} diagnoses missing events, retrieves references, and regenerates against those observations.
SoundscapeAgent~\citep{Audio_SoundscapeAgent2026} uses an executable scene plan to link source selection, temporal layout, rendering, aligned metadata, and human-editable state.
AudioGenie~\citep{Audio_AudioGenie2025} extends critic-guided refinement to multi-audio construction from heterogeneous source modalities.
Audio-Oscar~\citep{Audio_AudioOscar2026} coordinates character and voice state, a fine-grained event timeline, specialist generators, post-production, and feedback-driven refinement for complex audio scenes.
These links make an event, source, interval, or edit operation a repair unit for omissions, misalignment, or poor mixing.

\subsubsection*{Summary}

Audio separates inspectable symbolic or production state from qualities heard only after rendering.
Music represents notes, phrases, harmony, and form; spoken audio links source units to scripts, voices, and segments; and soundscapes link events, sources, edit programs, and rendered clips.
Structural checks can redirect upstream edits, but perceptual judgments support local repair only when they remain linked to an editable unit.

\subsection{Video Artifacts}
\label{sec:landscape-video}

Video artifacts are ordered visual or audiovisual sequences whose acceptance criteria extend across shots and time.
Represented state may include scripts, storyboards, shot lists, timelines, event graphs, and cross-shot memory, supporting edits to scenes, clips, transitions, motion, sound, or temporal constraints before their assembled consequences are visible.
We organize this family by communicative purpose rather than production technique.
Expository videos explain externally grounded material, narrative videos realize an evolving story world, and promotional videos present a subject or identity to an audience.
Editing, animation, personalization, and repair cut across these profiles, while hybrid cases follow the communicative obligation whose failure would invalidate the delivered video.

\subsubsection{Expository Videos}
Expository videos derive authority from an external source or subject matter, which they must cover accurately and communicate clearly after visuals, narration, subtitles, and timing are assembled.
Document- and script-grounded systems represent scenes together with aligned channels~\citep{Video_Paper2Video2025,Video_VideoAgentSci2025,Video_CommunicativeAgentsSlideshow2025}.
Code2Video~\citep{Video_Code2Video2025} and ManimAgent~\citep{Animation_ManimAgent2026} compile expository plans into executable code and inspect the render; a related multi-agent editor applies the same pattern to educational video~\citep{Video_BeyondE2E2026}.
Scientific-video and map-animation systems likewise retain paper-derived plans or editable graphic events that can be checked against the intended explanation~\citep{Animation_MapStory2025}.
The repair unit may be a claim--segment link, visual--narration pair, or timeline interval, with source links tracing claims back to the document.
Executable scenes make layout and timing reproducible, while playback feedback must identify both the faulty interval and its upstream channel when separately produced channels drift after assembly.

\subsubsection{Narrative Videos}
Narrative videos derive their primary authority from an evolving story world, even when a source narrative initializes that world.
Story intent, identity, causal order, and cinematic decisions must carry from one shot to the next.
Violations may become observable only after assembly.
StoryAgent~\citep{Video_StoryAgent2024}, DreamFactory~\citep{Video_DreamFactory2024}, MM-StoryAgent~\citep{Video_MMStoryAgent2025}, and MovieAgent~\citep{Video_MovieAgent2025} represent story, scene, shot, and modality plans hierarchically; Hollywood Town~\citep{Video_LongVideo2025} extends this pattern to long-video orchestration.
These levels make long-form work divisible but place consistency at their handoffs.
VideoMemory~\citep{Video_VideoMemory2026}, StoryBlender~\citep{Video_StoryBlender2026}, and StoryFlow~\citep{Video_ShotsStories2025} externalize longer-horizon dependencies as entity memory, editable spatial storyboards, or shot descriptions.
LogiStory~\citep{Video_LogiStory2026} provides adjacent evidence from multi-image story visualization by representing causal chains and checking story-level consistency.
Animated-story systems instantiate the same criteria through different production state.
AnimAgents~\citep{Video_AnimAgents2025} coordinates ideation, scripting, design, and storyboarding; AniME~\citep{Video_AniME2025} retains director-level memory across long-form production; AniMaker~\citep{Video_AniMaker2025} evaluates candidate clips in sequence context.
FilmWorld~\citep{Video_FilmWorld2026} carries entity and cinematic-world state from a source narrative through shot construction and closed-loop verification.
GenMAC~\citep{Video_GenMAC2024} and MUSE~\citep{Video_MUSE2026} revisit prompts, controls, scripts, or code after detecting a violation.
Role-specific systems make theatrical or cinematographic decisions inspectable~\citep{Video_HAMLET2026,Video_CameraArtist2026}.
Executable event graphs encode actors, actions, objects, and temporal relations in programmatically checked state~\citep{Video_EventGraphs2026}.
With this state, revision can target an event, entity state, shot, or cross-shot relation rather than regenerate the sequence indiscriminately.
Formal checks cover only encoded constraints, while prompt-level repair leaves unrepresented dependencies unenforced.

\subsubsection{Promotional Videos}
Promotional videos are outward-facing productions organized around how a destination, brand, persona, or performer is presented to an audience; direct advertising is one case rather than the entire profile.
Their acceptance depends on message fidelity, recognizable identity, audiovisual appeal, and fit to an intended audience or channel.
ATVG~\citep{Video_ATVG2026} grounds travel-advertisement scenes in destination facts and verifies visual artifacts before assembly, whereas BrandFusion~\citep{Video_BrandFusion2026} coordinates brand placement with scene-level generation.
PersonaVlog~\citep{Video_PersonaVlog2025} retains a personalized narrative and style across multimodal production.
AutoMV~\citep{Video_AutoMV2025} and GLANCE~\citep{Video_GLANCE2026} couple scene selection and sequencing with musical structure, using global plans and local timeline decisions to preserve an intended audiovisual identity.
Represented state may therefore combine a communication brief, audience or identity profile, factual source material, selected assets, scene plans, and an editable timeline.
The repair unit can be a claim, placement, shot, beat-aligned interval, or style decision; a locally polished segment can still fail if it weakens identity, factual grounding, or the global presentation.

\subsubsection*{Summary}

The profiles classify the delivered video by communicative purpose, while their construction mechanisms differ in the timeline state and edit operations they expose.
Clip- and tool-level editors map visual, audio, contextual, or physical observations to bounded revisions~\citep{D3_EditDuet2025,Video_UniVA2025,Video_RLVideoEditing2023,Video_PhyT2V2025,Video_VISTA2025}.
Task-spanning systems, trajectory benchmarks, and animation pipelines extend this state across modalities, procedural representations, and post-production stages~\citep{Video_Mora2024,Video_AgenticVBench2026,Animation_CoMA2024,Video_FantasyHSI2025,Video_VideoCoCo2026,Video_PhysAgent2026,Animation_LogoMotion2025,Animation_ParticleGen2026,Video_ShareVerse2026}.
Videos produced for downstream tasks remain inside the family only when the video itself is the accepted output~\citep{Video_Gen4Track2025,Video_ScalingInstructionBasedVideoEditing2025}.
Across the profiles, represented state binds sourced claims, identities, events, and audience-facing messages to shots and assembled channels.
Playback supports targeted repair only when a failure maps back to a responsible claim, shot, segment, transition, or motion unit; otherwise, feedback triggers broad regeneration.
The shared bottleneck is delayed and weakly localized temporal feedback.

\subsection{Spatial Artifacts}
\label{sec:landscape-3d}

Spatial artifacts are defined by persistent geometry and spatial relations, whether the result is rendered, simulated, fabricated, or later embedded in an application.
Their state may take the form of meshes, scene graphs, asset libraries, engine scenes, parametric histories, or CAD programs, with feedback arriving through rendering, geometry operations, traversal, simulation, or fabrication checks.
Across authoring environments, the independently accepted output remains either an asset or a scene, while engineering, simulation, and fabrication requirements strengthen its acceptance criteria.

\subsubsection{3D Assets}
3D assets are independently accepted objects whose geometry and edit handles must survive revision.
Visual assets expose semantic parts and appearance controls, whereas parametric models preserve dimensions, features, constraints, and precise boundary geometry.

\paragraph{Visual Assets.}
Visual assets must preserve correspondence among semantic parts, appearance, and edit handles through revision.
ShapeCraft~\citep{3D_ShapeCraft2025}, LL3M~\citep{D3_LL3M2025}, InfiniHuman~\citep{D3_InfiniHuman2025}, and SmartAvatar~\citep{D3_SmartAvatar2025} expose semantic parts or appearance controls as high-level edit units.
Procedural histories preserve modeling operations~\citep{D3_FilmSceneDesigner2025,D3_3Dify2025}, and progressive mesh construction records geometric state~\citep{D3_MeshLLM2025}.
A plausible surface may still lack recoverable design intent, and local edits can break topology, part relations, or downstream compatibility.
Links from controls and construction history to the resulting geometry and invalidated checks make targeted repair possible; without them, revision falls back to whole-asset regeneration.

\paragraph{Parametric Models.}
Parametric models carry engineering constraints in addition to visual and semantic commitments.
Seek-CAD~\citep{CAD_SeekCAD2025} and CADDesigner~\citep{CAD_CADDesigner2025} preserve dimensions, ordered features, and editable variables; LLM-driven FreeCAD automation provides related evidence for executable parametric modeling~\citep{CAD_FreeCADLLM2025}.
STEP- and NURBS-based generators provide precise boundary geometry~\citep{CAD_STEPLLM2026,CAD_NURBGen2026}.
CADIR~\citep{CAD_CADIR2026} adds a cross-backend construction graph with dependencies, constraints, topology, and diagnostics for reconstruction and later editing.
ArtisanCAD~\citep{CAD_ArtisanCAD2026} preserves expert-grounded operations in an executable intermediate representation.
Repair remains at the asset level while feedback changes a sketch, feature, parameter, operation, or boundary representation.
Construction histories can become brittle after upstream edits, and neither precise geometry nor renderability alone preserves design intent or guarantees solid validity and manufacturability.

\subsubsection{3D Scenes}
3D scenes combine assets into multi-object configurations whose relations must survive assembly and use.
Spatial worlds emphasize layout, traversal, and task compatibility, whereas engineered models make component interfaces and dimensional constraints part of the delivered artifact's acceptance criteria.

\paragraph{Spatial Worlds.}
Spatial worlds combine assets into layouts that must remain relationally consistent, traversable, task-compatible, and sometimes simulation-ready.
WorldGen~\citep{D3_WorldGen2025}, WorldGrow~\citep{D3_WorldGrow2025}, WorldCraft~\citep{D3_WorldCraft2025}, and SceneCraft~\citep{D3_SceneCraft2024} construct rendered, traversable, or extensible worlds; DreamScene~\citep{D3_DreamScene2024}, DreamScene360~\citep{D3_DreamScene3602024}, and LatticeWorld~\citep{D3_LatticeWorld2025} provide complementary end-to-end and panoramic formulations.
WorldAgents~\citep{D3_WorldAgents2026} uses agents to select views and verify the 3D reconstruction of an explorable spatial scene.
SAGE~\citep{3D_SAGE2026}, SceneSmith~\citep{3D_SceneSmith2026}, RoomPlanner~\citep{3D_RoomPlanner2025}, and MA3DSG~\citep{D3_MA3DSG2026} encode structural constraints in layouts, scene graphs, tasks, or simulation representations; spatially contextualized scene generation follows the same principle~\citep{D3_Agentic3DSceneGen2025}.
RAISECity~\citep{3D_RAISECity2025}, UnrealLLM~\citep{D3_UnrealLLM2025}, Scenethesis~\citep{D3_Scenethesis2025}, and Code2Worlds~\citep{D3_Code2Worlds2026} use engines, vision, or code execution to reveal construction consequences.
SceneWeaver~\citep{D3_SceneWeaver2025}, Edit3D~\citep{D3_Edit3D2024}, Chat-Edit-3D~\citep{D3_ChatEdit3D2024}, and iControl3D~\citep{D3_iControl3D2024} preserve scene state across critique or successive user edits.
Persistent authoring state carries evolving requirements and deployment constraints~\citep{D3_MUSEAuthoring2026}.
Agentic Designer~\citep{Spatial_AgenticDesigner2026} applies progressive proposal, verification, and adjustment to structurally constrained interior layouts.
Explicit graphs help locate relational failures, but render quality does not establish collision-free, task-compatible, or deployable behavior, which may become observable only during traversal or simulation.
Local repair therefore depends on tracing delayed execution feedback to the responsible relation, asset, or requirement.

\paragraph{Engineered Models.}
Engineered models add dimensional, topological, component, or building relations to multi-object acceptance criteria.
Sketch2BIM~\citep{D3_Sketch2BIM2025} links extracted wall and opening relations to an editable BIM model, SceneGenAgent~\citep{D3_SceneGenAgent2025} grounds industrial scene construction in executable code and rendered verification, and SPADA~\citep{CAD_SPADA2026} couples generated CAD code to deterministic assembly tests.
PhyScensis~\citep{D3_PhyScensis2026} and MapAgent~\citep{D3_MapAgent2026} expose solver or model feedback against persistent spatial state, while interactive interior design exposes user feedback against the same kind of state~\citep{D3_IIDRS2023}.
Here feedback repairs the delivered configuration or assembly; Section~\ref{sec:applications-engineering} examines workflows that revise engineering requirements, solver settings, performance objectives, manufacturing plans, or deployment decisions.

\subsubsection*{Summary}

The spatial profiles expand their represented state from semantic parts and editable geometry to parameters, constraints, layouts, traversal, and component relations.
Rendering exposes appearance, while traversal, simulation, and fabrication reveal failures missed by static inspection.
Repair remains local only when these observations trace back to a responsible part, relation, feature, constraint, or asset.

\subsection{Behavioral Artifacts}
\label{sec:landscape-software}

Behavioral artifacts are accepted through their state-dependent responses to inputs, actions, and environment interaction rather than through a static surface alone.
We distinguish two branches by what that behavior is meant to realize: software systems implement intended functionality or experience, while simulation models approximate a target environment or process.
Within software, the validation horizon widens from bounded repository execution to browser interaction and sampled gameplay trajectories.
Within simulation, the main reference shifts from specified virtual rules to observations and laws outside the artifact.
Operational representations retain the state against which execution or rollout is interpreted; the resulting observations can then direct later changes.
Family assignment follows the defining behavior and repair unit; code that produces a video or spatial model remains in those families.

\subsubsection{Software Systems}

The software profiles differ in how far behavior must unfold before a consequential failure becomes visible.
Repositories emphasize executable and integration checks, web applications add browser-visible action--state sequences, and games require evidence from longer, more open-ended trajectories.

\paragraph{Software repositories.}
Software repositories are accepted through consistency among requirements, implementation units, dependencies, integration behavior, and executable checks.
The repair unit may be a file, module, API, test, or dependency-spanning change.
Paper2Agent~\citep{Code_Paper2Agent2025} illustrates how a textual source can lead to an executable deliverable with a different acceptance criterion.
MapCoder, CodeTree, CodeAgent, and AgentCoder retain planning state, search branches, code, and tests so that execution can redirect later edits~\citep{Code_MapCoder2024,Code_CodeTree2025,Code_CodeAgent2024,Code_AgentCoder2024}.
CodeChain, L2MAC, and ReVeal preserve modules, evolving file stores, or verifier feedback across construction~\citep{Code_CodeChain2024,Code_L2MAC2024,Code_ReVeal2026}.
CodeCRDT coordinates concurrent edits~\citep{Code_CodeCRDT2025}, while EvoMAC and Lessons from the Past retain experience across tasks~\citep{Code_EvoMAC2025,Code_LessonsLearned2025}.
SWE-agent~\citep{Code_SWEAgent2024}, PatchPilot~\citep{Code_PatchPilot2025}, and INDICT~\citep{Code_INDICT2024} connect issue reproduction and critique to patches, safety constraints, and repeated validation.
Repository-scale construction adds dependencies across generated units.
DeNovoSWE and NEMO record requirement or execution relations, while ARC and RUCA-Coder provide test-driven and repository-usage evidence across files~\citep{Code_DeNovoSWE2026,Code_NEMO2026,Code_ARC2026,Code_RUCACoder2026}.
Data-science systems must also preserve the relation among data, experimental code, and measured outcomes~\citep{Code_DSAgent2024,Code_MARS2026,Code_RFAgent2025}.
For front-end repositories, WebDesignIter~\citep{Code_WebDesignIter2026} co-evolves architectural knowledge with differential patches under sandbox execution and testing.
CodeFlowBench~\citep{Code_CodeFlowBench2026} isolates multi-turn reuse and dependency complexity, while NL2Repo-Bench~\citep{Code_NL2RepoBench2026} tests construction of an installable repository from requirements and an empty workspace.
SWE-bench supplies a repository repair environment~\citep{Code_SWEBench2024}; MLAgentBench and PaperBench extend executable evaluation to scientific programs and research replication~\citep{Code_MLAgentBench2024,Code_PaperBench2025}.
BackendForge~\citep{Code_BackendForge2026} uses an OpenAPI specification as an executable contract and co-evolves tests with a reference backend during end-to-end service construction.
When a workflow or agent architecture is itself the delivered artifact, programs, operators, and allocation decisions become directly revisable~\citep{Code_AFlow2025,Code_Cognify2025,Code_ReFuGe2026,Code_ADAS2025}.
Across these cases, compilation, tests, and traces support local repair only when a failure can be traced to the affected file, node, or dependency; passing checks still does not establish the intended semantic result.

\paragraph{Web applications.}
Web applications extend verification beyond executable checks to browser-visible action--state sequences.
They are independently accepted for appearance and interaction, whereas text-first outputs remain textual artifacts when interaction is secondary.
Construction must keep source structure, rendered regions, and behavior synchronized as the artifact changes.
Document-to-web systems make the tension between content preservation and browser-native reorganization explicit~\citep{Website_FullStackAgent2026,Code_WebGenAgent2025}.
WebVIA~\citep{Code_WebVIA2025}, WebGen-Agent~\citep{Code_WebGenAgent2025}, and VisionRefine~\citep{Code_VisionRefine2026} add screenshots or runtime traces to construction feedback.
UI2Code$^N$~\citep{UI_UI2CodeN2026}, FullStack-Agent~\citep{Website_FullStackAgent2026}, and ComUICoder~\citep{Web_ComUICoder2026} combine rendered or development-oriented feedback with localized source changes across components and services.
Persistent web workflows retain cross-session state~\citep{Code_WebDesignIter2026}, while mixed-initiative systems expose user edits~\citep{UI_DashChat2025,UI_DuetUI2026}.
AutoWebWorld~\citep{Website_AutoWebWorld2026} first externalizes pages, actions, and transition rules as a finite-state machine, then uses coding agents to construct an interactive website.
Task success is checked programmatically against the transition graph.
InfiniteWeb~\citep{Website_InfiniteWeb2026} constructs functional multi-page sites from a unified specification and couples task-centric tests, automatic repair, and dense evaluators to the evolving implementation.
ProductWebGen~\citep{Web_ProductWebGen2026} and WebCode2M~\citep{Web_WebCode2M2025} provide complementary benchmark evidence for product-page and design-to-code construction.
Local repair requires links among components, rendered regions, interaction state, and user changes: screenshots expose symptoms without their causes, while source and traces do not settle aesthetics, accessibility, or usability.

\paragraph{Games.}
Games extend the behavioral horizon from individual interactions to sampled play trajectories.
Valid code and rendered assets therefore do not establish balance, narrative coherence, or robustness across states.
Gameplay traces observe the assembled experience~\citep{Game_MultiAgentGameGen2025}.
GameDevBench~\citep{Game_GameDevBench2026} adds agentic game-development tasks whose runtime image and video observations expose failures that code-only feedback misses.
DreamGarden~\citep{Game_DreamGarden2025}, UniGen~\citep{Game_ZeroCode3D2025}, and OpenGame~\citep{Game_OpenGame2026} expose plans, scenes, code, physics, and runtime state inside the game engine.
Story2Game~\citep{Game_Story2Game2025} connects story structures to rules and playable scenes.
AutomatedUnity~\citep{Game_AutomatedUnity2025}, RPGAgent~\citep{Game_RPGAgent2026}, and narrative-to-scene generation~\citep{Game_NarrativetoScene2025} cover related engine-authoring and construction stages.
Sampled playthroughs cover only part of the state space and may not identify whether a failure calls for revising a rule, asset, level, narrative dependency, or behavior policy.

\subsubsection{Simulation Models}
\label{sec:landscape-simulators}
Simulation models are accepted for how faithfully they represent the state and evolution of a target process or environment.
This separates them from games, whose defining output is the designed experience rather than a model used to reproduce or predict behavior.
The delivered model may be symbolic, code-driven, database-backed, learned, physics-based, or hybrid.
The two profiles differ mainly in their reference: virtual world simulators follow specified world and task rules, while physical world models must also agree with observations, laws, or processes outside the artifact.
An internal predictor belongs to a system's representation rather than its delivered artifact, while a simulation-ready 3D environment remains spatial when revision targets its geometry and layout.

\paragraph{Virtual World Simulators.}
Virtual world simulators construct an operational world whose entities, actions, state transitions, rewards, and termination conditions define what can occur.
Their primary reference is the specified world and its task rules.
Transitions should be legal and reproducible under that specification, tasks should remain reachable, and persistent state should evolve consistently across interaction.
Agent2World~\citep{WorldModel_Agent2World2025} constructs symbolic PDDL domains and executable simulators, then uses adaptive unit tests and simulation-based validation to revise behavioral errors.
WorldCoder~\citep{Simulation_WorldCoder2024} instead updates executable Python dynamics directly from environment interactions.
Code World Models~\citep{WorldModel_CodeWorldModels2025} translates natural-language game rules and trajectories into Python functions for state transition, legal-action enumeration, and termination.
The resulting artifact is a verifiable planning simulator rather than a gameplay experience.
Agent World Model~\citep{WorldModel_AgentWorldModel2026} instantiates the same criteria in code-driven, database-backed tool environments with explicit state access and reliable transition and reward execution.
Repair can target a state variable, action precondition, transition rule, reward function, task, or termination condition.
Consistency with the specified world does not settle interface quality or experiential value.
AutoWebWorld~\citep{Website_AutoWebWorld2026} and InfiniteWeb~\citep{Website_InfiniteWeb2026} independently evaluate browser-visible interaction and are therefore treated as web applications; LingBot-World 2.0~\citep{WorldModel_LingBotWorld2_2026} emphasizes interaction horizon and rendered experience, placing it near the game boundary.

\paragraph{Physical World Models.}
Physical world models reconstruct or approximate a process whose relevant behavior exists outside the artifact.
The external process supplies the reference for acceptance.
Geometry, parameters, trajectories, flows, and dynamics must agree with observations, domain constraints, or governing laws over the tested regime.
Perceptual self-reflection revises physics code from rendered animation rather than syntax alone~\citep{Simulation_SelfReflection2026}.
Coding Agent Is Good As World Simulator~\citep{Simulation_CodingAgentWorldSimulator2026} coordinates scene planning, code generation, visual review, and physics analysis to repair both requirement and physical-constraint violations.
Agentic Real2Sim~\citep{Simulation_AgenticReal2Sim2026} starts from a real interaction recording and assembles geometry, object state, physical parameters, poses, and trajectories into a runnable episodic twin.
Sketch2Simulation~\citep{Simulation_Sketch2Sim2026} constructs executable process flowsheets through a graph intermediate representation, simulator-interface code, execution, and structural validation.
SOCIA-$\nabla$~\citep{Simulation_SOCIANabla2025} optimizes simulator code against loss and constraints across user, behavioral, and mobility processes, fitting this profile when an external process supplies the fidelity target.
Repair may target a physical parameter, pose, contact relation, process connection, constraint, governing rule, or executable model component.
A visually plausible rollout or structurally valid flowsheet supports fidelity only over the observations, initial conditions, actions, horizons, and domain laws that were checked.

\subsubsection*{Summary}

The software profiles expand the behavioral horizon from bounded repository checks to browser interactions and sampled gameplay, while the simulation profiles differ in whether their main reference is a specified virtual world or an external process.
Across both branches, execution reveals consequential behavior but covers only part of the relevant state space.
As the horizon grows, obtaining representative observations becomes more expensive and a failed trajectory points less directly to its cause.
Agentic repair therefore depends on tracing observed behavior to the responsible file, component, rule, or model parameter and then revalidating the affected conditions.
Family assignment follows the delivered artifact and its repair unit; Section~\ref{sec:applications} examines workflows that coordinate several independently accepted deliverables.

Across the six families, useful observations range from static inspection through playback to runtime consequences.
Later-stage observations can broaden coverage but become harder to localize unless they remain linked to persistent, editable state.
The shared comparison is whether an observed failure can be traced to the representation and repaired at a bounded unit.

%% file: figures/artifact_taxonomy_tree.tex
\ifdefined\artifactpreprintlayout
\begin{figure}[p]
\else
\begin{figure}[!t]
\fi
\centering
\ifdefined\artifactpreprintlayout
  \newcommand{\taxrowbreak}{\taxsep}
  \newcommand{\taxextra}[1]{\taxsep#1}
  \def\taxyunit{1.17cm}
\else
  \newcommand{\taxrowbreak}{\\[-1pt]}
  \newcommand{\taxextra}[1]{}
  \def\taxyunit{1cm}
\fi
\begingroup
\definecolor{taxInk}{HTML}{26332F}
\definecolor{taxMuted}{HTML}{66716C}
\definecolor{taxText}{HTML}{3F7881}
\definecolor{taxVisual}{HTML}{4F78A5}
\definecolor{taxAudio}{HTML}{756397}
\definecolor{taxVideo}{HTML}{A9584C}
\definecolor{taxSpatial}{HTML}{916A34}
\definecolor{taxSoftware}{HTML}{59687B}

\newcommand{\taxcite}[2]{\mbox{#1~\citep{#2}}}
\newcommand{\taxsep}{%
  \discretionary{}{}{%
    \hbox{\hspace{0.08cm}\textcolor{taxMuted}{\textperiodcentered}\hspace{0.08cm}}%
  }%
}
\resizebox{\textwidth}{!}{%
\begin{tikzpicture}[
  x=1cm,
  y=\taxyunit,
  line cap=round,
  line join=round,
  family/.style={
    rounded corners=3pt,
    minimum width=1.90cm,
    minimum height=0.88cm,
    text width=1.68cm,
    inner xsep=3pt,
    inner ysep=3pt,
    font=\sffamily\scriptsize,
    align=center,
    line width=0.56pt
  },
  profile/.style={
    rounded corners=2pt,
    minimum width=2.45cm,
    minimum height=0.46cm,
    text width=2.20cm,
    inner xsep=4pt,
    inner ysep=1.8pt,
    font=\sffamily\tiny,
    align=left,
    line width=0.46pt
  },
  subprofile/.style={
    rounded corners=2pt,
    minimum width=2.35cm,
    minimum height=0.44cm,
    text width=2.10cm,
    inner xsep=4pt,
    inner ysep=1.6pt,
    fill=white,
    text=taxInk,
    font=\sffamily\tiny\itshape,
    align=left,
    line width=0.40pt
  },
  work/.style={
    rounded corners=2pt,
    minimum width=4.26cm,
    minimum height=0.44cm,
    text width=3.96cm,
    inner xsep=4pt,
    inner ysep=1.4pt,
    fill=white,
    text=taxInk,
    font=\sffamily\tiny,
    align=left,
    line width=0.38pt
  },
  branch/.style={line width=0.56pt}
]

\node[font=\sffamily\tiny\bfseries, text=taxMuted] at (2.55,0.58) {FAMILY};
\node[font=\sffamily\tiny\bfseries, text=taxMuted] at (5.20,0.58) {ANALYTICAL PROFILE};
\node[font=\sffamily\tiny\bfseries, text=taxMuted] at (8.00,0.58) {SECTION CATEGORY};
\node[font=\sffamily\tiny\bfseries, text=taxMuted] at (11.45,0.58) {REPRESENTATIVE WORKS};

\node[
  rounded corners=4pt,
  minimum width=1.52cm,
  minimum height=1.18cm,
  fill=taxInk,
  draw=taxInk,
  text=white,
  align=center,
  inner sep=4pt,
  font=\sffamily
] (root) at (0.55,-7.14) {
  \bfseries\scriptsize AGENTIC\\[-1pt]
  \bfseries\scriptsize ARTIFACT\\[-1pt]
  \bfseries\scriptsize CREATION
};

\node[family, draw=taxText!78, fill=taxText!9, text=taxInk]
  (fText) at (2.55,-1.12) {\textbf{Textual}};
\node[family, draw=taxVisual!78, fill=taxVisual!9, text=taxInk]
  (fVisual) at (2.55,-4.20) {\textbf{2D Visual}};
\node[family, draw=taxAudio!78, fill=taxAudio!9, text=taxInk]
  (fAudio) at (2.55,-6.44) {\textbf{Audio}};
\node[family, draw=taxVideo!78, fill=taxVideo!9, text=taxInk]
  (fVideo) at (2.55,-8.12) {\textbf{Video}};
\node[family, draw=taxSpatial!82, fill=taxSpatial!10, text=taxInk]
  (fSpatial) at (2.55,-10.36) {\textbf{Spatial}};
\node[family, draw=taxSoftware!78, fill=taxSoftware!9, text=taxInk]
  (fSoftware) at (2.55,-13.16) {\textbf{Behavioral}};

\node[profile, draw=taxText!66, fill=taxText!7, text=taxInk]
  (pCreative) at (5.20,-0.28) {Creative Writing};
\node[subprofile, draw=taxText!52]
  (sNarratives) at (8.00,0.00) {Narratives};
\node[work, draw=taxText!42]
  (wNarratives) at (11.45,0.00) {
    \taxcite{StoryBox}{Text_StoryBox2025}\taxsep
    \taxcite{Constella}{Text_Constella2025}\taxrowbreak
    \taxcite{BookWorld}{Text_BookWorld2025}\taxsep
    \taxcite{CreAgentive}{Text_CreAgentive2025}%
    \taxextra{\taxcite{Personas2Plot}{Text_PersonasToPlot2026}}
  };
\node[subprofile, draw=taxText!52]
  (sPerformative) at (8.00,-0.56) {Performative Texts};
\node[work, draw=taxText!42]
  (wPerformative) at (11.45,-0.56) {
    \taxcite{ComedyClub}{Text_ComedyClub2026}\taxsep
    \taxcite{OpenMic}{Text_OpenMic2026}
  };

\node[profile, draw=taxText!66, fill=taxText!7, text=taxInk]
  (pProfessional) at (5.20,-1.40) {Professional Documents};
\node[subprofile, draw=taxText!52]
  (sReports) at (8.00,-1.12) {Informational Reports};
\node[work, draw=taxText!42]
  (wReports) at (11.45,-1.12) {
    \taxcite{STORM}{Text_STORM2024}\taxsep
    \taxcite{FinRpt}{Report_FinRpt2026}\taxrowbreak
    \taxcite{DRACO}{DRACO2025}\taxsep
    \taxcite{ResearchAgent}{Text_ResearchAgent2024}%
    \taxextra{\taxcite{LateralReader}{Report_LateralReader2026}\taxsep
      \taxcite{DeepResearchAgent}{Report_DeepResearchAgent2026}}
  };
\node[subprofile, draw=taxText!52]
  (sFunctional) at (8.00,-1.68) {Functional Documents};
\node[work, draw=taxText!42]
  (wFunctional) at (11.45,-1.68) {
    \taxcite{MCQG-SRefine}{Text_MCQGSRefine2025}\taxsep
    \taxcite{AutoManual}{Text_AutoManual2024}\taxrowbreak
    \taxcite{CRMAgent}{Text_CRMAgent2025}\taxsep
    \taxcite{PAME-AI}{Data_PAMEAI2025}%
    \taxextra{\taxcite{FormAct}{Text_FormAct2026}\taxsep
      \taxcite{SheetAgent}{Text_SheetAgent2025}}
  };

\node[profile, draw=taxText!66, fill=taxText!7, text=taxInk]
  (pScholarly) at (5.20,-2.24) {Scholarly Manuscripts};
\node[work, draw=taxText!42]
  (wScholarly) at (11.45,-2.24) {
    \taxcite{PaperOrchestra}{Text_PaperOrchestra2026}\taxsep
    \taxcite{PaperDebugger}{Text_PaperDebugger2025}\taxrowbreak
    \taxcite{MARG}{Text_MARG2024}\taxsep
    \taxcite{DIAGPaper}{Text_DIAGPaper2026}%
    \taxextra{\taxcite{AI Scientist}{Text_AIScientist2024}\taxsep
      \taxcite{SurGE}{Text_SurGE2026}}
  };

\node[profile, draw=taxVisual!66, fill=taxVisual!7, text=taxInk]
  (pDataViz) at (5.20,-3.08) {Data Visualizations};
\node[work, draw=taxVisual!42]
  (wDataViz) at (11.45,-3.08) {
    \taxcite{AMACE}{DataViz_AMACE2025}\taxsep
    \taxcite{DV-World}{DataViz_DVWorld2026}\taxrowbreak
    \taxcite{DataWink}{Data_DataWink2025}\taxsep
    \taxcite{MultiVis-Agent}{DataVis_MultiVisAgent2026}%
    \taxextra{\taxcite{Data Formulator 2}{DataViz_DataFormulator2_2025}\taxsep
      \taxcite{PlotGen}{DataViz_PlotGen2025}}
  };

\node[profile, draw=taxVisual!66, fill=taxVisual!7, text=taxInk]
  (pIllustrative) at (5.20,-3.92) {Illustrative Graphics};
\node[subprofile, draw=taxVisual!52]
  (sImages) at (8.00,-3.64) {Images};
\node[work, draw=taxVisual!42]
  (wImages) at (11.45,-3.64) {
    \taxcite{T2I-Copilot}{Image_T2ICopilot2025}\taxsep
    \taxcite{CanvasAgent}{Image_CanvasAgent2026}\taxrowbreak
    \taxcite{ReDesign}{Image_ReDesign2026}\taxsep
    \taxcite{CAMEO}{Image_CAMEO2026}%
    \taxextra{\taxcite{GenArtist}{Image_GenArtist2024}\taxsep
      \taxcite{Idea2Img}{Image_Idea2Img2024}}
  };
\node[subprofile, draw=taxVisual!52]
  (sDiagrams) at (8.00,-4.20) {Diagrams};
\node[work, draw=taxVisual!42]
  (wDiagrams) at (11.45,-4.20) {
    \taxcite{Crafter}{Diagram_Crafter2026}\taxsep
    \taxcite{EvoDiagram}{Diagram_EvoDiagram2026}\taxrowbreak
    \taxcite{SciFig}{Diagram_SciFig2026}\taxsep
    \taxcite{DuetSVG}{Diagram_DuetSVG2025}%
    \taxextra{\taxcite{SketchAgent}{Diagram_SketchAgent2025}\taxsep
      \taxcite{PaperBanana}{Image_PaperBanana2026}}
  };

\node[profile, draw=taxVisual!66, fill=taxVisual!7, text=taxInk]
  (pVisualDocs) at (5.20,-5.04) {Visual Documents};
\node[subprofile, draw=taxVisual!52]
  (sPosters) at (8.00,-4.76) {Posters};
\node[work, draw=taxVisual!42]
  (wPosters) at (11.45,-4.76) {
    \taxcite{Paper2Poster}{Poster_Paper2Poster2025}\taxsep
    \taxcite{PosterForest}{Poster_PosterForest2025}\taxrowbreak
    \taxcite{PosterMELD}{Poster_PosterMELD2026}\taxsep
    \taxcite{PosterCraft}{Poster_PosterCraft2025}%
    \taxextra{\taxcite{PosterAgent}{Poster_PosterAgent2026}\taxsep
      \taxcite{AutoDesign}{Poster_AutoDesign2026}}
  };
\node[subprofile, draw=taxVisual!52]
  (sPresentations) at (8.00,-5.32) {Presentations};
\node[work, draw=taxVisual!42]
  (wPresentations) at (11.45,-5.32) {
    \taxcite{SlideGen}{Slide_SlideGen2025}\taxsep
    \taxcite{PPTAgent}{Slide_PPTAgent2025}\taxrowbreak
    \taxcite{ArcDeck}{Slide_ArcDeck2026}\taxsep
    \taxcite{EvoPresent}{Presentation_EvoPresent2026}%
    \taxextra{\taxcite{Auto-Slides}{Slide_AutoSlides2025}\taxsep
      \taxcite{SlideBot}{Slide_SlideBot2026}}
  };

\node[profile, draw=taxAudio!66, fill=taxAudio!7, text=taxInk]
  (pMusic) at (5.20,-5.92) {Music};
\node[work, draw=taxAudio!42]
  (wMusic) at (11.45,-5.92) {
    \taxcite{CoComposer}{Music_CoComposer2025}\taxsep
    \taxcite{Libretto}{Music_Libretto2026}\taxrowbreak
    \taxcite{MusicSwarm}{Music_MusicSwarm2025}\taxsep
    \taxcite{RIME}{Music_RIME2026}%
    \taxextra{\taxcite{WeaveMuse}{Music_WeaveMuse2025}}
  };
\node[profile, draw=taxAudio!66, fill=taxAudio!7, text=taxInk]
  (pSpoken) at (5.20,-6.44) {Spoken Audio};
\node[work, draw=taxAudio!42]
  (wSpoken) at (11.45,-6.44) {
    \taxcite{PodAgent}{Audio_PodAgent2025}\taxsep
    \taxcite{Audio-Agent}{Audio_AudioAgent2024}\taxrowbreak
    \taxcite{AudioToolAgent}{Audio_AudioToolAgent2025}
  };
\node[profile, draw=taxAudio!66, fill=taxAudio!7, text=taxInk]
  (pSoundscapes) at (5.20,-6.96) {Soundscapes};
\node[work, draw=taxAudio!42]
  (wSoundscapes) at (11.45,-6.96) {
    \taxcite{WavCraft}{Audio_WavCraft2024}\taxsep
    \taxcite{LVAS-Agent}{Audio_LVASAgent2025}\taxrowbreak
    \taxcite{AudioGenie}{Audio_AudioGenie2025}\taxsep
    \taxcite{Audio-Oscar}{Audio_AudioOscar2026}%
    \taxextra{\taxcite{AudioRAG+}{Audio_AudioRAGPlus2026}\taxsep
      \taxcite{SoundscapeAgent}{Audio_SoundscapeAgent2026}}
  };

\node[profile, draw=taxVideo!66, fill=taxVideo!7, text=taxInk]
  (pExpository) at (5.20,-7.56) {Expository Videos};
\node[work, draw=taxVideo!42]
  (wExpository) at (11.45,-7.56) {
    \taxcite{Paper2Video}{Video_Paper2Video2025}\taxsep
    \taxcite{Code2Video}{Video_Code2Video2025}\taxrowbreak
    \taxcite{ManimAgent}{Animation_ManimAgent2026}\taxsep
    \taxcite{VideoAgent-Sci}{Video_VideoAgentSci2025}%
    \taxextra{\taxcite{MapStory}{Animation_MapStory2025}\taxsep
      \taxcite{Beyond E2E}{Video_BeyondE2E2026}}
  };
\node[profile, draw=taxVideo!66, fill=taxVideo!7, text=taxInk]
  (pNarrativeVideo) at (5.20,-8.12) {Narrative Videos};
\node[work, draw=taxVideo!42]
  (wNarrativeVideo) at (11.45,-8.12) {
    \taxcite{VideoMemory}{Video_VideoMemory2026}\taxsep
    \taxcite{FilmWorld}{Video_FilmWorld2026}\taxrowbreak
    \taxcite{StoryAgent}{Video_StoryAgent2024}\taxsep
    \taxcite{GenMAC}{Video_GenMAC2024}%
    \taxextra{\taxcite{AniME}{Video_AniME2025}\taxsep
      \taxcite{EventGraphs}{Video_EventGraphs2026}}
  };
\node[profile, draw=taxVideo!66, fill=taxVideo!7, text=taxInk]
  (pPromotionalVideo) at (5.20,-8.68) {Promotional Videos};
\node[work, draw=taxVideo!42]
  (wPromotionalVideo) at (11.45,-8.68) {
    \taxcite{ATVG}{Video_ATVG2026}\taxsep
    \taxcite{BrandFusion}{Video_BrandFusion2026}\taxrowbreak
    \taxcite{PersonaVlog}{Video_PersonaVlog2025}\taxsep
    \taxcite{AutoMV}{Video_AutoMV2025}%
    \taxextra{\taxcite{GLANCE}{Video_GLANCE2026}}
  };

\node[profile, draw=taxSpatial!68, fill=taxSpatial!7, text=taxInk]
  (pAssets) at (5.20,-9.80) {3D Assets};
\node[subprofile, draw=taxSpatial!54]
  (sVisualAssets) at (8.00,-9.52) {Visual Assets};
\node[work, draw=taxSpatial!44]
  (wVisualAssets) at (11.45,-9.52) {
    \taxcite{ShapeCraft}{3D_ShapeCraft2025}\taxsep
    \taxcite{LL3M}{D3_LL3M2025}\taxrowbreak
    \taxcite{SmartAvatar}{D3_SmartAvatar2025}\taxsep
    \taxcite{MeshLLM}{D3_MeshLLM2025}
  };
\node[subprofile, draw=taxSpatial!54]
  (sParametric) at (8.00,-10.08) {Parametric Models};
\node[work, draw=taxSpatial!44]
  (wParametric) at (11.45,-10.08) {
    \taxcite{CADIR}{CAD_CADIR2026}\taxsep
    \taxcite{ArtisanCAD}{CAD_ArtisanCAD2026}\taxrowbreak
    \taxcite{STEP-LLM}{CAD_STEPLLM2026}\taxsep
    \taxcite{NURBGen}{CAD_NURBGen2026}%
    \taxextra{\taxcite{Seek-CAD}{CAD_SeekCAD2025}\taxsep
      \taxcite{CADDesigner}{CAD_CADDesigner2025}}
  };

\node[profile, draw=taxSpatial!68, fill=taxSpatial!7, text=taxInk]
  (pScenes) at (5.20,-10.92) {3D Scenes};
\node[subprofile, draw=taxSpatial!54]
  (sSpatialWorlds) at (8.00,-10.64) {Spatial Worlds};
\node[work, draw=taxSpatial!44]
  (wSpatialWorlds) at (11.45,-10.64) {
    \taxcite{SAGE}{3D_SAGE2026}\taxsep
    \taxcite{SceneSmith}{3D_SceneSmith2026}\taxrowbreak
    \taxcite{WorldGen}{D3_WorldGen2025}\taxsep
    \taxcite{WorldCraft}{D3_WorldCraft2025}%
    \taxextra{\taxcite{SceneWeaver}{D3_SceneWeaver2025}\taxsep
      \taxcite{RoomPlanner}{3D_RoomPlanner2025}}
  };
\node[subprofile, draw=taxSpatial!54]
  (sEngineered) at (8.00,-11.20) {Engineered Models};
\node[work, draw=taxSpatial!44]
  (wEngineered) at (11.45,-11.20) {
    \taxcite{Sketch2BIM}{D3_Sketch2BIM2025}\taxsep
    \taxcite{PhyScensis}{D3_PhyScensis2026}\taxrowbreak
    \taxcite{MapAgent}{D3_MapAgent2026}\taxsep
    \taxcite{SPADA}{CAD_SPADA2026}%
    \taxextra{\taxcite{SceneGenAgent}{D3_SceneGenAgent2025}\taxsep
      \taxcite{IIDRS}{D3_IIDRS2023}}
  };

\node[profile, draw=taxSoftware!66, fill=taxSoftware!7, text=taxInk]
  (pSoftware) at (5.20,-12.60) {Software Systems};
\node[subprofile, draw=taxSoftware!52]
  (sRepositories) at (8.00,-12.04) {Software Repositories};
\node[work, draw=taxSoftware!42]
  (wRepositories) at (11.45,-12.04) {
    \taxcite{ARC}{Code_ARC2026}\taxsep
    \taxcite{DeNovoSWE}{Code_DeNovoSWE2026}\taxrowbreak
    \taxcite{NL2Repo-Bench}{Code_NL2RepoBench2026}\taxsep
    \taxcite{BackendForge}{Code_BackendForge2026}%
    \taxextra{\taxcite{SWE-agent}{Code_SWEAgent2024}\taxsep
      \taxcite{CodeCRDT}{Code_CodeCRDT2025}}
  };
\node[subprofile, draw=taxSoftware!52]
  (sWebApps) at (8.00,-12.60) {Web Applications};
\node[work, draw=taxSoftware!42]
  (wWebApps) at (11.45,-12.60) {
    \taxcite{WebGen-Agent}{Code_WebGenAgent2025}\taxsep
    \taxcite{FullStack-Agent}{Website_FullStackAgent2026}\taxrowbreak
    \taxcite{DuetUI}{UI_DuetUI2026}\taxsep
    \taxcite{InfiniteWeb}{Website_InfiniteWeb2026}%
    \taxextra{\taxcite{WebVIA}{Code_WebVIA2025}\taxsep
      \taxcite{UI2Code}{UI_UI2CodeN2026}}
  };
\node[subprofile, draw=taxSoftware!52]
  (sGames) at (8.00,-13.16) {Games};
\node[work, draw=taxSoftware!42]
  (wGames) at (11.45,-13.16) {
    \taxcite{GameDevBench}{Game_GameDevBench2026}\taxsep
    \taxcite{DreamGarden}{Game_DreamGarden2025}\taxrowbreak
    \taxcite{OpenGame}{Game_OpenGame2026}\taxsep
    \taxcite{Story2Game}{Game_Story2Game2025}%
    \taxextra{\taxcite{UniGen}{Game_ZeroCode3D2025}\taxsep
      \taxcite{MultiAgentGameGen}{Game_MultiAgentGameGen2025}}
  };

\node[profile, draw=taxSoftware!66, fill=taxSoftware!7, text=taxInk]
  (pSimulation) at (5.20,-14.00) {Simulation Models};
\node[subprofile, draw=taxSoftware!52]
  (sVirtualWorlds) at (8.00,-13.72) {Virtual World\\[-1pt]Simulators};
\node[work, draw=taxSoftware!42]
  (wVirtualWorlds) at (11.45,-13.72) {
    \taxcite{Agent2World}{WorldModel_Agent2World2025}\taxsep
    \taxcite{WorldCoder}{Simulation_WorldCoder2024}\taxrowbreak
    \taxcite{CWM}{WorldModel_CodeWorldModels2025}\taxsep
    \taxcite{AWM}{WorldModel_AgentWorldModel2026}%
    \taxextra{\taxcite{LingBot-World 2.0}{WorldModel_LingBotWorld2_2026}}
  };
\node[subprofile, draw=taxSoftware!52]
  (sPhysicalWorlds) at (8.00,-14.28) {Physical World Models};
\node[work, draw=taxSoftware!42]
  (wPhysicalWorlds) at (11.45,-14.28) {
    \taxcite{Self-Reflection}{Simulation_SelfReflection2026}\taxsep
    \taxcite{SOCIA-Nabla}{Simulation_SOCIANabla2025}\taxrowbreak
    \taxcite{Real2Sim}{Simulation_AgenticReal2Sim2026}\taxsep
    \taxcite{Sketch2Sim}{Simulation_Sketch2Sim2026}%
    \taxextra{\taxcite{CodingAgentSim}{Simulation_CodingAgentWorldSimulator2026}}
  };

\begin{pgfonlayer}{background}
  \fill[taxText!2, rounded corners=3pt] (1.48,0.34) rectangle (13.65,-2.58);
  \fill[taxVisual!2, rounded corners=3pt] (1.48,-2.74) rectangle (13.65,-5.66);
  \fill[taxAudio!2, rounded corners=3pt] (1.48,-5.68) rectangle (13.65,-7.24);
  \fill[taxVideo!2, rounded corners=3pt] (1.48,-7.26) rectangle (13.65,-9.02);
  \fill[taxSpatial!2, rounded corners=3pt] (1.48,-9.18) rectangle (13.65,-11.54);
  \fill[taxSoftware!2, rounded corners=3pt] (1.48,-11.70) rectangle (13.65,-14.62);

  \draw[branch, draw=taxInk!44] (root.east) -- (1.42,-7.14);
  \draw[branch, draw=taxInk!44] (1.42,-1.12) -- (1.42,-13.16);
  \draw[branch, draw=taxText!80] (1.42,-1.12) -- (fText.west);
  \fill[taxText] (1.42,-1.12) circle (1.10pt);
  \draw[branch, draw=taxVisual!80] (1.42,-4.20) -- (fVisual.west);
  \fill[taxVisual] (1.42,-4.20) circle (1.10pt);
  \draw[branch, draw=taxAudio!80] (1.42,-6.44) -- (fAudio.west);
  \fill[taxAudio] (1.42,-6.44) circle (1.10pt);
  \draw[branch, draw=taxVideo!80] (1.42,-8.12) -- (fVideo.west);
  \fill[taxVideo] (1.42,-8.12) circle (1.10pt);
  \draw[branch, draw=taxSpatial!82] (1.42,-10.36) -- (fSpatial.west);
  \fill[taxSpatial] (1.42,-10.36) circle (1.10pt);
  \draw[branch, draw=taxSoftware!80] (1.42,-13.16) -- (fSoftware.west);
  \fill[taxSoftware] (1.42,-13.16) circle (1.10pt);

  \draw[branch, draw=taxText!74] (fText.east) -- (3.78,-1.12);
  \draw[branch, draw=taxText!74] (3.78,-0.28) -- (3.78,-2.24);
  \draw[branch, draw=taxText!74] (3.78,-0.28) -- (pCreative.west);
  \draw[branch, draw=taxText!74] (3.78,-1.40) -- (pProfessional.west);
  \draw[branch, draw=taxText!74] (3.78,-2.24) -- (pScholarly.west);

  \draw[branch, draw=taxVisual!74] (fVisual.east) -- (3.78,-4.20);
  \draw[branch, draw=taxVisual!74] (3.78,-3.08) -- (3.78,-5.04);
  \draw[branch, draw=taxVisual!74] (3.78,-3.08) -- (pDataViz.west);
  \draw[branch, draw=taxVisual!74] (3.78,-3.92) -- (pIllustrative.west);
  \draw[branch, draw=taxVisual!74] (3.78,-5.04) -- (pVisualDocs.west);

  \draw[branch, draw=taxAudio!74] (fAudio.east) -- (3.78,-6.44);
  \draw[branch, draw=taxAudio!74] (3.78,-5.92) -- (3.78,-6.96);
  \draw[branch, draw=taxAudio!74] (3.78,-5.92) -- (pMusic.west);
  \draw[branch, draw=taxAudio!74] (3.78,-6.44) -- (pSpoken.west);
  \draw[branch, draw=taxAudio!74] (3.78,-6.96) -- (pSoundscapes.west);

  \draw[branch, draw=taxVideo!74] (fVideo.east) -- (3.78,-8.12);
  \draw[branch, draw=taxVideo!74] (3.78,-7.56) -- (3.78,-8.68);
  \draw[branch, draw=taxVideo!74] (3.78,-7.56) -- (pExpository.west);
  \draw[branch, draw=taxVideo!74] (3.78,-8.12) -- (pNarrativeVideo.west);
  \draw[branch, draw=taxVideo!74] (3.78,-8.68) -- (pPromotionalVideo.west);

  \draw[branch, draw=taxSpatial!76] (fSpatial.east) -- (3.78,-10.36);
  \draw[branch, draw=taxSpatial!76] (3.78,-9.80) -- (3.78,-10.92);
  \draw[branch, draw=taxSpatial!76] (3.78,-9.80) -- (pAssets.west);
  \draw[branch, draw=taxSpatial!76] (3.78,-10.92) -- (pScenes.west);

  \draw[branch, draw=taxSoftware!74] (fSoftware.east) -- (3.78,-13.16);
  \draw[branch, draw=taxSoftware!74] (3.78,-12.60) -- (3.78,-14.00);
  \draw[branch, draw=taxSoftware!74] (3.78,-12.60) -- (pSoftware.west);
  \draw[branch, draw=taxSoftware!74] (3.78,-14.00) -- (pSimulation.west);

  \draw[branch, draw=taxText!62] (pCreative.east) -- (6.68,-0.28);
  \draw[branch, draw=taxText!62] (6.68,0.00) -- (6.68,-0.56);
  \draw[branch, draw=taxText!62] (6.68,0.00) -- (sNarratives.west);
  \draw[branch, draw=taxText!62] (6.68,-0.56) -- (sPerformative.west);
  \draw[branch, draw=taxText!62] (pProfessional.east) -- (6.68,-1.40);
  \draw[branch, draw=taxText!62] (6.68,-1.12) -- (6.68,-1.68);
  \draw[branch, draw=taxText!62] (6.68,-1.12) -- (sReports.west);
  \draw[branch, draw=taxText!62] (6.68,-1.68) -- (sFunctional.west);

  \draw[branch, draw=taxVisual!62] (pIllustrative.east) -- (6.68,-3.92);
  \draw[branch, draw=taxVisual!62] (6.68,-3.64) -- (6.68,-4.20);
  \draw[branch, draw=taxVisual!62] (6.68,-3.64) -- (sImages.west);
  \draw[branch, draw=taxVisual!62] (6.68,-4.20) -- (sDiagrams.west);
  \draw[branch, draw=taxVisual!62] (pVisualDocs.east) -- (6.68,-5.04);
  \draw[branch, draw=taxVisual!62] (6.68,-4.76) -- (6.68,-5.32);
  \draw[branch, draw=taxVisual!62] (6.68,-4.76) -- (sPosters.west);
  \draw[branch, draw=taxVisual!62] (6.68,-5.32) -- (sPresentations.west);

  \draw[branch, draw=taxSpatial!64] (pAssets.east) -- (6.68,-9.80);
  \draw[branch, draw=taxSpatial!64] (6.68,-9.52) -- (6.68,-10.08);
  \draw[branch, draw=taxSpatial!64] (6.68,-9.52) -- (sVisualAssets.west);
  \draw[branch, draw=taxSpatial!64] (6.68,-10.08) -- (sParametric.west);
  \draw[branch, draw=taxSpatial!64] (pScenes.east) -- (6.68,-10.92);
  \draw[branch, draw=taxSpatial!64] (6.68,-10.64) -- (6.68,-11.20);
  \draw[branch, draw=taxSpatial!64] (6.68,-10.64) -- (sSpatialWorlds.west);
  \draw[branch, draw=taxSpatial!64] (6.68,-11.20) -- (sEngineered.west);

  \draw[branch, draw=taxSoftware!62] (pSoftware.east) -- (6.68,-12.60);
  \draw[branch, draw=taxSoftware!62] (6.68,-12.04) -- (6.68,-13.16);
  \draw[branch, draw=taxSoftware!62] (6.68,-12.04) -- (sRepositories.west);
  \draw[branch, draw=taxSoftware!62] (6.68,-12.60) -- (sWebApps.west);
  \draw[branch, draw=taxSoftware!62] (6.68,-13.16) -- (sGames.west);
  \draw[branch, draw=taxSoftware!62] (pSimulation.east) -- (6.68,-14.00);
  \draw[branch, draw=taxSoftware!62] (6.68,-13.72) -- (6.68,-14.28);
  \draw[branch, draw=taxSoftware!62] (6.68,-13.72) -- (sVirtualWorlds.west);
  \draw[branch, draw=taxSoftware!62] (6.68,-14.28) -- (sPhysicalWorlds.west);

  \draw[branch, draw=taxText!54] (pScholarly.east) -- (wScholarly.west);
  \draw[branch, draw=taxVisual!54] (pDataViz.east) -- (wDataViz.west);
  \draw[branch, draw=taxAudio!54] (pMusic.east) -- (wMusic.west);
  \draw[branch, draw=taxAudio!54] (pSpoken.east) -- (wSpoken.west);
  \draw[branch, draw=taxAudio!54] (pSoundscapes.east) -- (wSoundscapes.west);
  \draw[branch, draw=taxVideo!54] (pExpository.east) -- (wExpository.west);
  \draw[branch, draw=taxVideo!54] (pNarrativeVideo.east) -- (wNarrativeVideo.west);
  \draw[branch, draw=taxVideo!54] (pPromotionalVideo.east) -- (wPromotionalVideo.west);

  \draw[branch, draw=taxText!54] (sNarratives.east) -- (wNarratives.west);
  \draw[branch, draw=taxText!54] (sPerformative.east) -- (wPerformative.west);
  \draw[branch, draw=taxText!54] (sReports.east) -- (wReports.west);
  \draw[branch, draw=taxText!54] (sFunctional.east) -- (wFunctional.west);
  \draw[branch, draw=taxVisual!54] (sImages.east) -- (wImages.west);
  \draw[branch, draw=taxVisual!54] (sDiagrams.east) -- (wDiagrams.west);
  \draw[branch, draw=taxVisual!54] (sPosters.east) -- (wPosters.west);
  \draw[branch, draw=taxVisual!54] (sPresentations.east) -- (wPresentations.west);
  \draw[branch, draw=taxSpatial!56] (sVisualAssets.east) -- (wVisualAssets.west);
  \draw[branch, draw=taxSpatial!56] (sParametric.east) -- (wParametric.west);
  \draw[branch, draw=taxSpatial!56] (sSpatialWorlds.east) -- (wSpatialWorlds.west);
  \draw[branch, draw=taxSpatial!56] (sEngineered.east) -- (wEngineered.west);
  \draw[branch, draw=taxSoftware!54] (sRepositories.east) -- (wRepositories.west);
  \draw[branch, draw=taxSoftware!54] (sWebApps.east) -- (wWebApps.west);
  \draw[branch, draw=taxSoftware!54] (sGames.east) -- (wGames.west);
  \draw[branch, draw=taxSoftware!54] (sVirtualWorlds.east) -- (wVirtualWorlds.west);
  \draw[branch, draw=taxSoftware!54] (sPhysicalWorlds.east) -- (wPhysicalWorlds.west);
\end{pgfonlayer}

\end{tikzpicture}%
}
\endgroup
\ifdefined\artifactpreprintlayout
\caption{Organization of the artifact-family landscape. The six artifact families are organized into 16 analytical profiles and, where needed, finer section categories. Each terminal category lists up to six representative works. Profiles without a finer subdivision connect directly to their examples.}
\Description{A left-to-right literature navigation map. Agentic artifact creation branches into six primary families and the 16 analytical profiles used in the section outline. Creative Writing, Professional Documents, Illustrative Graphics, Visual Documents, 3D Assets, 3D Scenes, Software Systems, and Simulation Models expand into their finer section-level categories. Every terminal category connects to up to six representative works.}
\else
\caption{Organization of the artifact-family landscape. The six artifact families are organized into 16 analytical profiles and, where needed, finer section categories. Each terminal category lists up to four representative works. Profiles without a finer subdivision connect directly to their examples.}
\Description{A left-to-right literature navigation map. Agentic artifact creation branches into six primary families and the 16 analytical profiles used in the section outline. Creative Writing, Professional Documents, Illustrative Graphics, Visual Documents, 3D Assets, 3D Scenes, Software Systems, and Simulation Models expand into their finer section-level categories. Every terminal category connects to up to four representative works.}
\fi
\label{fig:artifact-taxonomy-tree}
\end{figure}

%% file: sections/04b_applications.tex
\section{Applications}
\label{sec:applications}

\begin{figure}[!t]
\centering
\includegraphics[width=\textwidth]{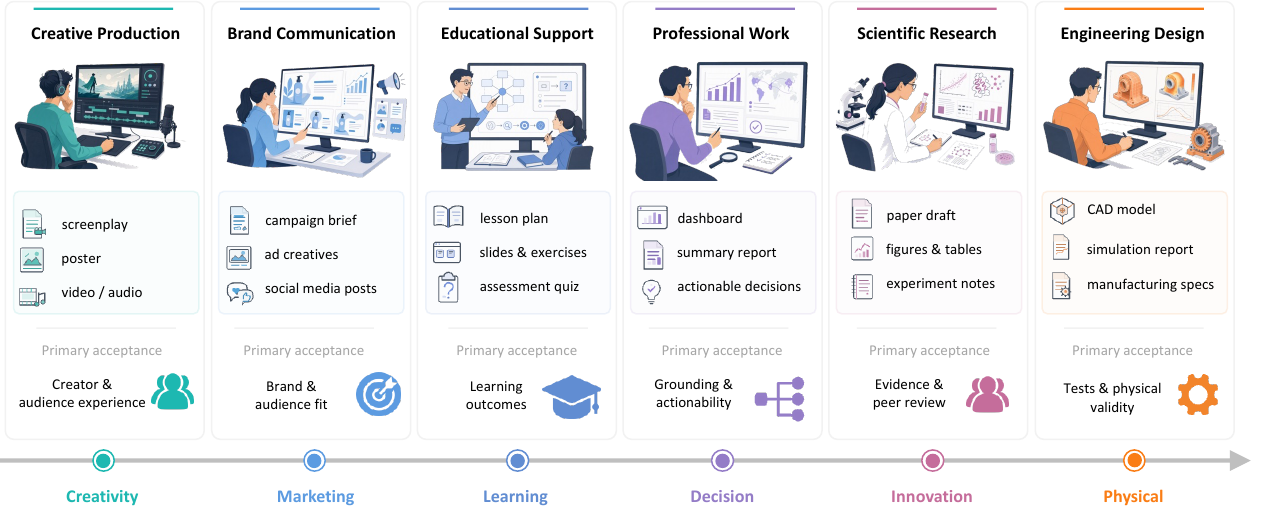}
\caption{Application contexts reconfigure shared artifact families through different objectives, artifact bundles, acceptance criteria, and human authority. The panels follow the shifting locus of primary acceptance from creator and audience experience through brand, learning, decision, and evidence concerns to executable or physical validity.}
\Description{Six application-context cards progress from creative production, brand communication, and educational support to professional work, scientific research, and engineering design. Each neutrally framed card uses a short colored top rail, an illustration, a lightly tinted module listing three characteristic artifacts, and a white primary-acceptance area with a contextual icon. A thin bottom axis orders the contexts from human and experiential acceptance on the left through experience, brand, learning, decision, and evidence to external and physical acceptance on the right. Category accents follow a continuous aqua-to-blue-to-violet-to-orange sequence along this progression.}
\label{fig:application-acceptance-horizon}
\end{figure}

Artifact family identifies the delivered object and revised state, whereas application context determines the domain objective, participating artifacts and tools, and authority over consequential use.
The operational boundary is what changes: artifact profiles track one independently accepted deliverable, while application profiles track workflow decisions that may propagate across artifacts or stages.
The same artifact can serve several domains, and one domain workflow may require several artifact families.
We organize the evidence into six contexts: creative production, brand communication, educational support, professional work, scientific research, and engineering design.
Multi-context systems are assigned by the downstream objective governing final acceptance.
Across these contexts, the primary locus of acceptance ranges from creator and audience experience through brand commitments, learning outcomes, professional judgment, and scientific evidence to executable or physical validity.
Figure~\ref{fig:application-acceptance-horizon} summarizes how these contexts reconfigure artifact bundles, specialized tools, acceptance evidence, and human authority.

Creators or users specify editorial and experiential preferences, domain experts contribute specialized judgment, and authorized decision-makers may approve consequential use.
One person may occupy several roles, but their information and decision rights differ.

\subsection{Creative Production}

Creative production aims to shape coherent audience experiences across interdependent narrative, visual, auditory, and interactive decisions.
Rendering, playback, and interaction can redirect editable plans and assets before delivery, whether the result is composed media or an experience that remains open to user action.

\textit{Media production.}
Media production coordinates narrative, character, pacing, sound, and visual style before delivery.
StoryAgent~\citep{Video_StoryAgent2024}, AniME~\citep{Video_AniME2025}, and MUSE~\citep{Video_MUSE2026} connect planning, scene construction, and assembly to audiovisual tools, while PodAgent~\citep{Audio_PodAgent2025} distributes conversational-audio production across roles.
Personalized media adds identity and preference constraints without making the delivered work interactive~\citep{Video_PersonaVlog2025}.
Narrative intentionality, dramatic coherence, style, and pacing require creator judgment alongside structural and rendering checks.

\textit{Interactive entertainment.}
In interactive entertainment, editable branching state persists after initial assembly.
Creator-facing tools~\citep{Game_CreatorCentric2026} expose process control, while DreamGarden~\citep{Game_DreamGarden2025} and LingBot-World 2.0~\citep{WorldModel_LingBotWorld2_2026} extend dependencies across user-conditioned interaction.
OpenGame~\citep{Game_OpenGame2026} adds execution- and vision-guided repair during end-to-end web-game construction.
Runtime validity and playability therefore join narrative and stylistic coherence among the acceptance criteria.
Creators retain editorial authority even when planning, production, and review are delegated.

\subsection{Brand Communication}

Brand communication aims to keep messages and identity consistent across channels while adapting to customer context and audience response.
Agentic workflows can carry shared commitments across assets and use observed responses to revise later work, whether the accepted result is a coordinated campaign or an ongoing communication process.

\textit{Campaign production.}
Campaign production brings copy, banners, posters, landing pages, short videos, and channel-specific variants under shared brand and message commitments.
BannerAgency~\citep{Marketing_BannerAgency2025} and MIMO~\citep{Image_MirrorAd2025} construct ad banners through role-specific or reflective processes, while BrandFusion~\citep{Video_BrandFusion2026}, AutoMV~\citep{Video_AutoMV2025}, and ATVG~\citep{Video_ATVG2026} add brand, music, or factual constraints to video production.
Rendering and format checks can reject broken assets; campaign-level consistency and release remain brand-owner decisions.

\textit{Customer engagement.}
Customer engagement conditions communication on records or observed audience response: CRMAgent~\citep{Text_CRMAgent2025} grounds text in CRM state, whereas AutoPP~\citep{Poster_AutoPP2026} uses click-through feedback to revise poster elements.
Response signals can redirect subsequent actions but remain noisy proxies for commercial goals.
Source-grounded material produced primarily for public understanding belongs instead to professional work.

\subsection{Educational Support}

Educational support aims to align instruction with learning goals while adapting to learners over time.
Assessment and interaction feedback can redirect the construction of either reusable instructional resources or an evolving learning trajectory.

\textit{Curriculum development.}
EduAgentQG~\citep{Education_EduAgentQG2025} and MCQG-SRefine~\citep{Text_MCQGSRefine2025} iteratively construct and revise assessment items.
TeachMaster~\citep{Edu_GenerativeTeaching2025} produces editable, curriculum-ready educational videos, while related systems generate multilingual materials and course structures~\citep{Education_AgenticEducational2025,Edu_CourseSyllabus2025}.
Auto-Slides~\citep{Slide_AutoSlides2025} constructs pedagogically structured presentations from research papers and supports interactive customization to learner knowledge and goals.
The acceptance criteria combine curricular alignment with structural, execution, accessibility, and assessment requirements.

\textit{Adaptive learning.}
Adaptive learning conditions explanation, interaction, or pacing on learner goals and responses.
GenMentor~\citep{Education_GoalOrientedLearning2025} and Learn Your Way~\citep{Education_AIAugmentedTextbook2025} extend maintained state from instructional content to the learner's evolving context.
DeepTutor~\citep{Education_DeepTutor2026} updates learner memory after each interaction to condition later tutoring and question generation.
SocialCoach~\citep{Education_SocialCoach2026} turns this state into a practice loop in which an RL-optimized scheduler selects role-play scenarios and session outcomes inform later tutoring and practice.
Teachers retain authority over learning commitments, while learner feedback may redirect presentation without silently changing those commitments.

\subsection{Professional Work}

Professional work turns evidence and operational state into reports, analyses, messages, and public information used in organizational decisions, professional practice, or public release.
Agentic workflows connect changing sources, tool results, and review to later revisions, while acceptance remains with an organizational decision-maker, regulated practitioner, or editor.

\textit{Business intelligence.}
Business intelligence connects reports, data, notebooks, dashboards, and enterprise knowledge to organizational decisions.
DatawiseAgent~\citep{Data_DatawiseAgent2025}, AMACE~\citep{DataViz_AMACE2025}, Data Formulator 2~\citep{DataViz_DataFormulator2_2025}, LightVA~\citep{DataViz_LightVA2024}, and Jupybara~\citep{Data_Jupybara2025} connect data and visual analysis to task-specific or user review; ProfiliTable~\citep{Text_ProfiliTable2026} and SheetAgent~\citep{Text_SheetAgent2025} retain execution feedback in table manipulation; AI Knowledge Assist~\citep{Data_AIKnowledgeAssist2025} constructs enterprise knowledge from service records.

\textit{Regulated services.}
LAW~\citep{Text_LAW2024} applies tool-assisted question answering and analysis to legal contracts, while PAME-AI~\citep{Data_PAMEAI2025} constructs and optimizes patient messages.
Both settings place domain context and professional review around consequential outputs.
Recorded provenance or a passing check can support auditability, but qualified professionals remain responsible for consequential interpretation and action.

\textit{Public information.}
Journalistic systems construct material for public understanding rather than persuasion, with acceptance centered on sourcing, correction, and editorial responsibility.
Data Journalist Agent~\citep{Data_DataJournalistAgent2026} coordinates data analysis and evidence-backed writing into verifiable multimodal articles.
LateralReader~\citep{Report_LateralReader2026} and the DRAGUN resources~\citep{Report_DRAGUNResources2026} expose evidence-sufficiency and attributed-report tasks; source lineage and review status aid inspection, but an editor still authorizes publication.

\subsection{Scientific Research}
\label{sec:applications-research}

Scientific research aims either to produce new evidence or to communicate established evidence faithfully.
Experimental results, executable checks, and review can redirect hypotheses, analyses, or presentation, distinguishing discovery workflows that change the evidence base from research communication that preserves it.

\textit{Discovery workflows.}
Discovery workflows coordinate questions, protocols, code, experiments, and analyses in pursuit of new evidence.
The AI Scientist~\citep{Text_AIScientist2024} connects ideation to computational experiments and manuscript production; Robin~\citep{Sci_Robin2026} extends this pattern to experimental biology by using analysed laboratory results to revise hypotheses and propose follow-up assays.
Autonomous-laboratory work demonstrates deterministic instrument control and multistep planning~\citep{Laboratory_AutonomousLaboratory2026}.
IdeaSynth~\citep{Sci_IdeaSynth2025} supports collaborative ideation, while Queryome~\citep{Sci_Queryome2025} and Elhuyar~\citep{Sci_ClosedLoopDiscovery2026} support iterative literature synthesis.
ChemCRAFT~\citep{Molecular_ChemCRAFT2026} optimizes molecules through chemical and protein--ligand feedback, while DrugAgent~\citep{Drug_DrugAgent2024} automates observation-guided ML programming for drug-discovery prediction tasks.
DS-Agent~\citep{Code_DSAgent2024}, AutoML-Agent~\citep{Code_AutoMLAgent2025}, and MARS~\citep{Code_MARS2026} extend executable analysis and model development across scientific tasks.
SciToolAgent~\citep{Sci_SciToolAgent2025} extends scientific tool orchestration; smart-microscopy work offers a conceptual roadmap for multi-agent adaptive instrumentation~\citep{Data_ReconceptualizingSmartMicroscopy2025}; the limitation-identification benchmark contributes evaluation evidence rather than another construction system~\citep{Data_CanLLMsIdentifyLimitations2025}.
These channels can test protocol consistency, execution, and reproducibility, but researchers remain responsible for methods, evidence, ethics, and claims.

\textit{Research communication.}
Research communication preserves an evidence base while revising its organization and presentation.
PaperDebugger~\citep{Text_PaperDebugger2025} revises manuscripts, while Paper2Agent~\citep{Code_Paper2Agent2025}, Paper2Poster~\citep{Poster_Paper2Poster2025}, SlideGen~\citep{Slide_SlideGen2025}, and Paper2Video~\citep{Video_Paper2Video2025} express shared evidence through different artifacts.
MLR-Bench~\citep{Text_MLRBench2025} and SurGE~\citep{Text_SurGE2026} make research-process and survey-construction obligations explicit evaluation targets.
Source fidelity, internal consistency, and traceability constrain these outputs; expert and community review assess novelty, significance, and interpretation.

\subsection{Engineering Design}
\label{sec:applications-engineering}

Engineering design aims to produce systems and models whose validity can be checked through execution, fabrication, simulation, or deployment.
Observed consequences can redirect requirements, geometry, control logic, or model parameters, with acceptance centered either on a deployable system or on a simulator intended to reproduce a target process or environment.

\textit{System development.}
System development coordinates requirements, architecture, geometry, control logic, and validation around a deployable or fabricable target.
CADIR~\citep{CAD_CADIR2026} and SPADA~\citep{CAD_SPADA2026} remain spatial-artifact cases in \S\ref{sec:landscape-3d}, with editable geometry and assembly tests; Sketch2BIM~\citep{D3_Sketch2BIM2025}, SceneGenAgent~\citep{D3_SceneGenAgent2025}, ArtisanCAD~\citep{CAD_ArtisanCAD2026}, and MapAgent~\citep{D3_MapAgent2026} become application-level when their outputs enter broader verified BIM, industrial-scene, CAD, or mapping workflows.
The C4 study~\citep{SoftArch_LLM2026} provides boundary evidence for architecture generation followed by evaluation, while the Agentic EDA survey~\citep{Hardware_EDA2025} organizes the broader toolchain without another construction episode.
TurboAgent~\citep{CAD_TurboAgent2026} and agentic TCAD~\citep{Sci_AgenticTCAD2025} connect design variables to prediction, optimization, or physics validation, while robot co-design~\citep{Robot_Debate2Create2025} lets simulated behavior redirect morphology or control.
Here simulation supplies feedback rather than defining the delivered result.

\textit{Simulation modeling.}
Simulation modeling instead treats the transition model or simulator as the delivered result.
Agent2World~\citep{WorldModel_Agent2World2025} and CWM~\citep{WorldModel_CodeWorldModels2025} revise symbolic or programmatic transition models through execution feedback.
SOCIA-$\nabla$~\citep{Simulation_SOCIANabla2025} and Sketch2Simulation~\citep{Simulation_Sketch2Sim2026} construct simulator code or flowsheets, while perceptual self-reflection~\citep{Simulation_SelfReflection2026}, coding-agent simulation~\citep{Simulation_CodingAgentWorldSimulator2026}, and Agentic Real2Sim~\citep{Simulation_AgenticReal2Sim2026} extend feedback to physical behavior and real-world-aligned episodic twins.
Simulation-ready scene construction remains a spatial artifact when the scene, rather than its dynamics, is the accepted output.
Because agentic world modeling spans several regimes, assignment follows the modeled process and downstream objective rather than the phrase \emph{world model} alone~\citep{Chu2026_AgenticWorldModeling}.
Behavioral fidelity, calibration beyond the tested regime, and engineering authorization govern deployment.

Two boundary cases clarify the application-level unit.
Some objectives require several independently accepted deliverables to remain aligned: Data Journalist Agent~\citep{Data_DataJournalistAgent2026} exhibits such dependencies within public information, whereas ResearchStudio-Reel~\citep{Workflow_ResearchStudioReel2026} coordinates separate posters, videos, and blogs through shared paper-derived state.
This is a cross-output dependency pattern rather than another artifact family or context.
Constructed environments can also serve as application infrastructure: Agent World Model~\citep{WorldModel_AgentWorldModel2026} builds executable tool environments, while AutoWebWorld~\citep{Website_AutoWebWorld2026} and InfiniteWeb~\citep{Website_InfiniteWeb2026} construct interactive websites with explicit transitions and evaluators.
Their downstream training role adds tasks, trajectories, rewards, datasets, and learned policies to the application bundle without replacing the original simulation or web acceptance criteria.
Both cases require evaluation to state the judged artifact and the application conditions governing its use.
Section~\ref{sec:verification} next separates these targets from the evidence and protocols used to judge them.

%% file: sections/04c_verification.tex
\section{Evaluation and Benchmarking}
\label{sec:verification}
\label{sec:evaluation}

Application context determines what counts as successful use and who may authorize it.
Evaluation turns those commitments into stated targets, criteria, evidence, and protocols, while Runtime Verification interprets observations of Action consequences and returns feedback to the current trajectory.
The distinction follows use rather than tool type: a compiler, renderer, simulator, specialist model, LM judge, or human reviewer becomes evaluation evidence when used within a stated task, criterion, metric, and protocol.
An artifact may compile yet fail its intended use, score highly yet violate source evidence, or satisfy one reviewer while remaining irreproducible.
An evaluation claim must therefore identify its target, criterion, metric, signal, and protocol.
Using the reviewed benchmark and system-evaluation subset, we characterize the protocols and gaps reported in the evidence.

\input{figures/verification_spectrum}

Figure~\ref{fig:verification-spectrum} assembles these elements: the target, criterion, and stage scope the claim; a channel--evaluator pairing produces a signal; and the metric, aggregation rule, and protocol determine what can be reported.

\subsection{Evaluation Targets}
\label{sec:evaluation-targets}

Evaluation can target the delivered artifact, a construction trajectory, or the agentic construction system itself.
These targets are related but not interchangeable.
Within each target, a dimension names a property of interest and a metric operationalizes it; the examples below depend on the artifact family and protocol.
Aggregating artifact or trajectory measurements across a task suite supports task-capability claims, but not by itself claims about system properties beyond capability.
The latter require contrasts across repeated runs, controlled operating conditions, resource tradeoffs, interventions, or system versions.

\paragraph{\textbf{\textup{Delivered artifacts.}}}

Artifact evaluation asks whether the delivered object satisfies stated requirements and serves its intended use.
Representative dimensions and metrics include the following.
\begin{itemize}
  \item \textit{Requirement satisfaction.} Hard-constraint pass rate, all-requirements success rate, and grounded-claim precision or recall measure validity, compliance, and grounding.
  \item \textit{Artifact quality.} Rubric scores, pairwise preferences, and artifact-specific semantic or perceptual scores measure quality and coherence.
  \item \textit{Use effectiveness.} Task completion, time, user error, and rated usefulness measure performance in the intended context.
  \item \textit{Internal consistency.} Cross-unit violation rates and continuity scores measure alignment across pages, shots, objects, modules, or other units.
\end{itemize}
A functional check can still miss intended use, while an aggregate score can hide unmet requirements.

\paragraph{\textbf{\textup{Construction trajectories.}}}

Trajectory evaluation asks how the result was reached: whether failures were detected early, repairs were localized, accepted work was preserved, and cost remained bounded.
Similar artifacts can arise from trajectories with very different rework, human intervention, and auditability.
Representative dimensions and metrics include the following.
\begin{itemize}
  \item \textit{Detection timeliness.} Actions or time from defect introduction to detection, together with downstream units affected before detection.
  \item \textit{Repair locality.} Defect-localization precision, the proportion of units modified, and repair success at the selected target.
  \item \textit{State preservation.} Accepted-state retention rate, regression rate, and the proportion of work recomputed after repair.
  \item \textit{Process traceability.} Record completeness and decision-to-evidence linkage for actions, checks, revisions, interventions, and stopping.
  \item \textit{Process efficiency.} Tool and verifier calls, time, compute, monetary cost, and human interventions per successful trajectory.
\end{itemize}
Process verification can operate at agent, step, or iteration level, as in MAS-ProVe and the dense step-level feedback of WebGen-Agent~\citep{MASProVe2025,Code_WebGenAgent2025}; context pruning and hierarchical memory help keep long trajectories interpretable~\citep{SWEPruner2025,ConfuciusCodeAgent2025,HiAgent2025}.
Mr.~Dre reports 16--27\% regression in previously covered content or citation quality during later revisions, providing a direct preservation measure~\citep{Text_MrDre2026}.
VideoMemory exposes cross-shot identity and continuity failures, while WorldGen makes navigability an explicit constraint checked during world construction~\citep{Video_VideoMemory2026,D3_WorldGen2025}.

\paragraph{\textbf{\textup{Agentic systems.}}}

Agentic-system evaluation begins beyond task capability.
A benchmark targets system properties beyond task capability when it probes reusable behavior not captured by aggregate artifact success across a task suite.
Representative dimensions and metrics include the following.
\begin{itemize}
  \item \textit{Run reliability.} Failure rate and score variance across repeated runs of the same tasks under fixed conditions.
  \item \textit{Condition robustness.} Performance degradation under perturbations, environment shifts, tool unavailability, or component failures with a fixed system version.
  \item \textit{Resource efficiency.} Success-normalized latency, compute, tool calls, monetary cost, and human effort aggregated over the task distribution.
  \item \textit{Goal controllability.} Correct handling of goal updates, approvals, interruptions, overrides, and rollbacks, including constraint violations and unauthorized actions.
  \item \textit{Update stability.} Regression and reproducibility under a fixed protocol across system, model, or tool updates.
\end{itemize}
Each dimension requires a distinct contrast: repeated fixed conditions for reliability, controlled perturbations for robustness, matched outcomes for efficiency, intervention tests for controllability, and versioned reruns for update stability.
Per-trajectory resource counts support system efficiency only after aggregation at matched success levels, while checkpointed or mixed-initiative authoring cannot be reduced to a binary human-in-the-loop label~\citep{DataViz_DataFormulator2_2025,Image_APPO2026,UI_DuetUI2026}.
A single success score, even averaged across a task suite, supports task-capability claims rather than claims about system properties beyond capability.

\subsection{Evaluation Evidence}
\label{sec:evaluation-evidence}
\label{sec:evaluation-framework}

Evaluation evidence connects a target and criterion to an observable signal.
An \emph{evaluation signal} pairs an \emph{evidence channel}, which specifies what is observed, with an \emph{evaluator}, which turns it into a judgment.
A metric and aggregation rule then map one or more signals to a reported quantity across requirements, tasks, or runs.
Because the same observation can support different claims under different evaluators, the pairing must be stated for a particular requirement and stage; neither component alone defines evidence strength.

We distinguish four evaluator types by how they form a judgment.
\begin{itemize}
  \item \textit{Rule-based checks} apply explicit executable conditions, including tests, validators, constraints, and exact comparisons.
  \item \textit{Specialist models} use a delimited evaluation function, as in detectors, OCR, embeddings, perceptual models, and reward models; unlike LM judges, they do not interpret an open-ended rubric at evaluation time.
  \item \textit{LM judges} use a language or multimodal language model to interpret criteria and task context at evaluation time.
  \item \textit{Human reviewers} include experts, creators, intended users, and crowd workers who assess the observed evidence.
\end{itemize}
An orchestrating agent inherits the types of the mechanisms that produce the final judgment, and a stack may combine several types.

The main evidence channels expose different parts of an artifact or construction process.
\begin{itemize}
  \item \textit{Reference inputs.} Specifications, sources, ground truth, reference artifacts, and standards support comparison, subject to their correctness and completeness.
  \item \textit{Structured state.} Document trees, hierarchies, graphs, parameters, and program structure expose explicit relations, but structural validity does not establish semantic adequacy or intended use.
  \item \textit{Rendered output.} Pages, images, audio, video, and interactive views expose perceptual qualities, although judgments may be unstable or hard to localize and LM judges carry known biases~\citep{Zheng2023_LLMJudge,Panickssery2024_SelfPreference,Hu2025_LengthBias}.
  \item \textit{Runtime behavior.} Compilation, tests, logs, traces, simulations, and measurements expose executed behavior only on checked paths and under the fidelity of the runtime or simulation.
  \item \textit{Construction history.} Actions, intermediate states, checks, revisions, rollback, interventions, and stopping records support measures of detection, repair, preservation, and cost, but incomplete records leave gaps.
  \item \textit{Use outcomes.} Task completion, user behavior, deployment records, and downstream or longitudinal effects expose contextual performance, but are costly and difficult to attribute.
\end{itemize}

A concrete signal therefore pairs a channel and evaluator for a stated requirement and stage; the same rendered page assessed by a layout rule or a human reviewer yields signals with different scopes and failure modes.
Its adequacy depends separately on coverage, validity against the intended criterion or a trusted reference, and stability across runs, presentations, evaluators, or perturbations; confidence-bearing signals also require calibration.
\label{sec:rw-judge}
\label{sec:verifier-strength}

Complex artifacts therefore benefit from complementary rather than merely numerous signals.
Posters and data stories may combine source, structural, runtime, rendered, and human evidence, yet multiple judges can still share a blind spot~\citep{Poster_PosterForest2025,Data_DataJournalistAgent2026}.
A \emph{requirement-to-check map} links requirements to checks, observed states, repairs, and shared errors; ARC and executable event graphs illustrate inspectable versions of this structure~\citep{Code_ARC2026,Video_EventGraphs2026}.
Variation across MAS-ProVe configurations and persistent factuality and sourcing errors in DRACO illustrate the risks of incomplete process verification and dependence on learned judges~\citep{MASProVe2025,DRACO2025}.

\subsection{Evaluation Protocols}
\label{sec:evaluation-protocols}

Evaluation protocols specify the conditions under which the targets and signals introduced above support a comparable claim.
An evaluation protocol fixes four groups of choices.
\begin{itemize}
  \item \textit{Task set.} Tasks, instances, inputs, initial states, and reference data specify what is tested.
  \item \textit{Evaluation metrics.} Metric definitions, scoring rules, and acceptance thresholds specify how each target and criterion is measured.
  \item \textit{Run configuration.} Model and prompt versions, tools, environment, feedback and human access, budgets, and stopping rules specify how runs proceed.
  \item \textit{Analysis plan.} Repeated trials, baselines, controlled contrasts, uncertainty, failure accounting, and resource or rework cost specify how results are compared.
\end{itemize}
Run configurations must be aligned across systems unless the contrast is itself part of the analysis plan; a fixed iteration limit indicates budget termination rather than autonomous completion.
Trajectory claims additionally require records of checks, revisions, intervention, and stopping, while system claims require observations matched to the stated dimension.
Missing fields remain evidence gaps rather than quantities to be inferred, and a final-artifact score does not establish planning quality, feedback use, repair locality, or trajectory efficiency.

\subsection{Benchmark Landscape}
\label{sec:benchmark-landscape}

Benchmarks operationalize the targets, signals, and protocol conditions introduced above through concrete tasks, measures, and comparisons.
Figure~\ref{fig:design-arena-task-profiles}, based on a snapshot of the public \href{https://www.designarena.ai/leaderboard}{Design Arena leaderboard} retrieved on August 19, 2026, provides a concrete example of why an overall arena score is insufficient~\citep{DesignArena2026}.
In the Models Arena, Kimi K3~\citep{KimiTeam2026KimiK3} occupies the top percentile on all five tasks, while Claude Fable 5, GPT-5.6 Sol, and GLM 5.2 cluster closely behind it; the clearest deviation is Gemini 3.5 Flash on data visualization.
The Agents Arena exhibits stronger task dependence: Kimi K3 remains at or near the top on web apps, full-stack development, slides, and mobile apps, but falls to the 54.3rd percentile on native Android, where Claude Fable 5 reaches the top percentile.
GPT-5.6 Sol similarly ranges from the 87.5th percentile on slides to the 34.2nd percentile on mobile apps.
Together, the profiles show that task category and execution configuration are part of the evaluated capability: a single aggregate rank can conceal task-specific differences in relative standing.
Agentic evaluations should therefore report task-stratified results and name the harness and environment configuration alongside the model.

\input{figures/design_arena_task_profiles}

Table~\ref{tab:artifact-evaluation} compares three representative protocols per artifact family using the targets and evidence defined above.
Selection prioritizes influence and publication record, then complementary target and evidence coverage.
Because standalone benchmarks remain sparse in audio, video, and spatial creation, the matrix also includes named protocols released with core systems, but not ordinary system evaluations.
The balanced display supports comparison rather than frequency estimates, and each row describes only protocol-supported claims.

\begin{table}[t]
\caption{Three representative benchmark protocols per artifact family, organized by evaluation target and evidence. Agentic-system entries report properties beyond task capability.}
\label{tab:artifact-evaluation}
\centering
\scriptsize
\setlength{\tabcolsep}{2.8pt}
\renewcommand{\arraystretch}{1.04}
\resizebox{\textwidth}{!}{%
\begin{tabular}{>{\raggedright\arraybackslash}p{3.05cm}>{\raggedright\arraybackslash}p{1.45cm}cc>{\centering\arraybackslash}p{2.65cm}>{\raggedright\arraybackslash}p{3.55cm}cccc}
\toprule
\multicolumn{1}{c}{\textbf{Benchmark}} & \multicolumn{1}{c}{\textbf{Artifact}} & \multicolumn{3}{c}{\textbf{Evaluation Target}} & \multicolumn{5}{c}{\textbf{Evaluation Evidence}} \\[-2pt]
\cmidrule(lr){3-5}\cmidrule(lr){6-10}
\multicolumn{1}{c}{\textbf{Name}} & \multicolumn{1}{c}{\textbf{Family}} & \multicolumn{1}{c}{\textbf{Artifact}} & \multicolumn{1}{c}{\textbf{Trajectory}} & \multicolumn{1}{c}{\shortstack{\textbf{Agentic System}\\\textbf{Beyond Capability}}} & \multicolumn{1}{c}{\textbf{Channels}} & \multicolumn{1}{c}{\textbf{Rules}} & \multicolumn{1}{c}{\textbf{Models}} & \multicolumn{1}{c}{\textbf{LM}} & \multicolumn{1}{c}{\textbf{Human}} \\
\midrule
MLR-Bench~\citep{Text_MLRBench2025} & Textual & \cmark & $\triangle$ & -- & Reference inputs, Structured state, Runtime behavior, Construction history & -- & -- & \cmark & \cmark \\
\rowcolor{cRowAlt}
DRACO~\citep{DRACO2025} & Textual & \cmark & -- & -- & Reference inputs, Structured state, Rendered output & -- & -- & \cmark & -- \\
Mr.~Dre~\citep{Text_MrDre2026} & Textual & \cmark & \cmark & Controllability & Reference inputs, Structured state, Construction history & -- & -- & \cmark & \cmark \\
\midrule
\rowcolor{cRowAlt}
DiagramGenBenchmark~\citep{Diagram_Text2Diagram2024} & 2D visual & \cmark & -- & -- & Reference inputs, Structured state, Rendered output & \cmark & \cmark & -- & \cmark \\
Paper2Poster~\citep{Poster_Paper2Poster2025} & 2D visual & \cmark & -- & -- & Reference inputs, Structured state, Rendered output & -- & \cmark & \cmark & -- \\
\rowcolor{cRowAlt}
DV-World~\citep{DataViz_DVWorld2026} & 2D visual & \cmark & $\triangle$ & -- & Structured state, Rendered output, Runtime behavior & \cmark & -- & \cmark & \cmark \\
\midrule
RIME~\citep{Music_RIME2026} & Audio & \cmark & \cmark & -- & Reference inputs, Structured state, Rendered output, Construction history & \cmark & \cmark & -- & -- \\
\rowcolor{cRowAlt}
LVAS-Bench~\citep{Audio_LVASAgent2025} & Audio & \cmark & -- & -- & Reference inputs, Rendered output & -- & \cmark & -- & -- \\
MA-Bench~\citep{Audio_AudioGenie2025} & Audio & \cmark & -- & -- & Reference inputs, Rendered output & -- & \cmark & -- & \cmark \\
\midrule
\rowcolor{cRowAlt}
AgenticVBench~\citep{Video_AgenticVBench2026} & Video & \cmark & \cmark & Reliability & Reference inputs, Rendered output, Runtime behavior, Construction history & \cmark & -- & -- & \cmark \\
Paper2Video~\citep{Video_Paper2Video2025} & Video & \cmark & -- & -- & Reference inputs, Structured state, Rendered output & -- & \cmark & \cmark & -- \\
\rowcolor{cRowAlt}
MMMC~\citep{Video_Code2Video2025} & Video & \cmark & $\triangle$ & -- & Reference inputs, Structured state, Rendered output, Runtime behavior & \cmark & \cmark & \cmark & -- \\
\midrule
AuthorBench~\citep{D3_MUSEAuthoring2026} & Spatial & \cmark & \cmark & Controllability & Structured state, Runtime behavior, Construction history & \cmark & -- & \cmark & \cmark \\
\rowcolor{cRowAlt}
3DCodeBench~\citep{D3_3DCodeBench2026} & Spatial & \cmark & $\triangle$ & -- & Reference inputs, Structured state, Rendered output, Runtime behavior & \cmark & \cmark & -- & \cmark \\
Code4D~\citep{D3_Code2Worlds2026} & Spatial & \cmark & $\triangle$ & -- & Reference inputs, Structured state, Rendered output, Runtime behavior & \cmark & \cmark & \cmark & -- \\
\midrule
\rowcolor{cRowAlt}
SWE-bench~\citep{Code_SWEBench2024} & Behavioral & \cmark & -- & -- & Reference inputs, Structured state, Runtime behavior & \cmark & -- & -- & -- \\
CodeFlowBench~\citep{Code_CodeFlowBench2026} & Behavioral & \cmark & \cmark & -- & Structured state, Runtime behavior, Construction history & \cmark & -- & -- & -- \\
\rowcolor{cRowAlt}
NL2Repo-Bench~\citep{Code_NL2RepoBench2026} & Behavioral & \cmark & -- & -- & Structured state, Runtime behavior & \cmark & -- & -- & -- \\
\bottomrule
\end{tabular}}
\vspace{2pt}

\begin{minipage}{0.98\textwidth}
\scriptsize
\emph{Legend:} For Artifact and Trajectory, \cmark{} denotes explicit coverage, $\triangle$ limited or indirect coverage, and -- no reported coverage. Agentic System cells name only properties evaluated beyond task capability: Reliability uses fixed-condition repetitions, Robustness compares controlled operating conditions, Efficiency aggregates resource use at matched outcomes, Controllability tests goal updates or interventions, and Update Stability compares system versions; -- denotes task-capability evidence only or no system-property claim. For evaluator types, \cmark{} denotes reported use and -- no reported use. Rules, Models, LM, and Human denote Rule-based Checks, Specialist Models, LM Judges, and Human Reviewers; channel labels follow Section~\ref{sec:evaluation-evidence}.
\end{minipage}
\end{table}

The representative set still reveals uneven maturity: dedicated benchmarks are most developed for 2D visual and behavioral artifacts, while audio, video, and spatial rows more often rely on protocols from system papers.
Complementary studies extend textual and visual evaluation to grounding, source quality, editability, and rendering~\citep{Text_LegalAgentBench2024,DRACO2025,Text_SurGE2026,Diagram_Crafter2026,Image_DrawAI2026}; audio and video to composition, listening, playback, physical checks, and repair~\citep{Music_CoComposer2025,Audio_SoundscapeAgent2026,Video_MUSE2026,Video_EventGraphs2026,Video_PhyT2V2025,Animation_LogoMotion2025,Video_PhysAgent2026}; and spatial and behavioral evaluation to geometry, traversal, CAD, repositories, interfaces, and simulation~\citep{D3_WorldGen2025,CAD_STEPLLM2026,CAD_NURBGen2026,CAD_CADIR2026,CAD_SPADA2026,Code_BackendForge2026,Code_WebDesignIter2026,Code_WebGenAgent2025,UI_ComputerUse2025,UI_UXAgent2025,UI_Vision2Web2026,WorldModel_Agent2World2025,Simulation_Sketch2Sim2026,Simulation_SelfReflection2026}.
Gameplay-agent benchmarks provide adjacent evidence about interactive environments rather than constructed artifacts~\citep{Game_GVGAILLM2025,Game_lmgameBench2025}.

Across the matrix and broader review, evidence is strongest for delivered artifacts and bounded executions; trajectories, use outcomes, and validity beyond tested cases remain less consistently covered.
Only three protocols evaluate a system property beyond task capability under these definitions, each covering one property. Inconsistent reporting of calibration, disagreement, tools, budgets, stopping, and cost further limits comparison.
\FloatBarrier

%% file: figures/verification_spectrum.tex
\begin{figure*}[!t]
  \centering
  \includegraphics[width=\textwidth]{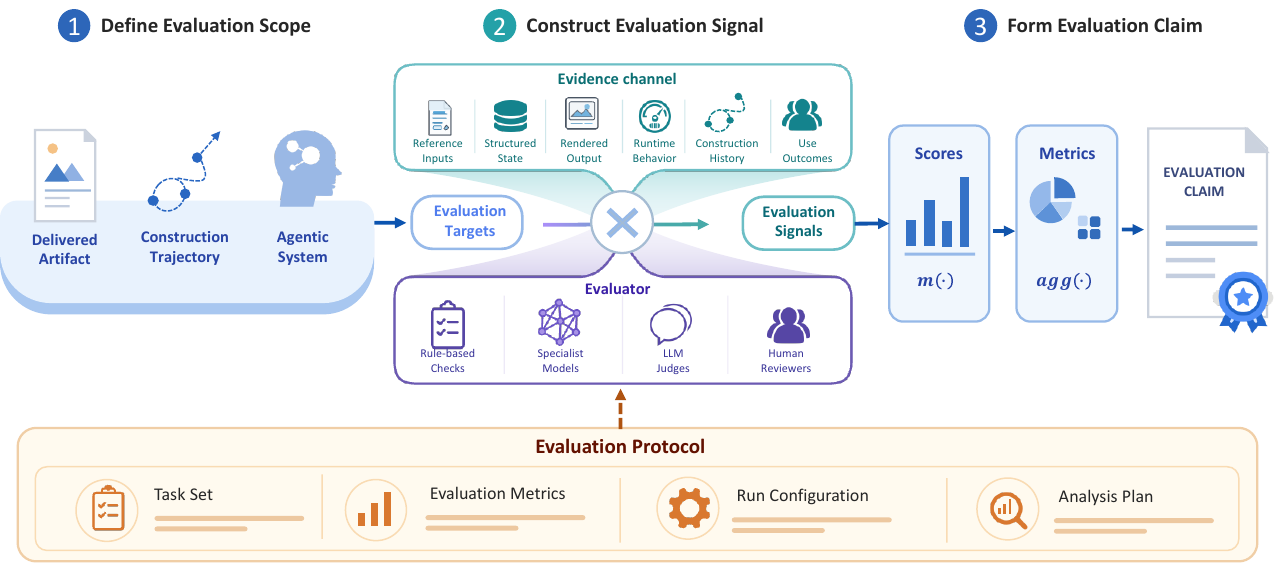}
  \caption{Framework for external evaluation. A target, criterion, and stage scope the claim. An evidence channel paired with an evaluator yields a signal, which a metric and aggregation rule turn into a reported claim. The task set, metrics, run configuration, and analysis plan define the evaluation protocol. Evidence channels and evaluator types are orthogonal and do not rank evidence strength.}
  \Description{A left-to-right framework for external evaluation. First, the evaluator scopes a claim around a delivered artifact, construction trajectory, or agentic system and states the criterion and evaluation stage. Second, one of six evidence channels is paired with one of four evaluator types to construct an evaluation signal. Third, a metric and aggregation rule turn one or more signals into an evaluation claim. A shared evaluation-protocol foundation contains the task set, evaluation metrics, run configuration, and analysis plan.}
  \label{fig:verification-spectrum}
\end{figure*}

%% file: figures/design_arena_task_profiles.tex
\begin{figure}[!t]
  \centering
  \includegraphics[width=\textwidth]{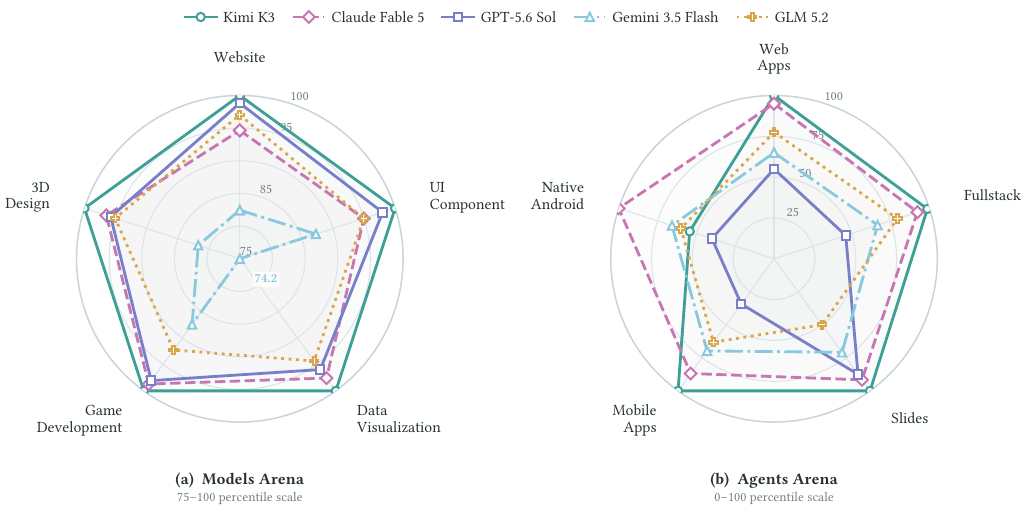}
  \caption{Taskwise model profiles in Design Arena (snapshot: August 19, 2026). Within each task, the Elo rank percentile is computed across leaderboard entries with at least 100 battles, mapping the lowest and highest eligible ranks to 0 and 100; thus, 100 denotes the top relative standing rather than a perfect task score.}
  \Description{Two side-by-side radar charts compare five models across Design Arena tasks. The Models Arena chart covers Website, UI Component, Data Visualization, Game Development, and 3D Design on a 75-to-100 percentile scale. The Agents Arena chart covers Web Apps, Fullstack, Slides, Mobile Apps, and Native Android on a 0-to-100 percentile scale. Colors, markers, and line styles identify Kimi K3, Claude Fable 5, GPT-5.6 Sol, Gemini 3.5 Flash, and GLM 5.2.}
  \label{fig:design-arena-task-profiles}
\end{figure}

%% file: sections/05_methodology.tex
\section{Principles of Agentic Creation}
\label{sec:patterns}
\label{sec:transfer}

The preceding analyses motivate one cross-cutting focus: whether acceptance criteria, artifact state, control responsibility, evaluation evidence, and repair actions remain inspectably connected across construction.
Across Sections~\ref{sec:paradigm}--\ref{sec:verification}, these relations appear through functional roles, artifact-specific states and edit units, objectives, authority, and evidence.
We therefore synthesize relations that should hold across contexts rather than prescribe one architecture or representation format.

We use \emph{decision interdependence}, \emph{failure observability}, and \emph{repairability} as a qualitative lens on construction difficulty: they ask how far decisions constrain other state, whether failures become visible and attributable in time, and whether a diagnosis maps to a bounded action that preserves accepted work.
Narrative, video, and world-construction systems illustrate delayed or incomplete observations~\citep{Text_NarrativeWorldModel2026,Video_VideoMemory2026,D3_WorldGen2025}, while repository benchmarks show how long-range dependencies challenge global coherence~\citep{Code_CodeFlowBench2026,Code_NL2RepoBench2026}.
These are recurring qualitative conditions rather than independent causal factors or a universal ranking of artifact families.
Using this lens, we formulate four principles:
\emph{Externalize Commitments} connects acceptance criteria to addressable state; \emph{Define Control Boundaries} connects dependencies and consequences to decision responsibility; \emph{Make Feedback Actionable} connects evidence to bounded repair; and \emph{Revalidate Affected State} connects change to selective revalidation.
The first is definition-derived, the middle two synthesize recurring qualitative design relations, and the fourth is a lifecycle implication supported by emerging and comparatively sparse evidence.
Figure~\ref{fig:construction-principles} summarizes the resulting structure.

\input{figures/construction_principles}

\subsection{Externalize Commitments}

Externalize the commitments that later decisions must preserve, change, or verify, and connect them to addressable artifact state.
Stateful construction requires consequential requirements and prior decisions to remain available to later actions and checks.
Depending on the artifact, the relevant state may include acceptance criteria, source links, constraints, semantic edit units, dependency edges, provenance, or review status.
Plans, agent memory, and tool traces remain policy-side context unless they are encoded as artifact-side commitments that govern later modification or acceptance.

A representation transfers by the relation it preserves, not by its format.
Crafter~\citep{Diagram_Crafter2026} carries typed revisions through an evolving figure specification, while CADIR~\citep{CAD_CADIR2026} preserves an editable construction graph that remains executable across geometric backends.
Conversely, narrative and video systems show how omitted event, identity, or temporal commitments can invalidate later construction~\citep{Text_NarrativeWorldModel2026,Video_VideoMemory2026}.
The aim is sufficient rather than maximal externalization: context pruning and hierarchical memory can remove information that no longer governs downstream decisions~\citep{SWEPruner2025,ConfuciusCodeAgent2025}, while short tasks with weak dependencies may need only a light representation or direct generation.

\subsection{Define Control Boundaries}

Define delegation, handoff, tool-permission, and review boundaries at points where artifact dependencies or decision consequences change.
A useful boundary specifies both who or what may act and who remains responsible for preserving commitments that cross the boundary.
Decomposition helps when modules follow meaningful artifact dependencies and handoffs carry the state required downstream.
Otherwise, locally successful outputs can fail when recomposed because shared constraints were lost or no decision-maker owned cross-boundary consistency.

Poster, slide, video, and music systems use specialized stages and coordination mechanisms~\citep{Poster_PosterForest2025,Slide_PPTAgent2025,Video_AniME2025,Music_CoComposer2025}.
Creator-facing systems expose distinct points for preference input, technical feedback, and interactive revision~\citep{Image_APPO2026,Animation_LogoMotion2025,UI_DuetUI2026}.
Review authority should therefore follow ownership and consequence rather than uncertainty alone.
Because additional boundaries can increase communication, rework, latency, and propagation cost, comparisons should report recomposed artifact quality and preserved commitments alongside local performance and coordination cost.

\subsection{Make Feedback Actionable}

Make feedback actionable by connecting a relevant acceptance criterion to timely evidence, a diagnosis, the affected state, and a feasible repair or escalation.
External evaluation supports a claim under a stated protocol, whereas Runtime Verification must inform a live construction decision.
For targeted repair, an observation must cover the relevant consequence, arrive while the implicated state is still actionable, and localize the failure at a scope exposed by the Edit Interface.

A global score may support candidate selection or stopping, but it cannot by itself identify a local repair; conversely, a precise diagnosis has little control value when the only available action regenerates the entire artifact.
DrawAI, ReDesign, ParticleGen, LogoMotion, and PhyT2V illustrate mappings from observed defects to editable raster structure, design components, procedural parameters, motion code, or prompts~\citep{Image_DrawAI2026,Image_ReDesign2026,Animation_ParticleGen2026,Animation_LogoMotion2025,Video_PhyT2V2025}.
MapCoder and CodeTree provide software counterparts, using iterative debugging or execution-guided search to revise code~\citep{Code_MapCoder2024,Code_CodeTree2025}.
Verifier stacks should therefore be judged by complementary criterion coverage and error dependence, with fine-grained evidence scoped to the consequence and repairability of the failure.

\subsection{Revalidate Affected State}

Treat acceptance evidence as version-scoped.
When a source, requirement, tool, artifact state, or authorization changes, invalidate dependent evidence and approvals before selectively revalidating the affected state.
Without explicit change dependencies, stale evidence can appear current; global regeneration can avoid selective invalidation only by discarding accepted work and repeating unaffected checks.

The relation spans both coordinated outputs and persistent versions: a revised claim, asset, style decision, or audience assumption can invalidate several deliverables~\citep{Data_DataJournalistAgent2026,Workflow_ResearchStudioReel2026}, while websites, repositories, documentation, and agent infrastructures must respond to changing sources, tools, users, and requirements~\citep{Code_Paper2Agent2025,Code_WebDesignIter2026,Infrastructure_JarvisHub2026}.
A versioned dependency and evidence graph is a candidate shared abstraction for invalidation, regression, rollback, and renewed review, but current cross-family evidence is too sparse to establish one mechanism.
This relation motivates the challenges of maintaining coherence across revision and system evolution in Section~\ref{sec:challenges}.

Together, the four principles define inspectable control relations, not a maturity ladder or mandatory pipeline.
Their realization remains artifact- and application-specific, so the appropriate system is the lightest process that can preserve consequential commitments, assign responsibility, direct repair, and re-establish validity at an acceptable cost.

%% file: figures/construction_principles.tex
\begin{figure*}[!t]
  \centering
  \includegraphics[width=\textwidth]{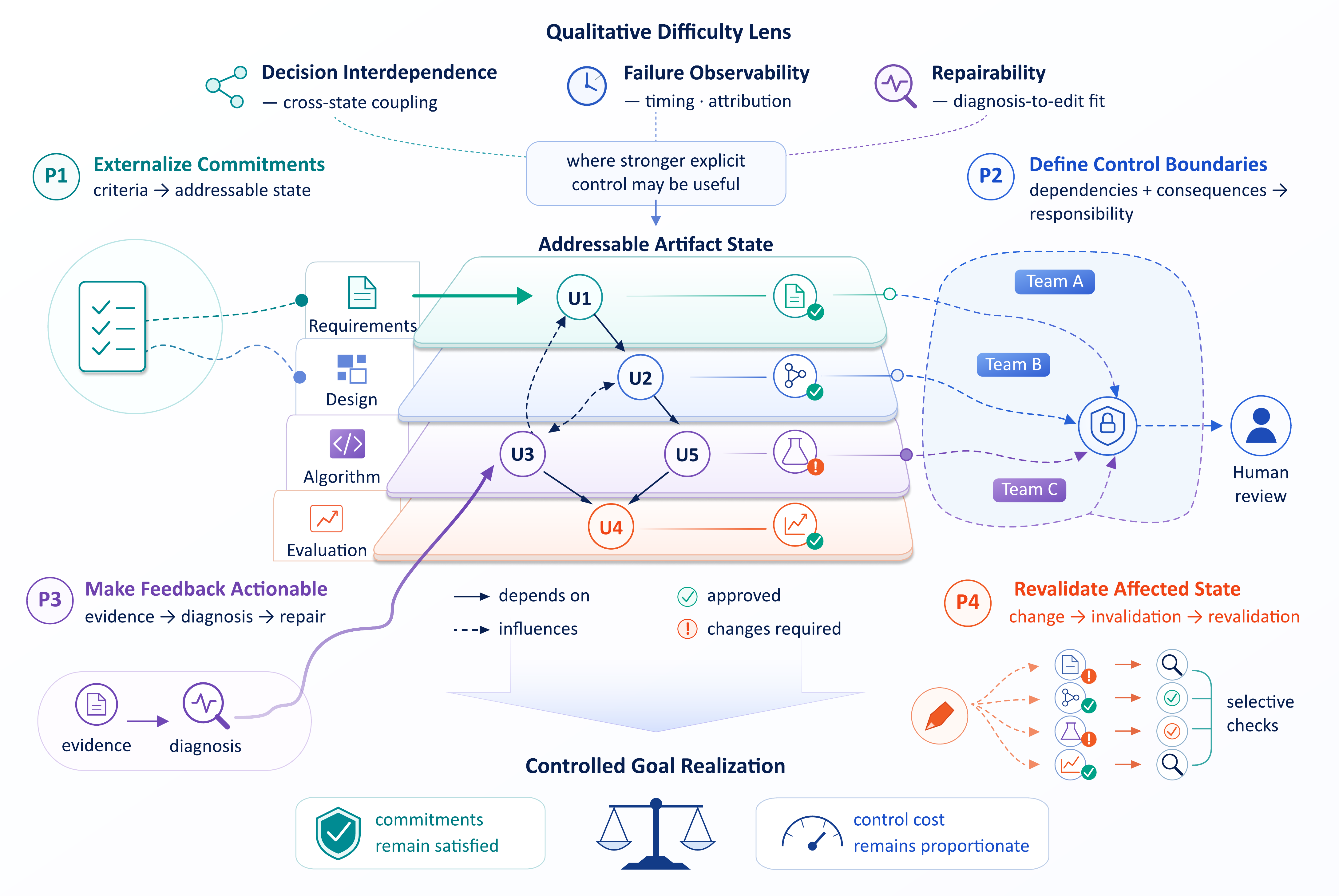}
  \caption{Principles of agentic creation. A qualitative lens identifies tasks that may benefit from stronger explicit control. Four principles connect criteria to state, consequences to responsibility, evidence to repair, and change to revalidation around addressable artifact state. Together, they support controlled goal realization with proportionate control. The relations are qualitative, not a sequential pipeline or maturity ladder.}
  \Description{A qualitative difficulty lens at the top contains decision interdependence, failure observability, and repairability. At the center, an addressable artifact state contains connected units, acceptance criteria, evidence, and approval markers. Four parallel principles act on this state: externalize commitments, define control boundaries, make feedback actionable, and revalidate affected state. They converge on controlled goal realization, where commitments remain satisfied and control cost remains proportionate.}
  \label{fig:construction-principles}
\end{figure*}

%% file: sections/06_challenges.tex
\section{Challenges and Opportunities}
\label{sec:challenges}

Taken together, the four principles in Section~\ref{sec:patterns} and the difficulty lens expose six unresolved control problems in agentic creation.
They concern coherence across dependent decisions, repair under diagnostic blind spots, durable system evolution, changing creator intent, authority under delegation, and evaluation when several outcomes may be valid.
The directions below are grounded in the reviewed systems but remain hypotheses; comparative claims should report and, where relevant, match model, tool, human, and resource budgets.

\subsection{Global Coherence under Decision Interdependence}

Agentic construction often decomposes an artifact into stages, modules, agents, or tool operations, yet decisions in one unit can constrain content, structure, behavior, or acceptance elsewhere.
Poster and slide systems coordinate specialized roles across multiple production stages~\citep{Poster_PosterForest2025,Slide_PPTAgent2025}.
Long animation and music composition expose similar coordination interfaces~\citep{Video_AniME2025,Music_CoComposer2025}.
Repository benchmarks show its counterpart in code: a locally plausible edit can still fail when the task spans dependencies~\citep{Code_CodeFlowBench2026,Code_NL2RepoBench2026}.
The challenge is preserving artifact-level commitments once work has been divided across allocation, sequencing, conflict resolution, and revision.
An Operational Representation must therefore expose consequential dependencies without drifting from rendered or executed behavior~\citep{Code_AgenticConsensus2026}.
Narrative memory preserves story relations~\citep{Text_NarrativeWorldModel2026}, while typed figure revisions and editable CAD graphs offer structured alternatives~\citep{Diagram_Crafter2026,CAD_CADIR2026}.

Dependency information belongs in the control state, not just in documentation.
Selective explicitness makes this representation adaptive: a typed commitment graph can link requirements to artifact units, decisions, owners, and evidence, but materialize only the links touched by a change or judged high-risk.
Verifiable data stories and research media connect claims to outputs~\citep{Data_DataJournalistAgent2026,Workflow_ResearchStudioReel2026}; paper-derived agents and repository web design link source commitments to executable states~\citep{Code_Paper2Agent2025,Code_WebDesignIter2026}.
The representation itself need not be fixed.
Automated representation search could choose what to externalize, retain, or refine for each task.
Held-out changes can then test whether a lighter candidate preserves coherence without excessive stale evidence, rework, revalidation, or coordination.

\subsection{Targeted Repair under Self-Diagnosis Limits}

Because agentic construction often uses the same or related models to generate, inspect, and revise, defects caused by shared knowledge, reasoning, or preference weaknesses may remain invisible.
Learned judges can likewise exhibit correlated preferences and self-preference~\citep{Zheng2023_LLMJudge,Panickssery2024_SelfPreference}.
Poster and slide workflows create composition points across generated components~\citep{Poster_PosterForest2025,Slide_PPTAgent2025}; long-video and simulation pipelines expose cross-stage consistency and render--oracle gaps~\citep{Video_LongVideo2025,Simulation_SelfReflection2026}.
Even then, a critic may identify a symptom without locating its origin or smallest safe edit.
Process-verification and research-evaluation studies show why aggregate judgments do not localize failures~\citep{MASProVe2025,DRACO2025}.
A precise diagnosis still cannot help if the edit interface hides the implicated structure.
Editable image workflows make the need for recoverable structure explicit~\citep{Image_DrawAI2026,Image_ReDesign2026}; multi-agent animation exposes a similar mismatch between diagnosis and edit granularity~\citep{Animation_ParticleGen2026}.
The central problem is a broken chain from observation to diagnosis to action: improving any one link does little when the others remain weak.

Diagnostic diversification pairs each requirement with evidence that need not share the generator's blind spots, such as executable checks, independent critics, or expert review.
These signals matter only when they fail differently; disagreement or an unresolved cause should trigger escalation rather than confident regeneration.
Dependency-aware fault localization then maps a failure to its source state and smallest safe edit.
Evaluation should follow the same chain, from calibrated detection and source localization to repair success, regression, and recovery cost.
Reporting these stages separately prevents accurate localization from masking weak detection or a repair that introduces regressions.

\subsection{System Self-Evolution beyond Task-Specific Gains}

Artifact revision changes state within one construction episode, whereas system self-evolution persists changes to the Operational Representation, Construction Policy, Runtime Verification, agent topology, tool routing, or policy-side memory across episodes~\citep{Hao2026_EvolveTeam,Liu2025_FoundationAgents,Xu2026_UnsupervisedPostTraining}.
VISTA and EvoPresent illustrate within-task refinement~\citep{Video_VISTA2025,Presentation_EvoPresent2026}.
ManimAgent retains experience across animation tasks~\citep{Animation_ManimAgent2026}; EvoMAC and ReVeal update software agents across tasks~\citep{Code_EvoMAC2025,Code_ReVeal2026}.
AutoDesign makes the distinction explicit by separating poster revision from a versioned harness update evaluated through a held-out development gate~\citep{Poster_AutoDesign2026}.
These mechanisms do not by themselves establish durable improvement: gains may reflect favorable task order, extra inference, evaluator bias, retrieval, or regressions outside the observed tasks.

Versioned self-evolution treats each retained change as a patch rather than silently absorbing it into the agent.
The patch should record its provenance, scope, and rollback path, then run in a shadow system on held-out episodes before promotion.
Changes to protected criteria and permissions should require independent authorization, since a system cannot validate improvement by weakening its standard.
Update attribution must separate persistent changes from retrieval, favorable task order, and matched extra inference, then report transfer, retention, regression, update faithfulness, and verification cost.
Only gains attributable to the retained patch count as durable improvement.

\subsection{Personalization under Evolving Creator Intent}

Creator intent is rarely exhausted by an initial prompt: later feedback may express a stable taste, a project commitment, a local correction, or a temporary reaction.
Treating every comment as persistent can overgeneralize from one revision, while discarding history forces creators to restate stable preferences~\citep{Lei2026_HumanLLM}.
PersonaVlog retains context for personalized vlogs~\citep{Video_PersonaVlog2025}; DuetUI maintains shared context during interface co-generation~\citep{UI_DuetUI2026}.
Creator-facing game and visual-analytics systems add interactive control~\citep{Game_CreatorCentric2026,DataViz_LightVA2024}, motivating explicit treatment of how preferences persist and change across revisions.

Scoped preference memory attaches both status and scope to what it retains.
A versioned creator-intent model can distinguish stable preferences, project commitments, provisional choices, and uncertain inferences, while leaving each open to inspection and correction.
A preference learned in one project should not silently govern another, and a withdrawal should take effect.
Conflicting feedback calls for clarification, but not every uncertainty does; the system should ask when new feedback contradicts retained intent instead of treating the latest comment as a silent override.
Over time, the Task Specification should follow the creator without false generalization, excessive clarification, or resistance to correction.

\subsection{Human Authority under Agentic Delegation}

A Construction Policy may select a technically useful action without having authority to make the underlying commitment; creative preference, domain validity, consequential action, and release approval can require different decision-makers.
Iterative visualization and interface systems expose revision points~\citep{DataViz_DataFormulator2_2025,UI_DuetUI2026}.
Preference-guided image generation and co-creation frameworks incorporate user feedback~\citep{Image_APPO2026,Data_HumanAICoCreation2025}.
Visual analytics and creator-centric storytelling add checkpoints~\citep{DataViz_LightVA2024,Game_CreatorCentric2026}.
These interaction mechanisms motivate separating decision authority from the timing of review.
Correctness and authority are therefore separate questions: evidence may support an action without granting the system permission to take it.

Authority-aware delegation represents permission at the level of decisions and actions, not with a single human-in-the-loop flag.
Review should follow ownership and consequence; approval gates can restrict capabilities and favor reversible actions when authority is incomplete.
Confidence cannot substitute for authority: a model may be certain about a decision it has no right to make.
Action-level provenance should let an independent reviewer reconstruct the source, decision, approval, and tool action after the fact.
These controls should prevent unauthorized or misrouted decisions without turning routine creation into constant interruption.

\subsection{Reliable Evaluation of Open-Ended Artifacts}

Open-ended artifacts admit multiple goal-satisfying realizations, while their criteria may combine fixed constraints, contextual obligations, and preferences that emerge only after inspection.
Open-endedness therefore makes criteria incomplete, heterogeneous, and evolving rather than absent.
DV-World evaluates underspecified visualization intent~\citep{DataViz_DVWorld2026}.
DuetUI and Crafter carry user updates or typed revisions forward~\citep{UI_DuetUI2026,Diagram_Crafter2026}.
Reference metrics may penalize legitimate variation; aggregate scores can hide criterion-specific failures.
Learned judges can share biases~\citep{Zheng2023_LLMJudge,Panickssery2024_SelfPreference}; stakeholder disagreement may reflect different acceptance conditions.

Plurality-preserving evaluation needs a versioned acceptance specification.
It should record fixed constraints, evolving preferences, supporting evidence, and evaluator roles without forcing all valid outcomes into one ordering.
Construction trajectories should also be budget-matched: a better final artifact says little about relative system quality when tools, model calls, feedback, compute, or human effort differ.
Matching model calls alone is insufficient if one system receives richer feedback or more expert review.
A protocol could start from an incomplete brief, reveal new constraints at controlled points, and compare evaluation methods under the same construction budget.
Reporting should keep criterion-level outcomes and evaluator stability visible alongside recovery behavior and total resource use.
The Task Specification, Operational Representation, trajectory, and review decisions then become part of evaluation rather than hidden conditions around the delivered artifact.

%% file: sections/08_conclusion.tex
\section{Conclusion}
\label{sec:conclusion}

This survey defines agentic artifact creation as a stateful construction process in which observations of an evolving artifact redirect later work.
The process connects an Operational Representation, a Construction Policy, and Runtime Verification.
Their alignment determines whether evidence can localize a failure and direct a feasible edit.
Across the six artifact families, recurring construction challenges extend beyond modality alone.
They include tightly coupled decisions, failures that appear late or without a clear cause, and edits that are too coarse for local repair.
The four principles in Section~\ref{sec:patterns} address these conditions by externalizing commitments, defining control boundaries, linking feedback to feasible repairs, and revalidating affected state.
Their implementation remains domain-specific, and direct generation remains preferable when there is little accepted state to preserve or the added control costs more than regeneration.
The six challenges ask whether this control can survive change.
A system may need to revise one part without breaking distant dependencies, repair flaws it cannot diagnose, or update itself without mistaking a local gain for durable improvement.
It must also learn from a creator without turning transient feedback into a permanent preference, delegate work while keeping decision authority explicit, and judge open-ended outcomes without collapsing them to a single score.
Comparative evidence should report and, where relevant, match model, tool, compute, feedback, and human budgets, pairing delivered-artifact results with trajectory evidence relevant to the claim.
Agentic construction is worth its added complexity when observations improve later decisions enough to justify the control cost; bounded revision and state preservation are important benefits.